\documentclass[preprint,3p,12pt]{elsarticle}
\usepackage{mathrsfs}
\usepackage{amsmath}
\usepackage{stmaryrd}
\usepackage{bbding}
\usepackage{dcolumn}
\usepackage{graphicx}
\usepackage{amsfonts}
\usepackage{amssymb}
\usepackage{psfrag}
\usepackage{wrapfig}
\usepackage{subfigure}
\usepackage{makeidx}
\usepackage{bm}
\usepackage{epsf}
\usepackage{color}
\usepackage{epsfig}
\usepackage{setspace}
\usepackage{graphicx}
\usepackage{epstopdf}
\usepackage{psfrag}
\usepackage{subfigure}
\usepackage{multirow}
\usepackage{diagbox}
\usepackage{verbatim}

\usepackage{makecell}
\usepackage{float}
\usepackage{esint}
\usepackage{comment}
\usepackage{movie15}
\usepackage{hyperref}
\usepackage[ruled,linesnumbered]{algorithm2e}

\newcommand{\mathsym}[1]{{}}
\newcommand{\unicode}[1]{{}}
\begin{document}
	
\title{A Compact Gas-Kinetic Scheme with Scalable Geometric Multigrid Acceleration for Steady-State Computation on 3D Unstructured Meshes}

	\author[XJTU,HKUST1]{Hongyu Liu}
    \ead{hliudv@connect.ust.hk}
	
	\author[XJTU]{Xing Ji\corref{cor1}}
	\ead{xjiad@connect.ust.hk}

	\author[XJTU]{Yunpeng Mao}
	\ead{m691810014@stu.xjtu.edu.cn}
		
	\author[HKUST1]{Yuan Ding}
    \ead{1219948011@qq.com}
	
	\author[HKUST1,HKUST2,HKUST3]{Kun Xu}
	\ead{makxu@ust.hk}

	\address[XJTU]{Shaanxi Key Laboratory of Environment and Control for Flight Vehicle, Xi'an Jiaotong University, Xi'an, China}
	\address[HKUST1]{Department of Mathematics, Hong Kong University of Science and Technology, Clear Water Bay, Kowloon, Hong Kong}
	\address[HKUST2]{Department of Mechanical and Aerospace Engineering, Hong Kong University of Science and Technology, Clear Water Bay, Kowloon, Hong Kong}
	\address[HKUST3]{Shenzhen Research Institute, Hong Kong University of Science and Technology, Shenzhen, China}
	\cortext[cor1]{Corresponding author}

\begin{abstract}

In this paper, we present an advanced high-order compact gas-kinetic scheme (CGKS) for 3D unstructured mixed-element meshes, augmented with a geometric multigrid technique to accelerate steady-state convergence. The scheme evolves cell-averaged flow variables and their gradients on the original mesh. Mesh coarsening employs a two-step parallel agglomeration algorithm using a random hash for cell interface selection and a geometric skewness metric for deletion confirmation, ensuring both efficiency and robustness. For the coarser meshes, first-order kinetic flux vector splitting (KFVS) schemes with explicit or implicit time-stepping are used. The proposed multigrid CGKS is tested across various flow regimes on hybrid unstructured meshes, demonstrating significant improvements. A three-layer V-cycle multigrid strategy, coupled with an explicit forward Euler method on coarser levels, results in a convergence rate up to ten times faster than standard CGKS. In contrast, the implicit lower-upper symmetric Gauss-Seidel (LU-SGS) method offers limited convergence acceleration. Our findings indicate that the explicit multigrid CGKS is highly scalable and effective for large-scale computations, marking a substantial step forward in computational fluid dynamics.

\end{abstract}


\maketitle

\section{Introduction}

In the past few decades, the high-order numerical methods in computational fluid dynamics (CFD) have achieved great success \cite{HUYNH2014209,abgrall2018high}. These methods, such as Weighted Essentially Non-Oscillatory (WENO) \cite{gao2012high}, Discontinuous Galerkin (DG) \cite{shu2003fd-fv-weno-dg-review,shu2016weno-dg-review}, Correction Procedure Via Reconstruction (CPR) \cite{CPR2011,wang2017towards}, and Variational Finite Volume (VFV) \cite{cances2020variational}, have played a crucial role in improving the accuracy of numerical simulations \cite{bhatti2020recent}. They are well-suited for handling complex flow phenomena, including turbulent flows \cite{li2010direct}, shock propagation \cite{romick2017high}, and multi-physics coupling \cite{ferrer2023high}.

In recent years, the high-order compact gas-kinetic scheme(CGKS) \cite{ji2018compact,zhao2023high,zhao2019compact} has developed based on gas-kinetic theory.
CGKS uses the time-dependent distribution function which has an accurate analytical integral solution at the cell interface of the Bhatnagar--Gross--Krook equation \cite{BGK}.
By using the accurate time-dependent distribution function at the cell interface, not only the Navier-Stokes flux functions can be obtained, but also the time accurate macroscopic flow variables will be evaluated. It means while cell-averaged flow variables are being updated under the finite volume framework, the cell-averaged gradients can also be updated by the Gauss-Green theorem.
With the cell-averaged flow variables and their gradients, a Hermite Weighted Essentially Non-Oscillatory (HWENO) \cite{li2021multi} method can be employed for the reconstruction, which can be found in our previous work of third-order CGKS \cite{zhang2023high}. As for temporal discretization, explicit two-stage fourth-order and other multi-stage multi-derivative time marching schemes can be used in CGKS for high-order temporal discretization \cite{li2016twostage}.
Due to the benefits of the more reliable evolution process based on mesoscopic gas kinetic theory, the CGKS has excellent performance on both smooth and discontinuous flow regimes. The CGKS also shows good performance in the regime of unsteady compressible flow, such as in computational aeroacoustics \cite{zhao2019acoustic} and implicit large eddy simulation \cite{ji2021compact}.
In CGKS, the compression factor extends beyond the assumptions of the finite volume framework by determining the presence of discontinuities within the cell for the upcoming time step. Thus, it further improves the robustness of the CGKS when facing strong discontinuity.
 As a result, CGKS also performs well in supersonic and hypersonic flow simulations on the three-dimensional hybrid unstructured mesh, such as YF-17 fighter jet and X-38 type spaceship \cite{ji2021gradient}. In summary, the CGKS demonstrates strong grid adaptability and robustness, effectively handling the complexities of low-quality meshes and pronounced discontinuities.

However, for the practical engineering problems with large-scale simulations, the explicit time marching approach has strong time-step size limitation for steady-state flow simulation. To speed up the convergence, acceleration techniques need to be employed. There are two mainstream ways for steady state acceleration, implicit temporal discretization and multigrid strategy. On one side, the implicit method is usually based on the lower-upper symmetric Gauss-Seidel (LU-SGS) method \cite{antoniadis2017assessment} or the Generalized minimal residual (GMRES) method \cite{yang2019robust,saad1986gmres,wienands2000fourier}. Both methods above use the fully implicit temporal discretization to solve the linear system derived from the original equations through a linearization process. The implicit techniques have also been constructed for GKS and the unified gas kinetic scheme (UGKS) \cite{tan2017time,li2017dualtime,zhu2016implicit}. Due to the highly nonlinear evolution function of CGKS at the cell interface, it is nontrivial to get the approximate or even accurate Jacobian matrix for the cell-averaged gradients in the development of implicit CGKS. Because LU-SGS method has difficulties in dealing with the parallel boundary, it's difficult to apply the large-scale simulation method \cite{WANG201233}. Compared with LU-SGS method, GMRES solves the linear system more accurately and can have a faster convergence rate. More investigations have been done to improve the performance and portability of the GMRES algorithm for unstructured mesh in industrial CFD solvers \cite{NEJAT20082582}. However, due to the inherent sequential properties, GMRES encountered difficulties in achieving high parallel computing performance. Diverse HPC architecture trends also introduce challenges in algorithm migration of GMRES \cite{zhang2023highgmres}. More importantly, preconditioners play a crucial role in influencing the stability and convergence rate of GMRES. Therefore, the selection of suitable preconditioners for GMRES in solving various compressible flow scenarios remains an unresolved and significant question. \cite{John2020kylov}.

On the other hand, the geometric multigrid strategy has been developed for decades and is widely used in the second order schemes \cite{briggs2000multigrid,stuben2001review,wesseling2001geometric,blazek2005comparison}. Meanwhile, a series of p-multigrid methods have been developed for compact methods with multiple degrees of freedom in each cell \cite{ronquist1987spectral,luo2006pmultigrid,bassi2009high,liang2009pmultigrid}. The equations are solved by different orders of accuracy in p-multigrid methods. Each order in p-multigrid is taken as a level of grid in geometric multigrid. The restriction and correct operators used in p-multigrid are the same as geometric-multigrid.

In this paper, an efficient and robust hybrid method of geometric-multigrid and p-multigrid for accelerating the CGKS in steady-state simulation is developed. On the finest level of mesh, the third-order explicit CGKS is employed to evolve both cell-averaged flow variables and cell-averaged gradients. Then the solution of flow variables and residuals are restricted to the coarse-level mesh. To guarantee the solution's accuracy, a force term will be constructed on the right-hand side of the equation. On the coarse levels of mesh, the first-order explicit  kinetic flux vector splitting (KFVS) scheme is used only to update the cell-averaged flow variables. Then the solution will be prolonged back to the finest level. In this paper, a three-layer geometric multigrid is used. Benefiting from the explicit time marching scheme, geometric multigrid CGKS has the advantage of having consistent results between serial and parallel computations. Moreover, the local time-stepping strategy is very easy to apply in the geometric multigrid CGKS.  As a result,
it is convenient to apply geometric multigrid CGKS to large-scale computations.
The computational cost between the explicit and geometric multigrid iteration is compared through the computations for nearly incompressible and supersonic flows under three-dimensional hybrid unstructured meshes, and the results indicate the algorithm this paper developed has a good convergence rate and grid adaptability.

The paper is organized as follows. In Section 2, the basic BGK equation, the finite volume framework and the construction method of GKS on three-dimensional hybrid unstructured mesh will be introduced. In Section 3, the third-order nonlinear spatial reconstruction and limiting procedure will be introduced. In Section 4, the mesh coarsening method and the strategy of the geometric multigrid method will be introduced. In Section 5, numerical examples including both inviscid and viscous flow computations will be given. The last section is conclusion.

\section{High order compact gas kinetic scheme framework}
\subsection{3-D BGK equation}
The Boltzmann equation \cite{cercignani1988boltzmann} describes the evolution of molecules at the mesoscopic scale. It indicates that each particle will either transport with a constant velocity or encounter a two-body collision. The BGK \cite{BGK} model simplifies the Boltzmann equation by replacing the full collision term with a relaxation model.
The 3-D gas-kinetic BGK equation \cite{BGK} is
\begin{equation}\label{bgk}
	f_t+\textbf{u}\cdot\nabla f=\frac{g-f}{\tau},
\end{equation}
where $f=f(\textbf{x},t,\textbf{u},\xi)$ is the gas distribution function, which is a function of space $\textbf{x}$, time $t$, phase space velocity $\textbf{u}$, and internal variable $\xi$.
$g$ is the equilibrium state
and $\tau$ is the collision time, which means an averaged time interval between two sequential collisions.  $g$ is expressed as a Maxwellian distribution function.
Meanwhile, the collision term on the right-hand side (RHS) of Eq.~\eqref{bgk} should satisfy the compatibility condition
\begin{equation*}\label{compatibility}
	\int \frac{g-f}{\tau} \pmb{\psi} \text{d}\Xi=0,
\end{equation*}
where $\pmb{\psi}=(1,\textbf{u},\displaystyle \frac{1}{2}(\textbf{u}^2+\xi^2))^T$,
$\text{d}\Xi=\text{d}u_1\text{d}u_2\text{d}u_3\text{d}\xi_1...\text{d}\xi_{K}$,
$K$ is the number of internal degrees of freedom, i.e.
$K=(5-3\gamma)/(\gamma-1)$ in the 3-D case, and $\gamma$
is the specific heat ratio. The details of the BGK equation can be found in \cite{xu2014directchapter2}.

In the continuous flow regime, distribution function $f$ can be taken as a small-scale expansion of Maxwellian distribution. Based on the Chapman-Enskog expansion \cite{CE-expansion}, the gas distribution function can be expressed as \cite{xu2014directchapter2},
\begin{align*}
	f=g-\tau D_{\textbf{u}}g+\tau D_{\textbf{u}}(\tau
	D_{\textbf{u}})g-\tau D_{\textbf{u}}[\tau D_{\textbf{u}}(\tau
	D_{\textbf{u}})g]+...,
\end{align*}
where $D_{\textbf{u}}={\partial}/{\partial t}+\textbf{u}\cdot \nabla$.
Through zeroth-order truncation $f=g$, the Euler equation can be obtained. The Navier-Stokes (NS) equations,
\begin{equation*}\label{ns-conservation}
	\begin{split}
		\textbf{W}_t+ \nabla \cdot \textbf{F}(\textbf{W},\nabla \textbf{W} )=0,
	\end{split}
\end{equation*}
 can be obtained by taking first-order truncation, i.e.,
\begin{align} \label{ce-ns}
	f=g-\tau (\textbf{u} \cdot \nabla g + g_t),
\end{align}
with $\tau = \mu / p$ and $Pr=1$.

Benefiting from the time-accurate gas distribution function, a time evolution solution at the cell interface is provided by the gas kinetic solver, which is distinguishable from the Riemann solver with a constant solution. This is a crucial point to construct the compact high-order gas kinetic scheme.
\begin{align}\label{point}
	\textbf{W}(\textbf{x},t)=\int \pmb{\psi} f(\textbf{x},t,\textbf{u},\xi)\text{d}\Xi,
\end{align}
and the flux at the cell interface can also be obtained
\begin{equation}\label{f-to-flux}
	\textbf{F}(\textbf{x},t)=
	\int \textbf{u} \pmb{\psi} f(\textbf{x},t,\textbf{u},\xi)\text{d}\Xi.
\end{equation}

\subsection{Finite volume framework}
The boundary of a three-dimensional arbitrary polyhedral cell $\Omega_i$ can be expressed as
\begin{equation*}
	\partial \Omega_i=\bigcup_{p=1}^{N_f}\Gamma_{ip},
\end{equation*}
where $N_f$ is the number of cell interfaces for cell $\Omega_i$.
$N_f=4$ for tetrahedron, $N_f=5$ for prism and pyramid, $N_f=6$ for hexahedron.
The semi-discretized form of the finite volume method for conservation laws can be written as
\begin{equation}\label{semidiscrete}
	\frac{\text{d} \textbf{W}_{i}}{\text{d}t}=\mathcal{L}(\textbf{W}_i)=-\frac{1}{\left| \Omega_i \right|} \sum_{p=1}^{N_f} \int_{\Gamma_{ip}}
	\textbf{F}(\textbf{W}(\textbf{x},t))\cdot\textbf{n}_p \text{d}s,
\end{equation}
with
\begin{equation*}\label{f-to-flux-in-normal-direction}
	\textbf{F}(\textbf{W}(\textbf{x},t))\cdot \textbf{n}_p=\int\pmb{\psi}  f(\textbf{x},t,\textbf{u},\xi) \textbf{u}\cdot \textbf{n}_p \text{d}\Xi,
\end{equation*}
where $\textbf{W}_{i}$ is the cell averaged values over cell $\Omega_i$, $\left|
\Omega_i \right|$ is the volume of $\Omega_i$, $\textbf{F}$ is the interface fluxes, and $\textbf{n}_p=(n_1,n_2,n_3)^T$ is the unit vector representing the outer normal direction of $\Gamma_{ip}$.
Through the iso-parametric transformation,
the Gaussian quadrature points can be determined and $\textbf{F}_{ip}(t)$ can be approximated by the numerical quadrature
\begin{equation*}\label{fv-3d-general-quadrature}
	\int_{\Gamma_{ip}}
	\textbf{F}(\textbf{W}(\textbf{x},t))\cdot\textbf{n}_p \text{d}s =  S_{i,p} \sum_{k=1}^{M} \omega_k
	\textbf{F}(\textbf{x}_{p,k},t)\cdot\textbf{n}_p,
\end{equation*}
where $S_{i,p}$ is the area of $\Gamma_{ip}$. Through the iso-parametric transformation,
in the current study the linear element is considered.
To meet the requirement of a third-order spatial accuracy,
three Gaussian points are used for a triangular face and four Gaussian points are used for a quadrilateral face.
In the computation, the fluxes are obtained under the local coordinates.
The details can be found in \cite{ji2021gradient,pan2020high,JI2024112590}.

\subsection{Gas kinetic solver}
In order to obtain the numerical flux at the cell interface, the integration solution based on the BGK equation is used
\begin{equation}\label{integral1}
	f(\textbf{x},t,\textbf{u},\xi)=\frac{1}{\tau}\int_0^t g(\textbf{x}',t',\textbf{u},\xi)e^{-(t-t')/\tau}\text{d}t'
	+e^{-t/\tau}f_0(\textbf{x}-\textbf{u}t,\textbf{u},\xi),
\end{equation}
where $\textbf{x}=\textbf{x}'+\textbf{u}(t-t')$ is the particle trajectory. $f_0$ is the initial gas distribution function, $g$ is the corresponding
equilibrium state in the local space and time.
This integration solution describes the physical picture of the particle evolution. Starting with an initial state $f_0$, the particle will take free transport with a probability of $e^{-\Delta t/\tau}$. Otherwise, it will suffer a series of collisions. The effect of collisions is driving the system to the local Maxwellian distribution $g$, and the particles from the equilibrium propagate along the characteristics with a surviving probability of $e^{-(t-t')/\tau}$.
The components of the numerical fluxes at the cell interface can be categorized as the upwinding free transport from $f_0$ and the central difference from the integration of the equilibrium distribution.

In order to construct a time-evolving gas distribution function at a cell interface,
the following notations are introduced first
\begin{align*}
	a_{x_i} \equiv  (\partial g/\partial x_i)/g=g_{x_i}/g,
	A \equiv (\partial g/\partial t)/g=g_t/g,
\end{align*}
where $g$ is the equilibrium state.  The partial derivatives $a_{x_i}$ and $A$, denoted by $s$,
have the form of
\begin{align*}
	s=s_j\psi_j =s_{1}+s_{2}u_1+s_{3}u_2+s_{4}u_3
	+s_{5}\displaystyle \frac{1}{2}(u_1^2+u_2^2+u_3^2+\xi^2).
\end{align*}
The initial gas distribution function in Eq.~\eqref{integral1} can be modeled as
\begin{equation*}
	f_0=f_0^l(\textbf{x},\textbf{u})\mathbb{H} (x_1)+f_0^r(\textbf{x},\textbf{u})(1- \mathbb{H}(x_1)),
\end{equation*}
where $\mathbb{H}(x_1)$ is the Heaviside function. Here $f_0^l$ and $f_0^r$ are the
initial gas distribution functions on the left and right sides of a cell
interface, which can be fully determined by the initially
reconstructed macroscopic variables. The first-order
Taylor expansion for the gas distribution function in space around
$\textbf{x}=\textbf{0}$ can be expressed as
\begin{align}\label{flux-3d-1}
	f_0^k(\textbf{x})=f_G^k(\textbf{0})+\frac{\partial f_G^k}{\partial x_i}(\textbf{0})x_i
	=f_G^k(\textbf{0})+\frac{\partial f_G^k}{\partial x_1}(\textbf{0})x_1
	+\frac{\partial f_G^k}{\partial x_2}(\textbf{0})x_2
	+\frac{\partial f_G^k}{\partial x_3}(\textbf{0})x_3,
\end{align}
for $k=l,r$.
According to Eq.~\eqref{ce-ns}, $f_{G}^k$ has the form
\begin{align}\label{flux-3d-2}
	f_{G}^k(\textbf{0})=g^k(\textbf{0})-\tau(u_ig_{x_i}^{k}(\textbf{0})+g_t^k(\textbf{0})),
\end{align}
where $g^k$ is the equilibrium state with the form of a Maxwell distribution.
$g^k$ can be fully determined from the
reconstructed macroscopic variables $\textbf{W}
^l, \textbf{W}
^r$ at the left and right sides of a cell interface
\begin{align}\label{get-glr}
	\int\pmb{\psi} g^{l}\text{d}\Xi=\textbf{W}
	^l,\int\pmb{\psi} g^{r}\text{d}\Xi=\textbf{W}
	^r.
\end{align}
Substituting Eq.~\eqref{flux-3d-1} and Eq.~\eqref{flux-3d-2} into Eq.~\eqref{integral1},
the kinetic part for the integral solution can be written as
\begin{equation}\label{dis1}
	\begin{aligned}
		e^{-t/\tau}f_0^k(-\textbf{u}t,\textbf{u},\xi)
		=e^{-t/\tau}g^k[1-\tau(a_{x_i}^{k}u_i+A^k)-ta^{k}_{x_i}u_i],
	\end{aligned}
\end{equation}
where the coefficients $a_{x_1}^{k},...,A^k, k=l,r$ are defined according
to the expansion of $g^{k}$.
After determining the kinetic part
$f_0$, the equilibrium state $g$ in the integral solution
Eq.~\eqref{integral1} can be expanded in space and time as follows
\begin{align}\label{equli}
	g(\textbf{x},t)= g^{c}(\textbf{0},0)+\frac{\partial  g^{c}}{\partial x_i}(\textbf{0},0)x_i+\frac{\partial  g^{c}}{\partial t}(\textbf{0},0)t,
\end{align}
where $ g^{c}$ is the Maxwellian equilibrium state located at an interface.
Similarly, $\textbf{W}^c$ are the macroscopic flow variables for the determination of the
equilibrium state $ g^{c}$
\begin{align}\label{compatibility2}
	\int\pmb{\psi} g^{c}\text{d}\Xi=
	\int_{u>0}\pmb{\psi} g^{l}\text{d}\Xi+
	\int_{u<0}\pmb{\psi} g^{r}\text{d}\Xi=\textbf{W}^c.
\end{align}
Substituting Eq.~\eqref{equli} into Eq.~\eqref{integral1}, the collision part in the integral solution
can be written as
\begin{equation}\label{dis2}
	\begin{aligned}
		\frac{1}{\tau}\int_0^t
		g&(\textbf{x}',t',\textbf{u},\xi)e^{-(t-t')/\tau}\text{d}t'
		=C_1 g^{c}+C_2 a_{x_i}^{c} u_i g^{c} +C_3 A^{c} g^{c} ,
	\end{aligned}
\end{equation}
where the coefficients
$a_{x_i}^{c},A^{c}$ are
defined from the expansion of the equilibrium state $ g^{c}$. The
coefficients $C_m, m=1,2,3$ in Eq.~\eqref{dis2}
are given by
\begin{align*}
	C_1=1-&e^{-t/\tau}, C_2=(t+\tau)e^{-t/\tau}-\tau, C_3=t-\tau+\tau e^{-t/\tau}.
\end{align*}
The coefficients in Eq.~\eqref{dis1} and Eq.~\eqref{dis2}
can be determined by the spatial derivatives of macroscopic flow
variables and the compatibility condition as follows
\begin{align}\label{co}
	&\langle a_{x_1}\rangle =\frac{\partial \textbf{W} }{\partial x_1}=\textbf{W}_{x_1},
	\langle a_{x_2}\rangle =\frac{\partial \textbf{W} }{\partial x_2}=\textbf{W}_{x_2},
	\langle a_{x_3}\rangle =\frac{\partial \textbf{W} }{\partial x_3}=\textbf{W}_{x_3},\nonumber\\
	&\langle A+a_{x_1}u_1+a_{x_2}u_2+a_{x_3}u_3\rangle=0,
\end{align}
where $\left\langle ... \right\rangle$ are the moments of a gas distribution function defined by
\begin{align}\label{co-moment}
	\langle (...) \rangle  = \int \pmb{\psi} (...) g \text{d} \Xi .
\end{align}

For the viscous flow, the physical relaxation time is determined by
\begin{align*}
	\tau=\mu/p,
\end{align*}
where $\mu$ is the dynamic viscosity coefficient and $p$ is the pressure at the cell interface.
In order to properly capture the unresolved shock structure, additional numerical dissipation is needed.
The physical collision time $\tau$ in the exponential function part can be replaced by a numerical collision time $\tau_n$,
\begin{align*}
	\tau_n=\frac{\mu}{p}+C \displaystyle|\frac{p_l-p_r}{p_l+p_r}|\Delta t,
\end{align*}
where $p_l$ and $p_r$ denote the pressure on the left and right
sides of the cell interface.
The inclusion of the pressure jump term is to implement the non-equilibrium transport mechanism in the flux function to mimic the physical process in the shock layer.
For the
inviscid flow, the collision time $\tau_n$ is modified as
\begin{align*}
	\tau_n=\varepsilon \Delta t+C\displaystyle|\frac{p_l-p_r}{p_l+p_r}|\Delta
	t,
\end{align*}
where $\varepsilon=0.01$ and $C=1$.
Then substitute Eq.~\eqref{dis1} and Eq.~\eqref{dis2} into Eq.~\eqref{integral1} with $\tau$ and $\tau_n$, the final second-order time-dependent gas distribution function becomes
\begin{align}\label{2nd-flux}
	f(\textbf{0},t,\textbf{u},\xi)
	=&(1-e^{-t/\tau_n}) g^{c}+[(t+\tau)e^{-t/\tau_n}-\tau]a_{x_i}^{c}u_i g^{c}\nonumber
	+(t-\tau+\tau e^{-t/\tau_n})A^{c}  g^{c}\nonumber\\
	+&e^{-t/\tau_n}g^l[1-(\tau+t)a_{x_i}^{l}u_i-\tau A^l]H(u_1)\nonumber\\
	+&e^{-t/\tau_n}g^r[1-(\tau+t)a_{x_i}^{r}u_i-\tau A^r] (1-H(u_1)).
\end{align}

\subsection{Direct evolution of the cell averaged first-order spatial derivatives} \label{slope-section}

As shown in Eq.~\eqref{2nd-flux}, a time evolution solution at a cell interface is provided by the gas-kinetic solver, which is distinguished from the Riemann solvers with a constant solution.
By recalling Eq.~\eqref{point}, the conservative variables at the Gaussian point  $\textbf{x}_{p,k}$ can be updated through the moments $\pmb{\psi}$
of the gas distribution function,
\begin{equation*}\label{point-interface}
	\begin{aligned}
		\textbf{W}_{p,k}(t^{n+1})=\int \pmb{\psi} f^n(\textbf{x}_{p,k},t^{n+1},\textbf{u},\xi) \text{d}\Xi,~ k=1,...,M.
	\end{aligned}
\end{equation*}

Then, the cell-averaged first-order derivatives within each element at $t^{n+1}$ can be evaluated based on the divergence theorem,

\begin{equation*}\label{gauss-formula}
	\begin{aligned}
		\nabla \overline{W}^{n+1} \left| \Omega \right|
		&=\int_{\Omega} \nabla \overline{W}(t^{n+1}) \text{d}V
		=\int_{\partial \Omega} \overline{W}(t^{n+1}) \textbf{n} \text{d}S
		= \sum_{p=1}^{N_f}\sum_{k=1}^{M_p} \omega_{p,k} W^{n+1}_{p,k} \textbf{n}_{p,k} \Delta S_p,
	\end{aligned}
\end{equation*}
where $\textbf{n}_{p,k}=((n_{1})_{p,k},(n_{2})_{p,k},(n_{3})_{p,k})$ is the outer unit normal direction at each Gaussian point $\textbf{x}_{p,k}$.

\section{Spatial reconstruction}
The 3rd-order compact reconstruction \cite{ji2020hweno}  is adopted here with cell-averaged values and cell-averaged first-order spatial derivative. To maintain both shock-capturing ability and robustness, WENO weights and gradient compression factor (CF) are used \cite{ji2021gradient}. Further improvement has been made to make reconstruction simple and more robust \cite{zhang2023high}. Only one large stencil and one sub-stencil are involved in the WENO procedure.
\subsection{3rd-order compact reconstruction for large stencil}
Firstly, a linear reconstruction approach is presented. To achieve a third-order accuracy in space, a quadratic polynomial $p^2$ is constructed as follows
\begin{equation*}
	\begin{aligned}
		p^2&=a_0+a_1(x-x_0)+a_2(y-y_0)+a_3(z-z_0)\\
		&+a_4(x-x_0)^2+a_5(y-y_0)^2+a_6(z-z_0)^2\\
		&+a_7(x-x_0)(y-y_0)+a_8(y-y_0)(z-z_0)+a_9(x-x_0)(z-z_0),
	\end{aligned}
\end{equation*}

The $p^2$ on $\Omega_0$  is constructed on the compact stencil $S$ including $\Omega_0$ and all von Neumann neighbors $\Omega_m$($m=1,\cdots,N_f$, where $N_f=6$ for hexahedron cell or $N_f=5$  for triangular prism) of it. The cell averages $\overline{Q}$ on $\Omega_0$ and $\Omega_m$ together with cell averages of space partial derivatives $\overline{Q}_x,\overline{Q}_y $ and $\overline{Q}_z$ on $\Omega_m$ are used to obtain $p^2$.

The polynomial $p^2$ is required to exactly satisfy cell averages over both $\Omega_0$ and $\Omega_m$ ($m=1,\cdots,N_f$)
\begin{equation*}
	\iiint_{\Omega_0} p^2 \text{d}V = \overline{Q}_0|\Omega_0|,\iiint_{\Omega_m} p^2 \text{d}V = \overline{Q}_m|\Omega_m| ,
\end{equation*}
with the following condition satisfied in a least-square sense
\begin{equation*}
	\begin{aligned}
		\iiint_{\Omega_m}\frac{\partial}{\partial x} p^2 \text{d}V = \left(\overline{Q}_x\right)_m|\Omega_m|\\
		\iiint_{\Omega_m}\frac{\partial}{\partial y} p^2 \text{d}V = \left(\overline{Q}_y\right)_m|\Omega_m|\\
		\iiint_{\Omega_m}\frac{\partial}{\partial z} p^2 \text{d}V = \left(\overline{Q}_z\right)_m|\Omega_m|.\\
	\end{aligned}
\end{equation*}
To solve the above system, the constrained least-square method is used.
\subsection{Green-Gauss reconstruction for the sub stencil}
The classical Green-Gauss reconstruction \cite{shima2013green} with only cell-averaged values is adopted to provide the linear polynomial $p^1$ for the sub stencil.
\begin{equation*}
	p^1=\overline{Q}+\boldsymbol{x}\cdot \sum_{m=1}^{N_f}\frac{\overline{Q}_m+\overline{Q}_0}{2}S_m\boldsymbol{n} _m,
\end{equation*}
where $S_m$ is the area of the cell's surface and $\boldsymbol{n}_m$ is the surface's normal vector. In most cases, Green-Gaussian reconstruction has only first-order precision.
\subsection{Gradient compression Factor}
The CF was first proposed in \cite{ji2021gradient}. Here several improvements have been made: there is no $\epsilon$ in the improved expression of CF; the difference of Mach number is added to improve the robustness under strong rarefaction waves. Denote $\alpha_i \in[0,1]$ as gradient compression factor at targeted cell $\Omega_i$
\begin{equation*}
	\alpha_i = \prod_{p=1}^m\prod_{k=0}^{M_p}\alpha_{p,k},
\end{equation*}
where $\alpha_{p,k}$ is the CF obtained by the $k$th Gaussian point at the interface $p$ around cell $\Omega_i$, which can be calculated by
\begin{equation*}
	\begin{aligned}
		&\alpha_{p,k}=\frac{1}{1+A^2}, \\
		&A=\frac{|p^l-p^r|}{p^l} +\frac{|p^l-p^r|}{p^r}+(\text{Ma}^{l}_n-\text{Ma}^{r}_n)^2+(\text{Ma}^{l}_t-\text{Ma}^{r}_t)^2,
	\end{aligned}
\end{equation*}
where $p$ is pressure, $\text{Ma}_n$ and $\text{Ma}_t$ are the Mach numbers defined by normal and tangential velocity, and superscripts $l,r$ denote the left and right values of the Gaussian points.

Then, the updated slope is modified by
\begin{equation*}
	\widetilde{\overline{\nabla \boldsymbol{W}}}_i^{n+1} = \alpha_i\overline{\nabla \boldsymbol{W}}_i^{n+1},
\end{equation*}
and the Green-Gauss reconstruction is modified as
\begin{equation*}
	p^1=\overline{Q}+\alpha \boldsymbol{x}\cdot \sum_{m=1}^{N_f}\frac{\overline{Q}_m+\overline{Q}_0}{2}S_m\boldsymbol{n} _m.
\end{equation*}
\subsection{Non-linear WENO weights}
In order to deal with discontinuity, the idea of multi-resolution WENO reconstruction is adopted \cite{ji2021gradient,zhu2020new}.
Here only two polynomials are chosen
\begin{equation*}
	P_2=\frac{1}{\gamma_2}p^2-\frac{\gamma_1}{\gamma_2}p^1  ,P_1 =p^1.
\end{equation*}
where $\gamma_1=\gamma_2=0.5$. So the quadratic polynomial $p^2$ can be written as
\begin{equation}\label{weno-linear}
	p^2 = \gamma_1P_1 + \gamma_2 P_2.
\end{equation}
Then, we can define the smoothness indicators
\begin{equation*}
	\beta_{j}=\sum_{|\alpha|=1}^{r_{j}}\Omega^{\frac{2}{3}|\alpha|-1} \iiint_{\Omega}\left(D^{\alpha} p^{j}(\mathbf{x})\right)^{2} \mathrm{~d} V,
\end{equation*}
where $\alpha$ is a multi-index and $D$ is the derivative operator, $r_1=1,r_2=2$. Special care is given for $\beta_1$ for better robustness
\begin{equation*}
	\beta_1=\min(\beta_{1,\text{Green-Gauss}},\beta_{1,\text{least-square}}),
\end{equation*}
where $\beta_{1,\text{Green-Gauss}}$ is the smoothness indicator defined by Green-Gauss reconstruction, and $\beta_{1,\text{least-square}}$ is the smoothness indicator defined by second-order least-square reconstruction. Then, the smoothness indicators $\beta_i$ are non-dimensionalized by
\begin{equation*}
	\tilde{\beta}_i=\frac{\beta_i}{Q_0^2+\beta_1+10^{-40}}.
\end{equation*}
The nondimensionalized global smoothness indicator $\tilde{\sigma}$ can be defined as
\begin{equation*}
	\tilde{\sigma}=\left|\tilde{\beta}_{1}-\tilde{\beta}_{0}\right|^{}.
\end{equation*}
Therefore, the corresponding non-linear weights are given by
\begin{equation*}
	\tilde{\omega}_{m}=\gamma_{m}\left(1+\left(\frac{\tilde{\sigma}}{\epsilon+\tilde{\beta}_{m}}\right)^{2}\right),\epsilon=10^{-5},
\end{equation*}
\begin{equation*}
	\bar{\omega}_{m}=\frac{\tilde{\omega}_{m}}{\sum \tilde{\omega}_{m}}, m=1,2.
\end{equation*}
Replacing $\gamma_m$ in equation (\ref{weno-linear}) by $\bar{\omega}_{m}$ , the final non-linear reconstruction can be obtained
\begin{equation*}
	R(\boldsymbol{x})=\bar{\omega}_2P_2+\bar{\omega}_1 P_1.
\end{equation*}
The desired non-equilibrium states at Gaussian points become
\begin{equation*}
	Q_{p, k}^{l, r}=R^{l, r}\left(\boldsymbol{x}_{p, k}\right), \left(Q_{x_{i}}^{l, r}\right)_{p, k}=\frac{\partial R^{l, r}}{\partial x_{i}}\left(\boldsymbol{x}_{p, k}\right).
\end{equation*}

\section{Mesh coarsening algorithm and geometric multigrid strategy}
In this section, the geometric multigrid strategy is introduced. To construct the geometric multigrid strategy with nice process-level parallel computation performance, firstly, a mesh coarsening algorithm which can guarantee a fast enough speed of the coarse mesh generating is introduced. Then, the implementation of the geometric multigrid and p-multigrid CGKS is introduced. Finally, the parallel communication strategy of geometric multigrid CGKS is introduced and the convergence histories under different conditions of parallel computation are shown to indicate the potential of the geometric multigrid CGKS in large-scale parallel computation.
\subsection{Mesh coarsening algorithm}
Since the number of cells in the large-scale simulation's mesh is usually very large, it is important to design an efficient algorithm for mesh coarsening. In this paper, a fast parallel mesh agglomeration algorithm is developed.

In this section, the face removal algorithm \cite{zhang20133d} is introduced first. To implement boundary conditions easily and transfer the information at the parallel interface efficiently, the boundary face and parallel interface will not be deleted. Considering a face with its left cell and right cell, its hash value is
\begin{equation} \label{hash value}
	h_f=[23(N_l+N_r)+N_lN_r]\%N_f
\end{equation}
where $h_f$ is the face's hash value, $N_l$ is the index of the face's left cell, $N_r$ is the index of the face's right cell, $N_f$ is the total interior face number of the mesh.  For example, if the face's left cell ID is 10, the face's right cell ID is 12, and the total face number of the mesh is 100, then the hash value of the face will be 5.

With the face's hash value calculation method, interior faces will be iterated and selected for the collection of faces to be deleted. When iterating at a face, the face's hash value will be calculated. Then, whether the hash value is already inserted in the built hash table will be determined. If the face's hash value has already been inserted in the hash table, the face will not be selected for the collection.

\begin{figure}[htp]	
	\centering
	\includegraphics[width=0.35\textwidth]
	{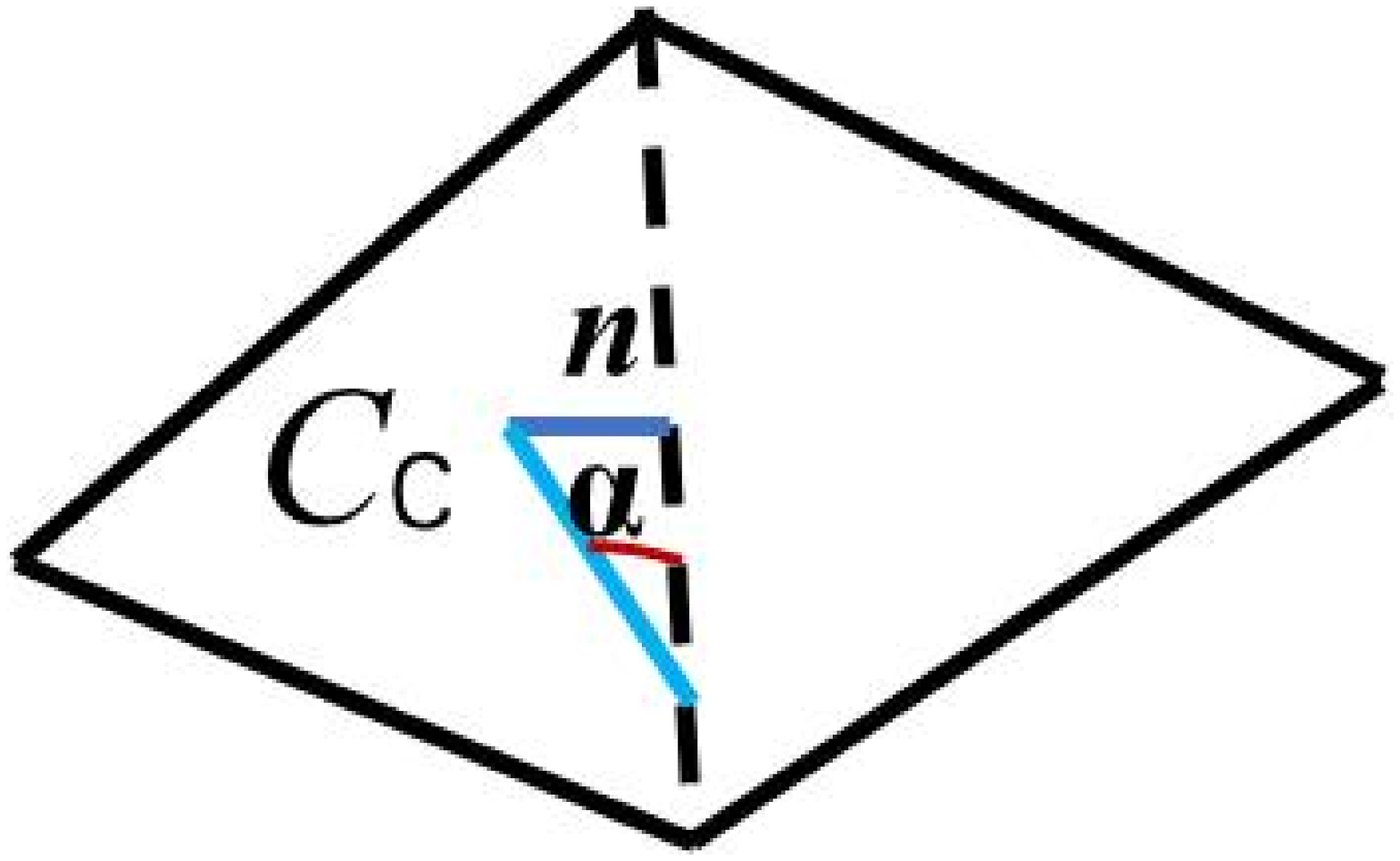}
	\caption{\label{skewsness}
		Skews angle factor .}
\end{figure}

To guarantee coarse meshes' quality, the skewness factor will be introduced to indicate the quality of the mesh. When a face in the above collection is selected to be deleted, the virtual center of the cell which will be constructed by the face's left and right cells is
\begin{equation} \label{virtual center}
	C_c=\frac{V_lC_l+V_rC_r}{V_l+V_r}
\end{equation}
where $V_l$ is the volume of the left cell, $C_l$ is the center of the left cell, $V_r$ is the volume of the right cell, and $C_r$ is the center of the right cell.
The angle between the unit normal vector of the face and the vector pointed from the virtual cell center to the face center is
\begin{equation} \label{skewness factor}
	\sin \alpha=\frac{d\cdot n}{|n|}
\end{equation}
where $\alpha$ is the skewness angle representing the skewness factor, d is the distance vector pointed from the virtual cell center to the face center, and n is the face's unit normal vector shown in Fig.~\ref{skewsness}.
If the angle is smaller than a value that has been set previously, which means the meshes' quality is bad, the face will not be deleted.
To sum up, the cell merge algorithm can be summarized as Algorithm.~\ref{coarsening algorithm} shows.

\begin{algorithm}\label{coarsening algorithm}
	\caption{Cell coarsening method}
	\label{cell-merge}
	Initialization: the collection of faces to be deleted $V_d$, the collection's volume at begin is zero.\;
	\For{j=1,..,$N_f$}
	{
		
		Calculate the $face_j$ 's hash value by Eq.~\eqref{hash value}\;
		If the hash value hasn't appeared in the previous hash table, insert the face into the collection.
	}
	\For{i=1,..,$N_i$}
	{
		where $N_i$ is the element in the collection of faces to be deleted\;
		Calculate the skewness of the face and its neighbor cells by Eq.~\eqref{skewness factor}\;
		If the skewness factor is bigger than the limit, then merge the left and right cells of the face\;
	}
\end{algorithm}	

After coarsening the original mesh for the first time, the geometric variables of the coarse cell should be recalculated. The volume of the coarse cell is
\begin{equation}
	V_c=\sum_{i=0}^n V_i
\end{equation}
where $V_c$ is the volume of the coarse cell, $V_i$ is the volume of the fine cell constructing the coarse cell. The center of the cell is
\begin{equation}
	C_c=\frac{\sum_{i=0}^n V_iC_i}{V_c}
\end{equation}

\subsection{Geometric multigrid and p-multigrid algorithm for third-order compact gas kinetic scheme}
 In this section, a hybrid method of geometric multigrid  and p-multigrid method is employed in high-order compact gas kinetic scheme. For the finest mesh, the third-order gas kinetic scheme solver is used as the pre-smoother and post-smoother \cite{mavriplis1990multigrid}. For the other coarse meshes, the first-order KFVS solver is employed to be the pre-smoother and post-smoother. For the finest mesh, the semi-discretization equation can be written as
 \begin{equation}\label{smo}
 	W_h^{n+1}=W_h^{n}+\frac{\delta V}{\delta t}\sum_{j} Res(W_h^n)
 \end{equation}
where $W_h^{n+1}$ is the cell average conservative variable at step $n+1$, $W_h^{n}$ is the cell average conservative variable at step $n$, $Res(W_h^n)$ is the residual calculated at step $n$ using cell average conservative variable $W_h^n$, $h$ means fine level, and $j$ means the $j^{th}$ face of the current cell.

\begin{figure}[htp]
	\centering
	\includegraphics[width=1.0\textwidth]
	{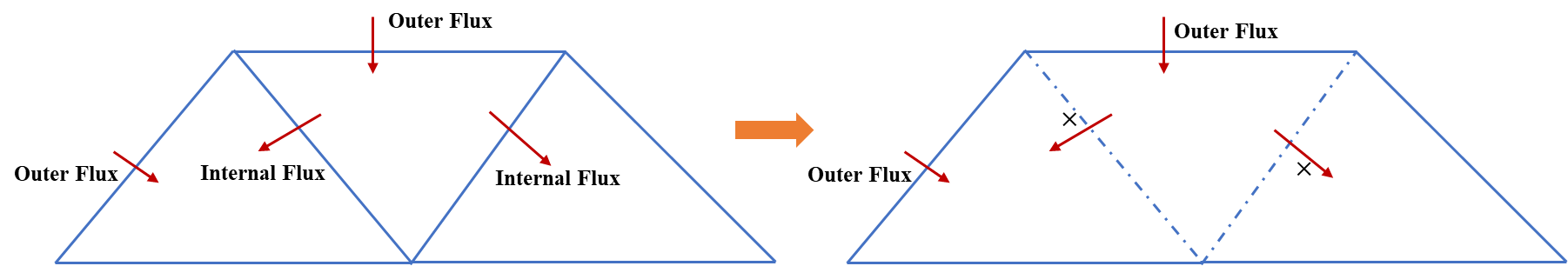}
	\caption{\label{Forcing term}
		Physical picture of the flux relationship.}
\end{figure}

Having cell average value updated once means doing smooth once. After smoothing cell average value, residual used for restricting the coarse meshes residual can be updated by Eq. \eqref{smo}. Cell average conservative variable  is restricted to coarse mesh by
\begin{equation}
	W_{2h}^0=\frac{\sum_{i\in {cell2h}} V_{ifc}W_{ifc}}{V_{2h}}
\end{equation}
where $W_{2h}^0$ is the initial conservative variable in the first step of coarse mesh calculation, $V_{2h}$ is the coarse cell's volume, $W_{ifc}$ is the $i^{th}$ fine cell's conservative variable.
Residual restricted from fine mesh to coarse mesh is  guided by
\begin{equation}
	Res_{2h}^{*}=\sum_{i\in{cell2h}} Res_{ifc}^{n+1}
\end{equation}
where $Res_{2h}^{*}$ is the residual restricted from fine mesh, $Res_{ifc}^{n+1}$ is the $i^{th}$ fine cell's residual. It's important to notice that the physical meaning of the residual restriction process is the total flux of the coarse cell using the flux calculated from fine mesh. It is shown in Fig.~\ref{Forcing term}.

In order to guarantee the final result can still keep high order accuracy, a forcing term is given by
\begin{equation}
	F=Res_{2h}^{*}-Res_{2h}(W_{2h}^{0})
\end{equation}

For the coarse mesh, the macroscopic variable is updated using forcing term by
\begin{equation}
	W_{2h}^{k+1}=W_{2h}^{k}+\frac{\delta t}{\delta V}(Res_{2h}^{k}+F)
\end{equation}

After updating the macroscopic variable on coarse mesh with some explicit steps, a correction process can be used to correct fine mesh solution by
\begin{equation}
	W_h^{*}=W_h^{n+1}+(W_{2h}^{k+1}-W_{2h}^{0})
\end{equation}
It is important to note that the sum of the number of pre-smoothing and post-smoothing should not be less than 2. In this paper, without specified claiming the pre-smooth number is set to 1, and the post-smooth number is set to 0.

\begin{figure}[htp]
	\centering
	\includegraphics[width=0.8\textwidth]
	{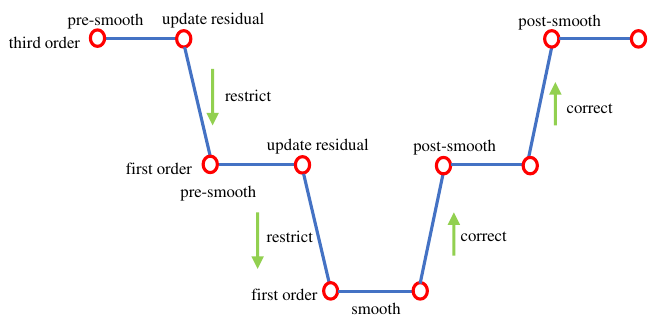}
	\caption{\label{algorithm pic}
		Hybrid method of geometric multigrid and p-multigrid.}
\end{figure}

A three level V cycle geometric grid algorithm's picture is given by Fig.~\ref{algorithm pic}.

\subsection{Implicit time marching scheme on coarse mesh}
In this section, we further implement the lower-upper symmetric Gauss-Seidel implicit relaxation on the coarse mesh. By comparing it with the explicit time marching scheme, the more efficient way for the time marching scheme of the geometric multigrid CGKS is explored.
The implicit backward Euler method is used as the time discretization method, which is introduced as follows.

By using first-order backward Euler as time discretization method, the conservation laws become
\begin{equation}
	\left|\Omega_i\right| \frac{{\overline{\mathbf{U}_i}}^{n+1}-{\overline{\mathbf{U}_i}}^n}{\Delta t}+\oint_{\partial \Omega}\left(\mathbf{F}_n^{n+1}-\mathbf{F}_n^n\right) \mathrm{d} s=-\oint_{\partial \Omega} \mathbf{F}_n^n \mathrm{~d} s:=\mathbf{R}_i^n
\end{equation}
where $\mathbf{F}_n=\mathbf{F} \cdot \mathbf{n}$, $\mathbf{n}$ is the outer unit normal direction for each interface, and $\mathbf{F}_n^n$ is the explicit flux at $t^{n}$. Denote $\Delta \mathbf{U}=\overline{\mathbf{U}}^{n+1}-\overline{\mathbf{U}}^n$, the equation is simplified to be

\begin{equation}\label{semi implicit form}
	\left|\Omega_i\right| \frac{\Delta \mathbf{U}_i}{\Delta t}+\oint_{\partial \Omega}\left(\mathbf{F}_n^{n+1}-\mathbf{F}_n^n\right) \mathrm{d} s=\mathbf{R}_i^n
\end{equation}
Linearize the term by first-order Taylor expansion
\begin{equation}
	\mathbf{F}_n^{n+1}-\mathbf{F}_n^n \approx\left(\frac{\partial \mathbf{F}_n}{\partial \mathbf{U}}\right)^n\left(\mathbf{U}^{n+1}-\mathbf{U}^n\right)=\mathbf{A}^n \Delta \mathbf{U}
\end{equation}
where $\mathbf{A}=\left(\frac{\partial \mathbf{F}_n}{\partial \mathbf{U}}\right)=\left(\frac{\partial(\mathbf{F} \cdot \mathbf{n})}{\partial \mathbf{U}}\right)$ is the Jacobian matrix of the normal flux. The Jacobian matrix of the first-order L-F flux is adopted, which is given by
\begin{equation}
	\begin{aligned}
		& (\mathbf{A} \Delta \mathbf{U})_{i, p}=\mathbf{A}_{i, p}^{+} \Delta \mathbf{U}_i+\mathbf{A}_{i, p}^{-} \Delta \mathbf{U}_{i, p} \\
		& \mathbf{A}^{ \pm}=\frac{1}{2}(\mathbf{A} \pm \lambda \mathbf{I})
	\end{aligned}
\end{equation}
where $\Delta \mathbf{U}_{i, p}$ is the increment of the conservative variable in the neighboring cell sharing the same interface $\Gamma_{i p}$. In computation, the modified spectral radius $\lambda$ is adopted by
\begin{equation}
	\lambda=\left|U_n\right|+c+2 \mu \frac{S}{|\Omega|}
\end{equation}

Then we have
\begin{equation}
	\oint_{\partial \Omega} \mathbf{A}^n \Delta \mathbf{U d} s=\sum_{p=1}^{N_f}\left(\mathbf{A}_i^{+} \Delta \mathbf{U}_i+\mathbf{A}_{i, p}^{-} \Delta \mathbf{U}_{i, p}\right) S_{i, p}
\end{equation}
Substitute it into Eq. \eqref{semi implicit form}, we have
\begin{equation}\label{simplify implicit}
	\Delta \mathbf{U}_i\left[\frac{\left|\Omega_i\right|}{\Delta t}+\frac{1}{2} \sum_{p=1}^{N_f}(\lambda S)_{i, p}\right]+\sum_{p=1}^{N_f} \mathbf{A}_{i, p}^{-} \Delta \mathbf{U}_{i, p} S_{i, p}=\mathbf{R}_{i, j}^n
\end{equation}
where $\lambda_{i+1 / 2, j}=\operatorname{Max}\left(\lambda_i, \lambda_{i, p}\right)$.

The delta flux $\mathbf{A}_{i, p}^{-} \Delta \mathbf{U}_{i, p}$ is approximated as

\begin{equation}
	\mathbf{A}_{i, p}^{-} \Delta \mathbf{U}_{i, p} \approx \Delta\left(\mathbf{F}_n\right)_{i, p}^{-}=\mathbf{F}_n^{-}\left(\mathbf{U}_{i, p}^n+\Delta \mathbf{U}_{i, p}\right)-\mathbf{F}_n^{-}\left(\mathbf{U}_{i, p}^n\right)
\end{equation}
The local time step for each cell is used for both explicit and implicit time marching schemes. Rewrite Eq. \eqref{simplify implicit}
 as
 \begin{equation}\label{final implicit form}
 	\Delta \mathbf{U}_i\left[\frac{\left|\Omega_i\right|}{\Delta t_i}+\frac{1}{2} \sum_{p=1}^{N_f}(\lambda S)_{i, p}\right]+\sum_{p=1}^{N_f}\left[\mathbf{F}_n^{-}\left(\mathbf{U}_{i, p}^n+\Delta \mathbf{U}_{i, p}\right)-\mathbf{F}_n^{-}\left(\mathbf{U}_{i, p}^n\right)\right]=\mathbf{R}_{i, j}^n
 \end{equation}
where $\Delta t_i$ is the local time step. Eq. \eqref{final implicit form} is the final form of the linear system. In this paper, a matrix-free point relaxation method is adopted to solve the linear system.

\begin{figure}[htp]	
	\centering	
	\includegraphics[height=0.25\textwidth]{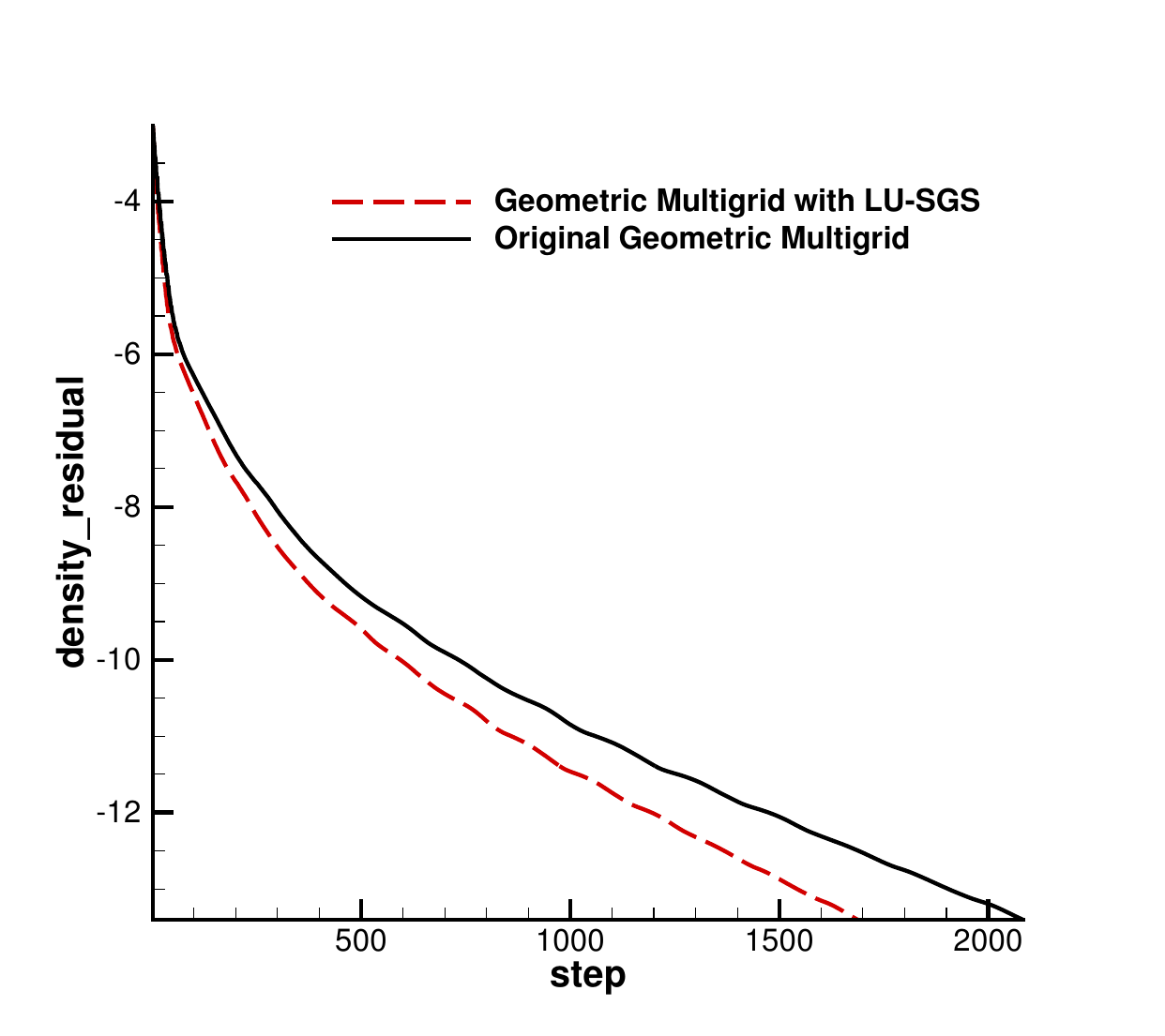}
	\includegraphics[height=0.25\textwidth]{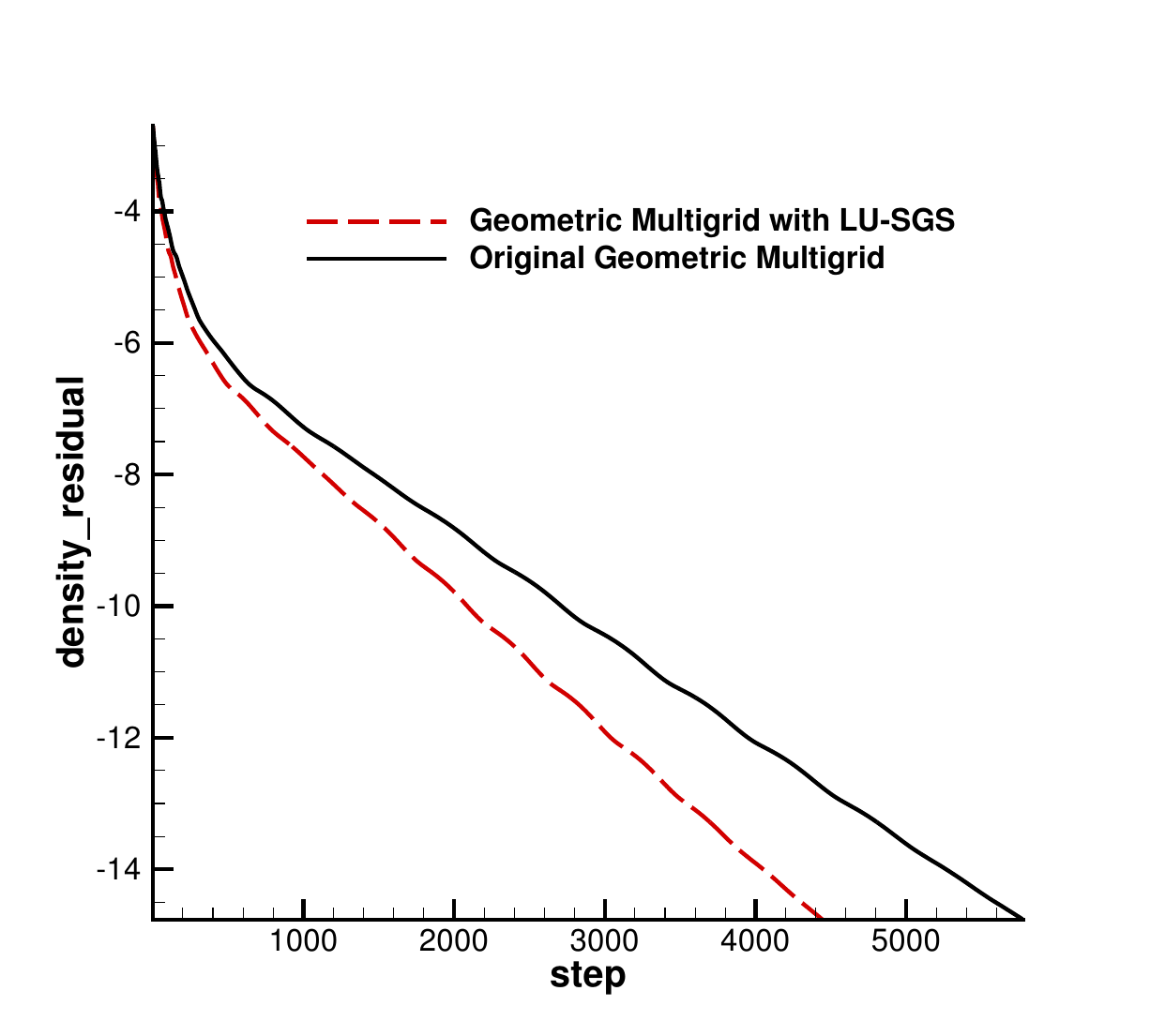}
	\includegraphics[height=0.25\textwidth]{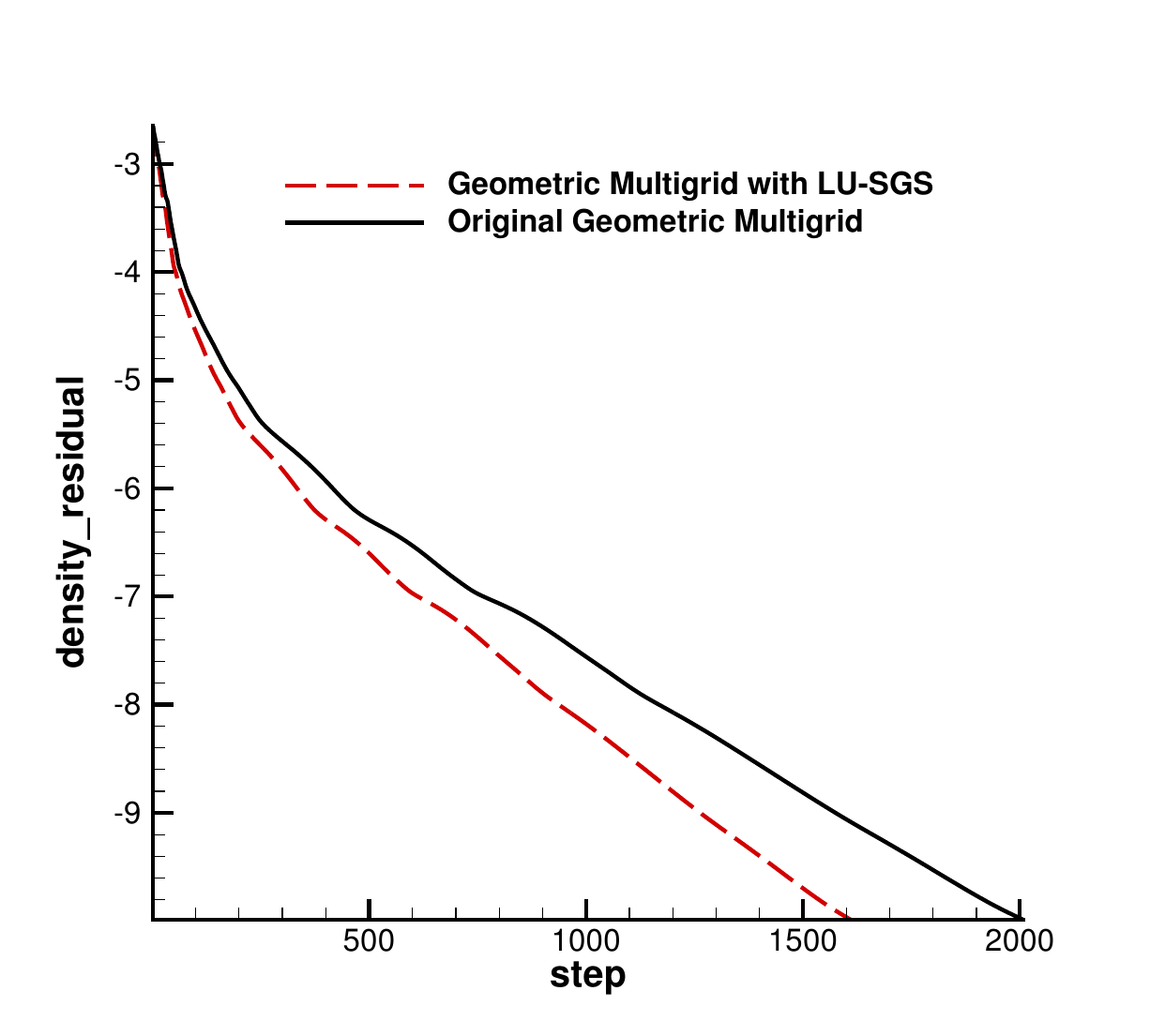}
	\caption{\label{LUSGSVSExplicit}
		Convergence history of geometric multigrid with LU-SGS and original geometric multigrid. Left: Subsonic flow around a sphere, Mach number is 0.2535, Reynolds number is 118.0. Middle: Transonic flow around a sphere, Mach number is 0.95, Reynolds number is 300.0. Right: Supersonic flow around a sphere, Mach number is 2.0, Reynolds number is 300.0.}
\end{figure}

The convergence histories obtained from the numerical tests, such as subsonic flow around a sphere, transonic flow around a sphere and supersonic flow around a sphere, show that slight convergence speedup has been made by adopted LU-SGS method on coarse-level mesh, as Fig.~\ref{LUSGSVSExplicit} shows. The improvement is insignificant and may destroy the performance of large-scale simulations.

\subsection{Parallel strategy of geometric multigrid}
The convergence rate of implicit iterations like LU-SGS faces challenges when employing in large-scale simulations. Therefore, the scalability of implicit iterations for high-performance parallel computation is relatively poor. In this work, a parallel strategy of geometric multigrid is proposed, which has the advantage of having almost the same results between serial and parallel computations. As a result, it is convenient to apply multigrid algorithms to large-scale computations.

\begin{figure}[htp]
	\centering
	\includegraphics[width=0.8\textwidth]
	{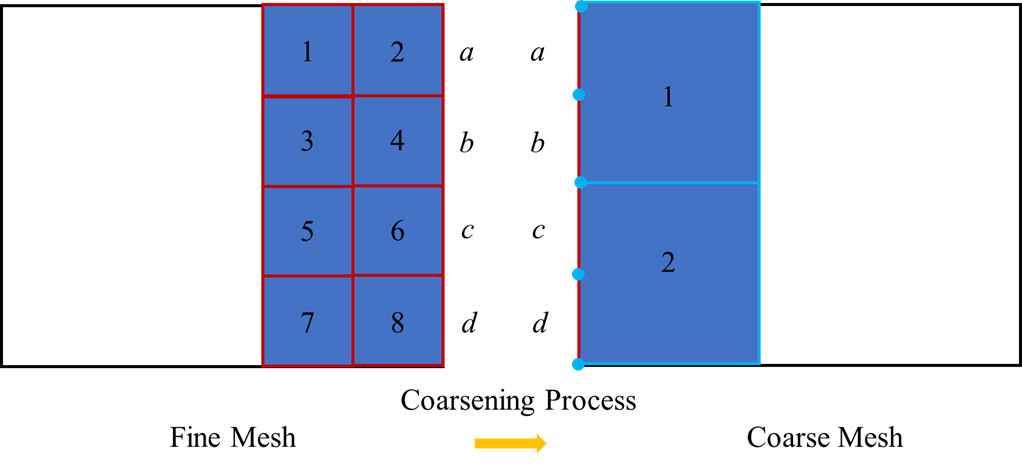}
	\caption{\label{parallel}
		Hybrid method of geometric multigrid and p-multigrid.}
\end{figure}

As mentioned before, for the finest mesh, this paper uses an explicit time-marching scheme and the parallel interfaces will not be merged when coarsening the mesh. These are the key reasons why the geometric multigrid CGKS has the advantage of having consistent results between serial and parallel computations. As Fig.~\ref{parallel} shows, when the fine mesh being coarsened to the coarse mesh, cell 1, cell 2, cell 3 and cell 4 of the fine mesh are merged into the new cell 1 of the coarse mesh. The parallel interfaces a and b of cell 2 and cell 4 of the fine mesh become the parallel interfaces a and b of cell 1 of the coarse mesh. This keeps the coarse meshes' parallel interface data structure the same as the one of fine meshes. As a result, this algorithm guarantees the non-docking faces will never appear at the parallel interfaces, preventing potential damage to the consistency of results in serial and parallel computation.

\begin{figure}[htp]	
	\centering	
	\includegraphics[height=0.25\textwidth]{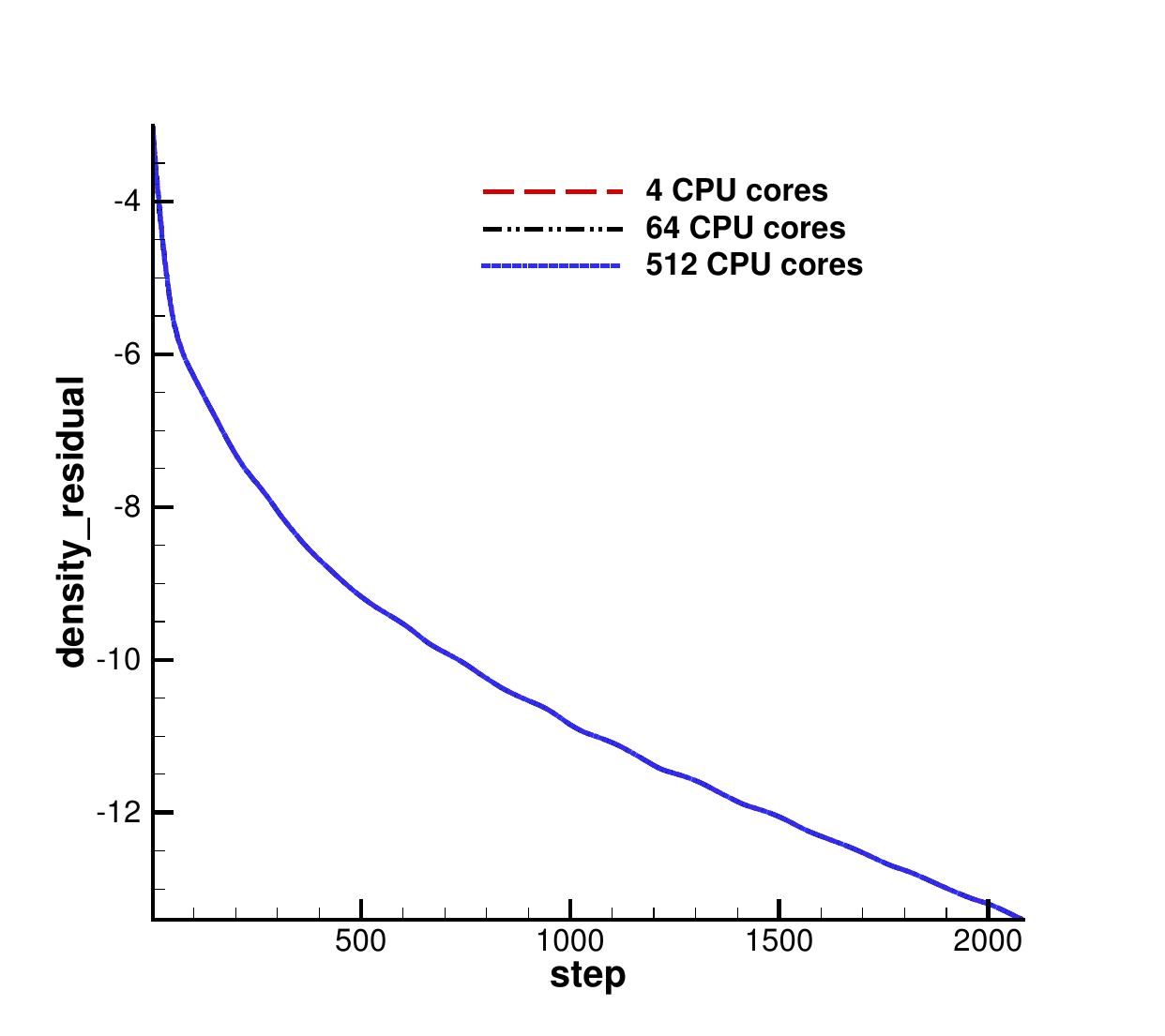}
	\includegraphics[height=0.25\textwidth]{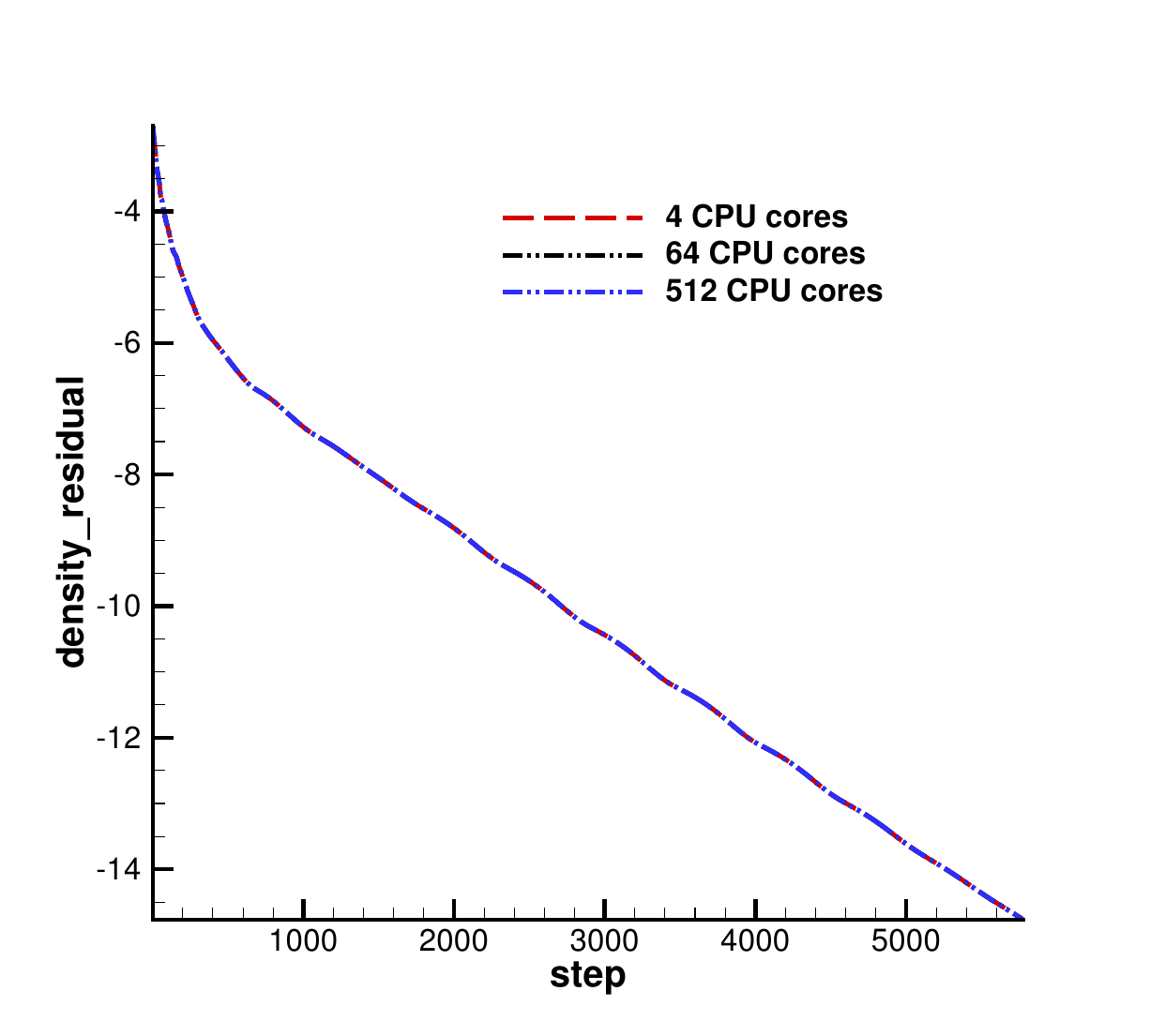}
	\includegraphics[height=0.25\textwidth]{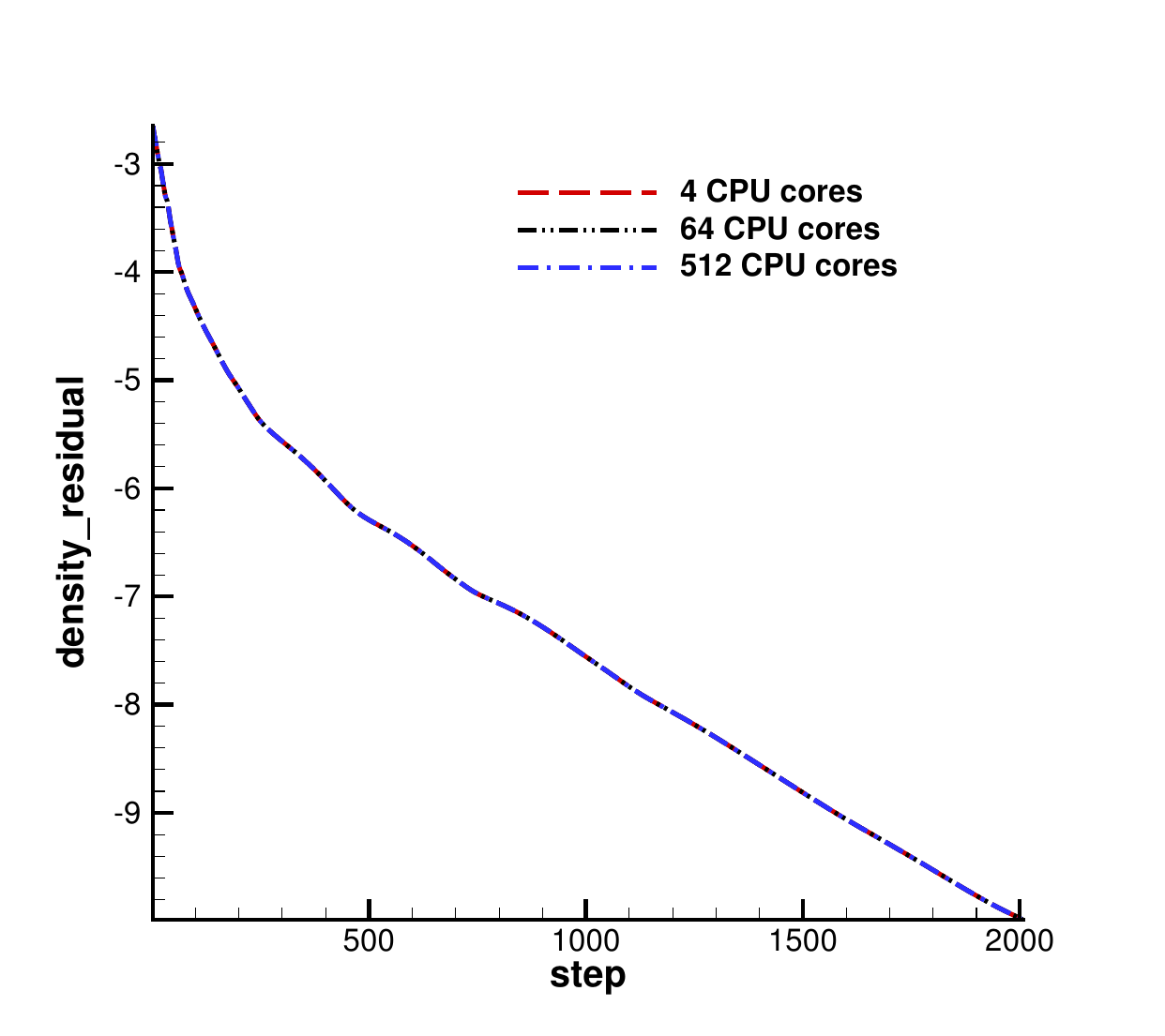}
	\caption{\label{sphere-parallel}
		Convergence history under the computation condition of 4, 32, 512 CPU cores. Left: Subsonic flow around a sphere. Mid: Transonic flow around a sphere. Right: Supersonic flow around a sphere.}
\end{figure}

\begin{figure}[htp]	
	\centering	
	\includegraphics[height=0.25\textwidth]{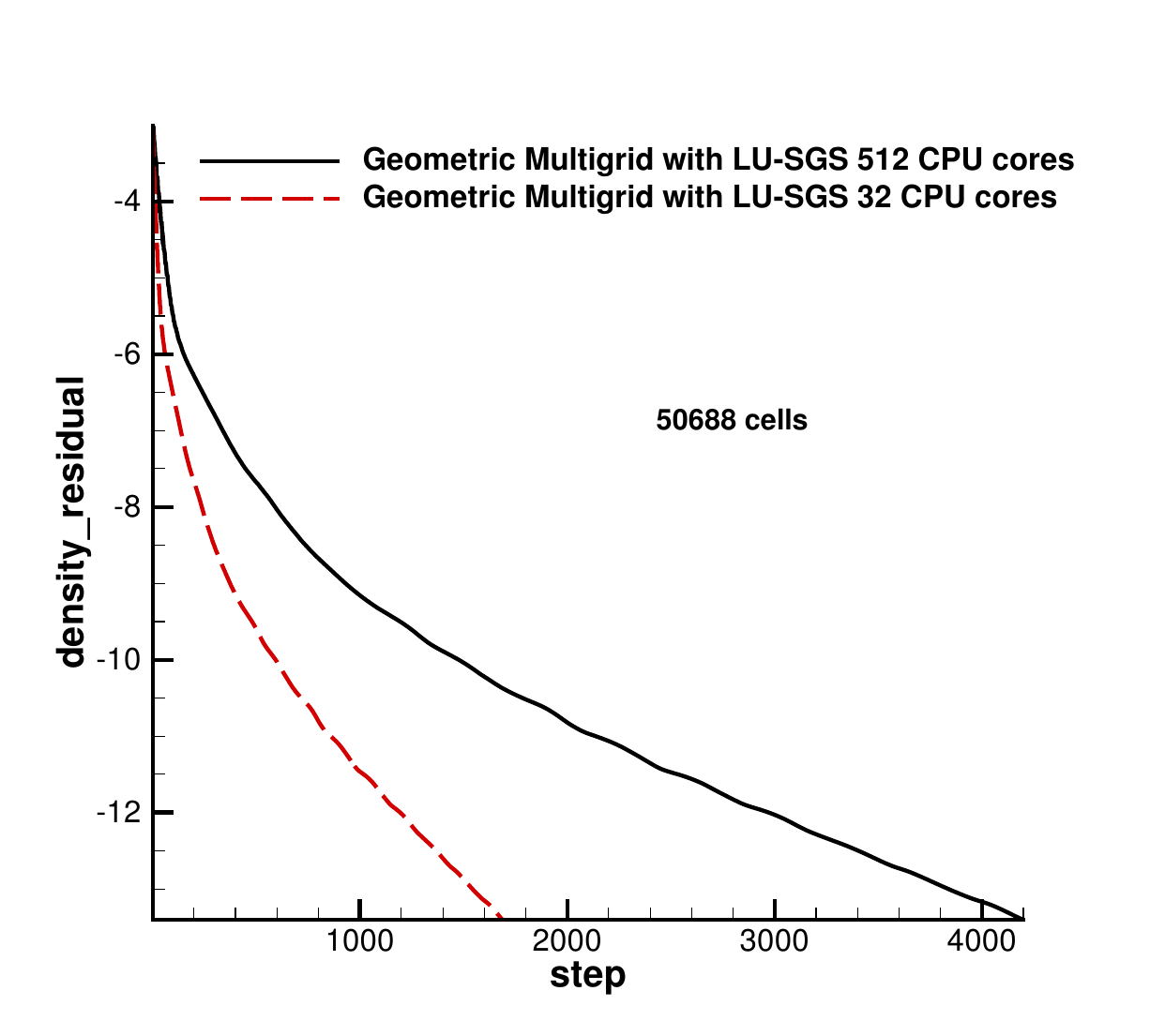}
	\includegraphics[height=0.25\textwidth]{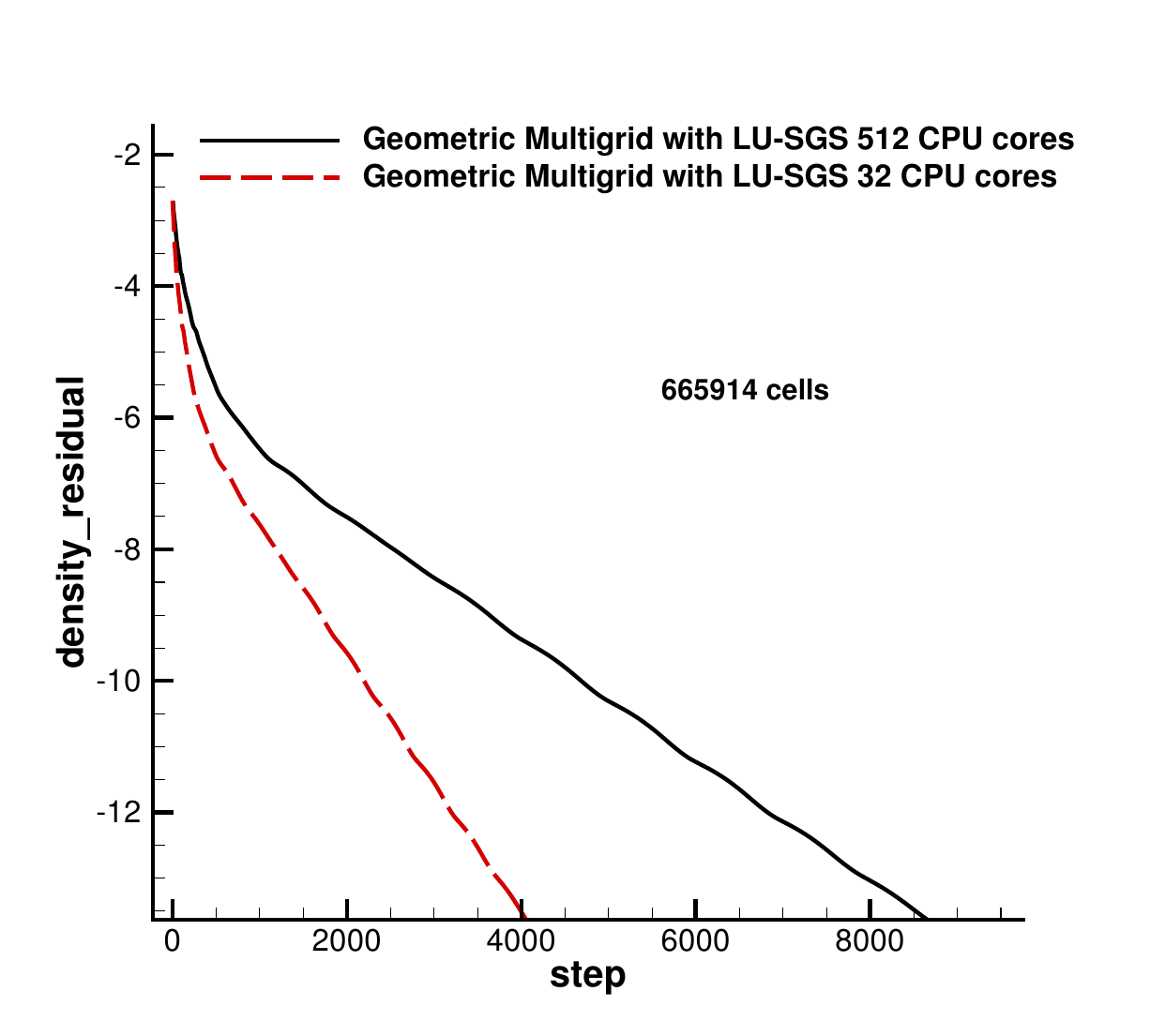}
	\includegraphics[height=0.25\textwidth]{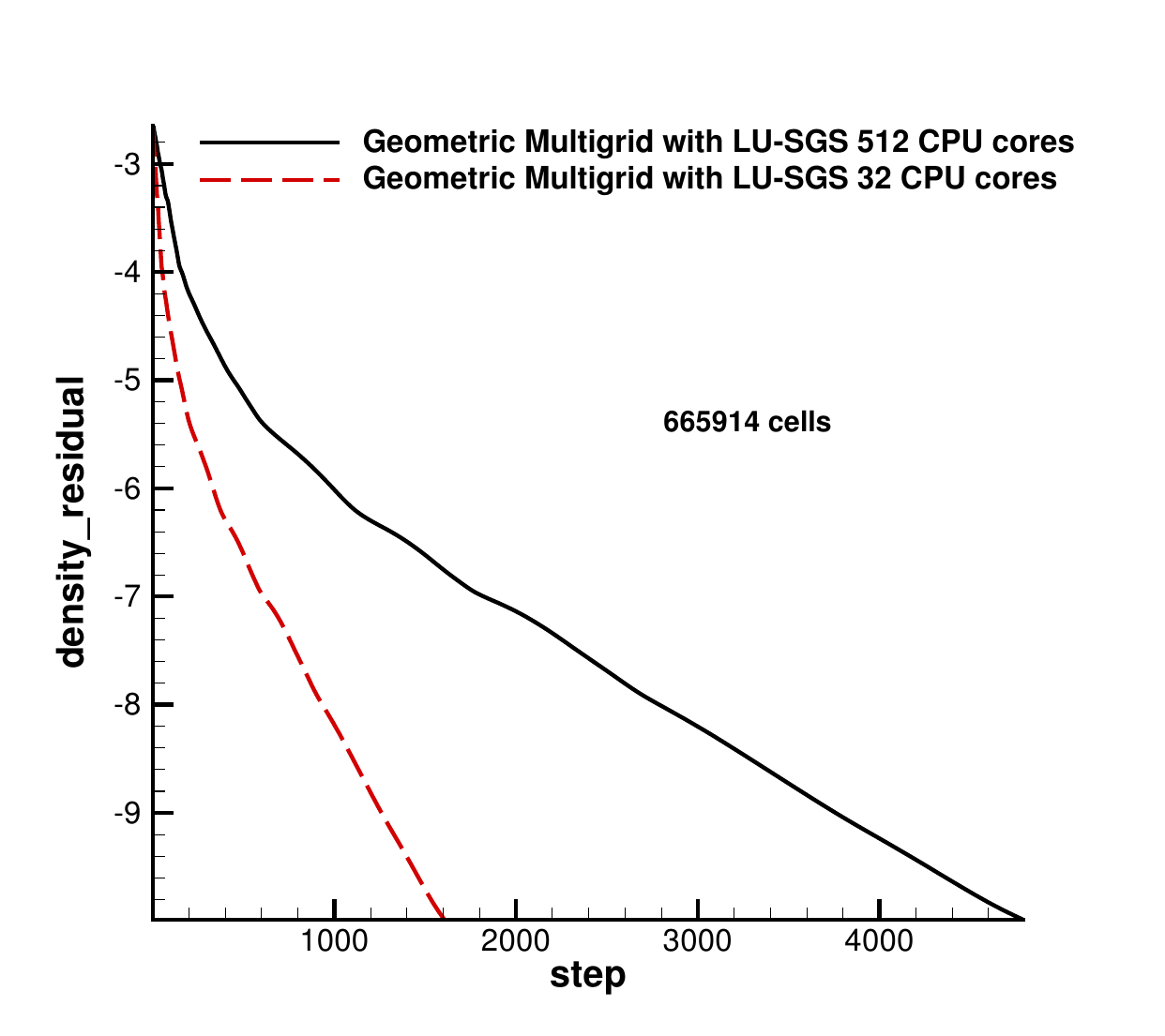}
	\caption{\label{sphere-parallel-lusgs}
		Convergence history of geometric multigrid CGKS with LU-SGS under the computation condition of 32 and 512 CPU cores. Left: Subsonic flow around a sphere. Mid: Transonic flow around a sphere. Right: Supersonic flow around a sphere.}
\end{figure}

The convergence histories obtained from the numerical tests, such as subsonic flow around a sphere, transonic flow around a sphere and supersonic flow around a sphere, validate the consistency of results in serial and parallel computation, as Fig.~\ref{sphere-parallel} shows. Moreover, geometric multigrid CGKS with LU-SGS can't be extended to large-scale computation, because as Fig.~\ref{sphere-parallel-lusgs} shows, its convergence rate decreases with the CPU cores increase.
Therefore,  the explicit time marching scheme on each level mesh with the local time stepping technique is suggested for the proposed method.

\begin{figure}[htp]	
	\centering	
	\includegraphics[height=0.55\textwidth]{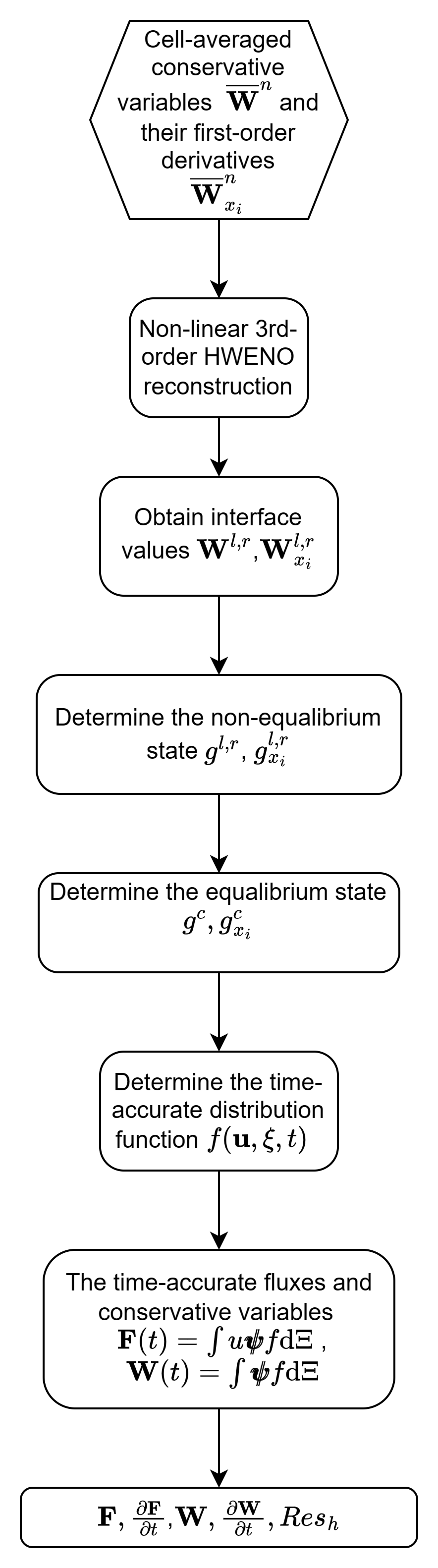}
	\includegraphics[height=0.55\textwidth]{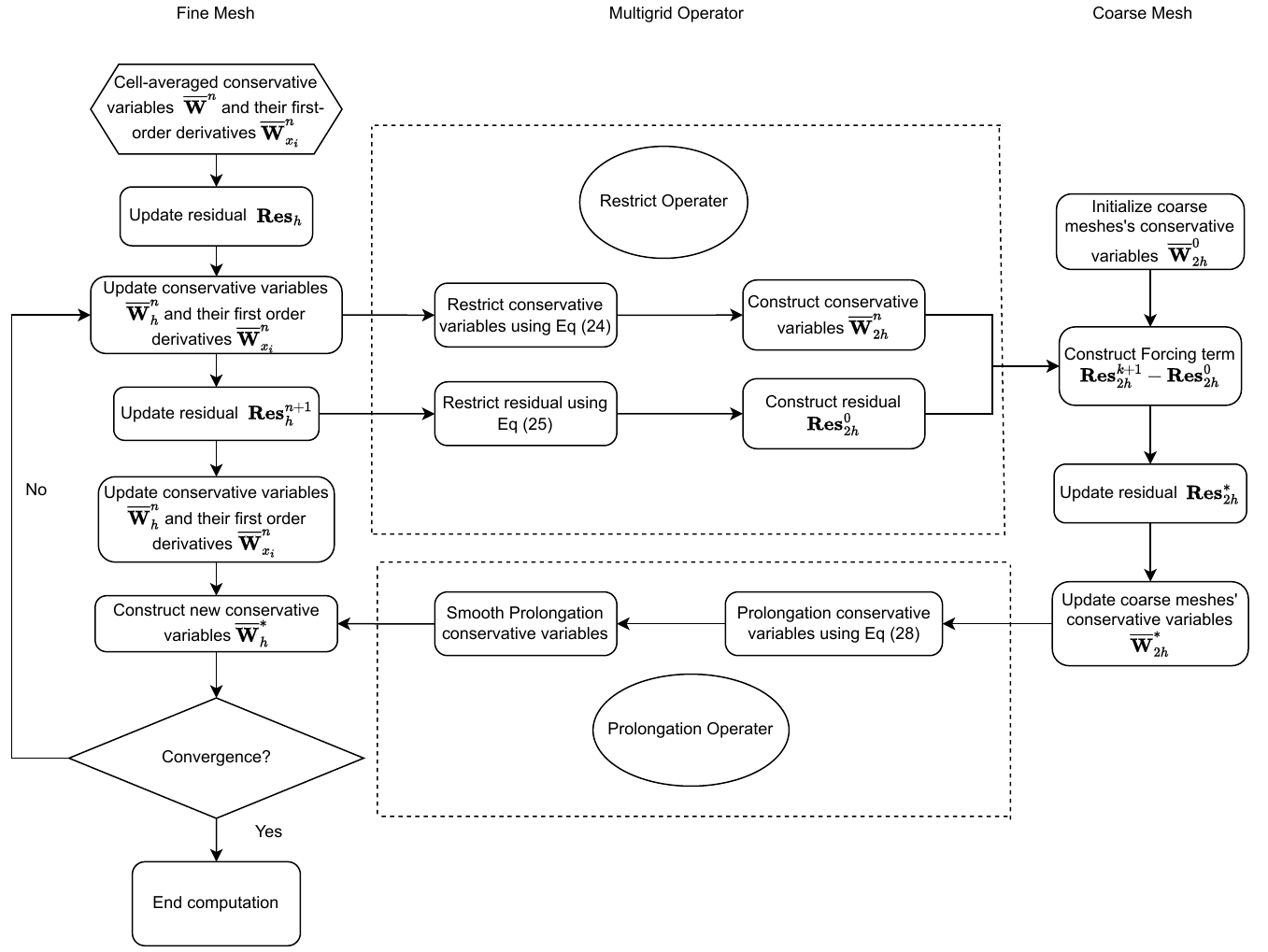}
	\caption{\label{Procedure}
		 Geometric multigrid p-multigrid CGKS algorithm flowchart. Left: Residual updating procedure of Explicit third-order CGKS. Right: Multigrid procedure with pre-smooth number is 1 and post smooth number is 0 if not specified.}
\end{figure}

The whole geometric multigrid p-multigrid CGKS algorithm flowchart is shown as Fig.~\ref{Procedure}.

\section{Numerical examples} \label{test-case}
In this section, numerical tests are presented to validate the proposed scheme. All the simulations are performed on 3-D hybrid unstructured mesh. The simulations are conducted by our in-house C++ solver, where MPI is used for parallel computation, and METIS is used for mesh partitioning.

\subsection{Flow around a sphere}
\noindent{\sl{(a) subsonic viscous flow around a sphere}}

\begin{figure}[htp]
	\centering	
	\includegraphics[height=0.35\textwidth]{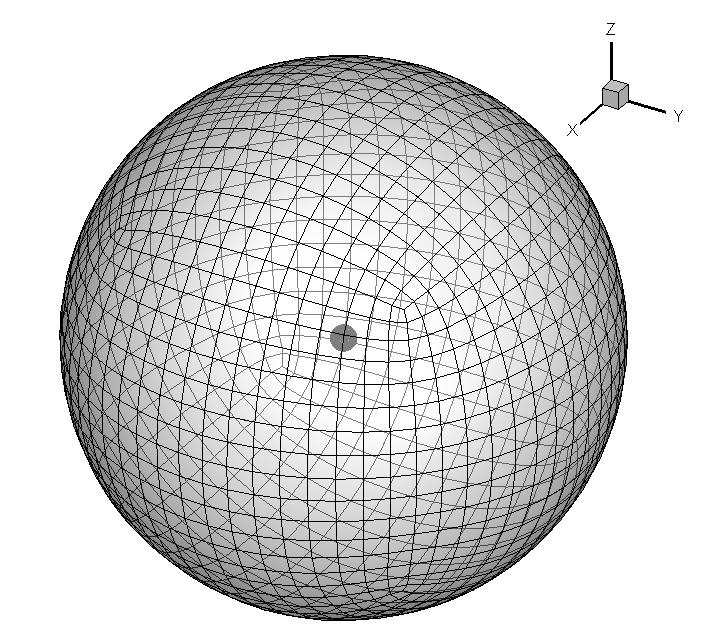}
	\includegraphics[height=0.35\textwidth]{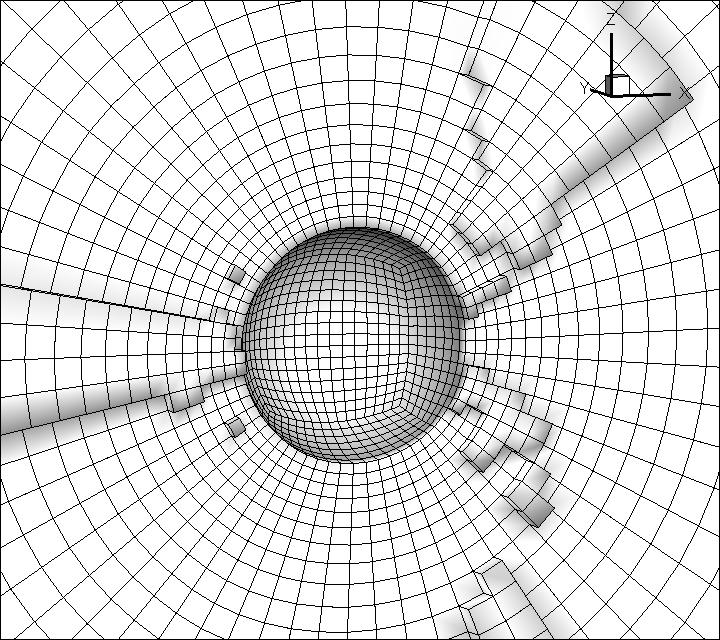}
	\caption{\label{viscous sphere mesh}
		Subsonic flow around  a sphere. Mesh sample.}
\end{figure}

\begin{figure}[htp]	
	\centering	
	\includegraphics[height=0.35\textwidth]{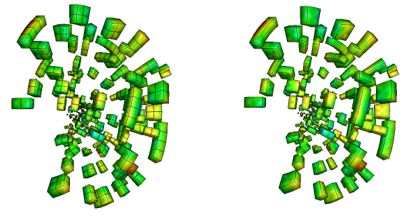}
	\caption{\label{viscous-sphere-coarse-mesh1}
		Subsonic flow around a sphere. The second layer coarse mesh.}
\end{figure}

\begin{figure}[htp]	
	\centering	
	\includegraphics[height=0.35\textwidth]{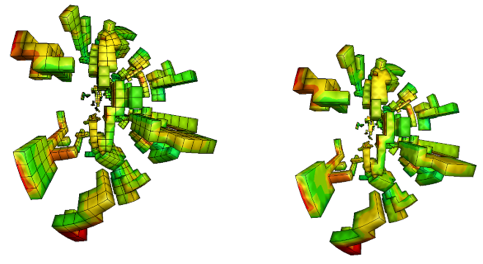}
	\caption{\label{viscous-sphere-coarse-mesh2}
		Subsonic flow around a sphere. The third layer coarse mesh.}
\end{figure}

A subsonic flow around a sphere is simulated in this case. The Mach number is set to be 0.2535 and the Reynolds number is set to be 118.0. The first mesh off the wall has the size $h = 4.5 \times 10^{-2}D$, and the total cell number is 50688. A three-layer ``V" cycle p-multigrid geometric multigrid is used to solve the case. The finest mesh is shown in Fig.~\ref{viscous sphere mesh}. The coarse meshes at the second level and the third level are shown in Fig.~\ref{viscous-sphere-coarse-mesh1} and Fig.~\ref{viscous-sphere-coarse-mesh2}.

\begin{figure}[htp]	
	\centering	
	\includegraphics[height=0.35\textwidth]{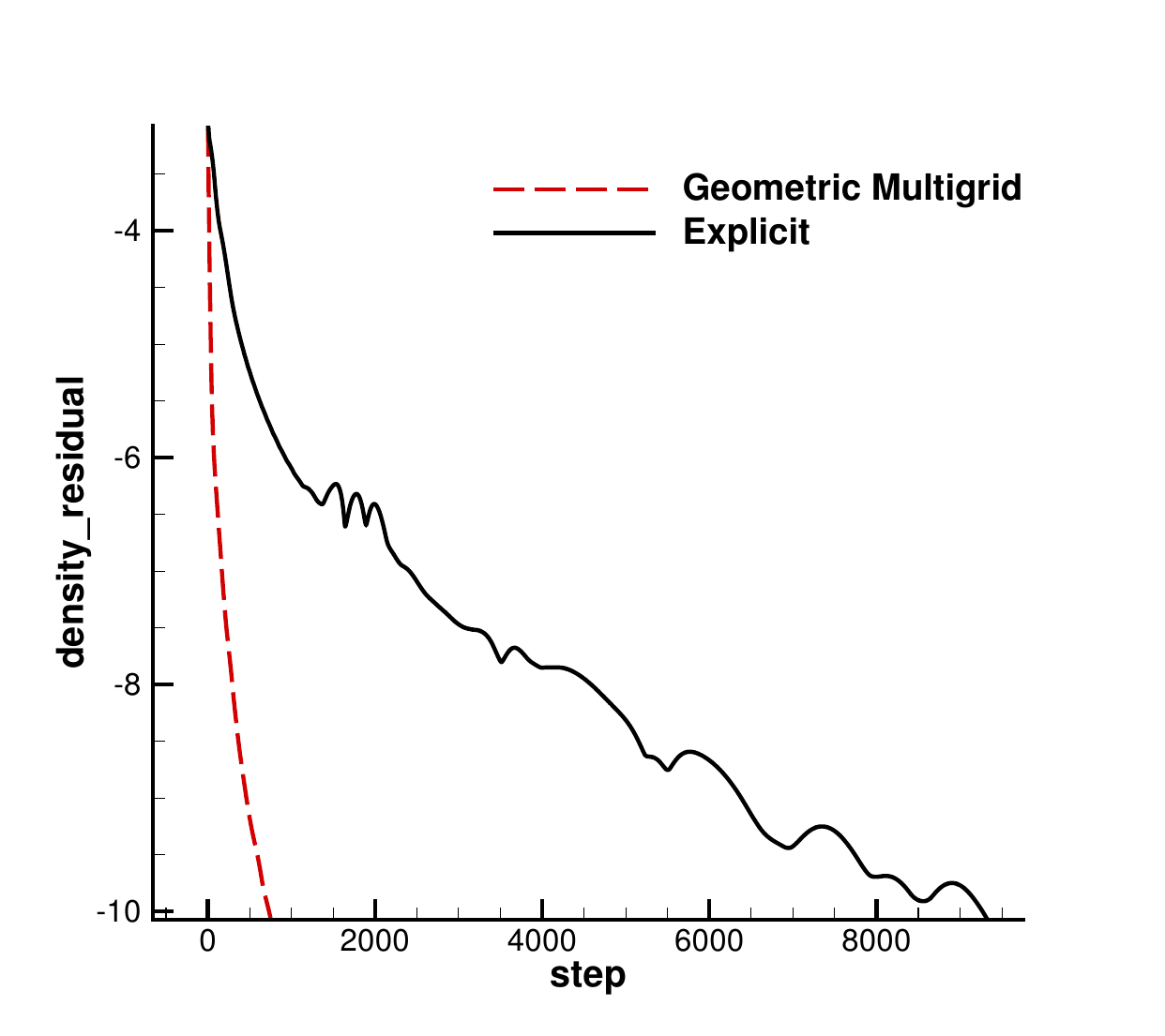}
	\includegraphics[height=0.35\textwidth]{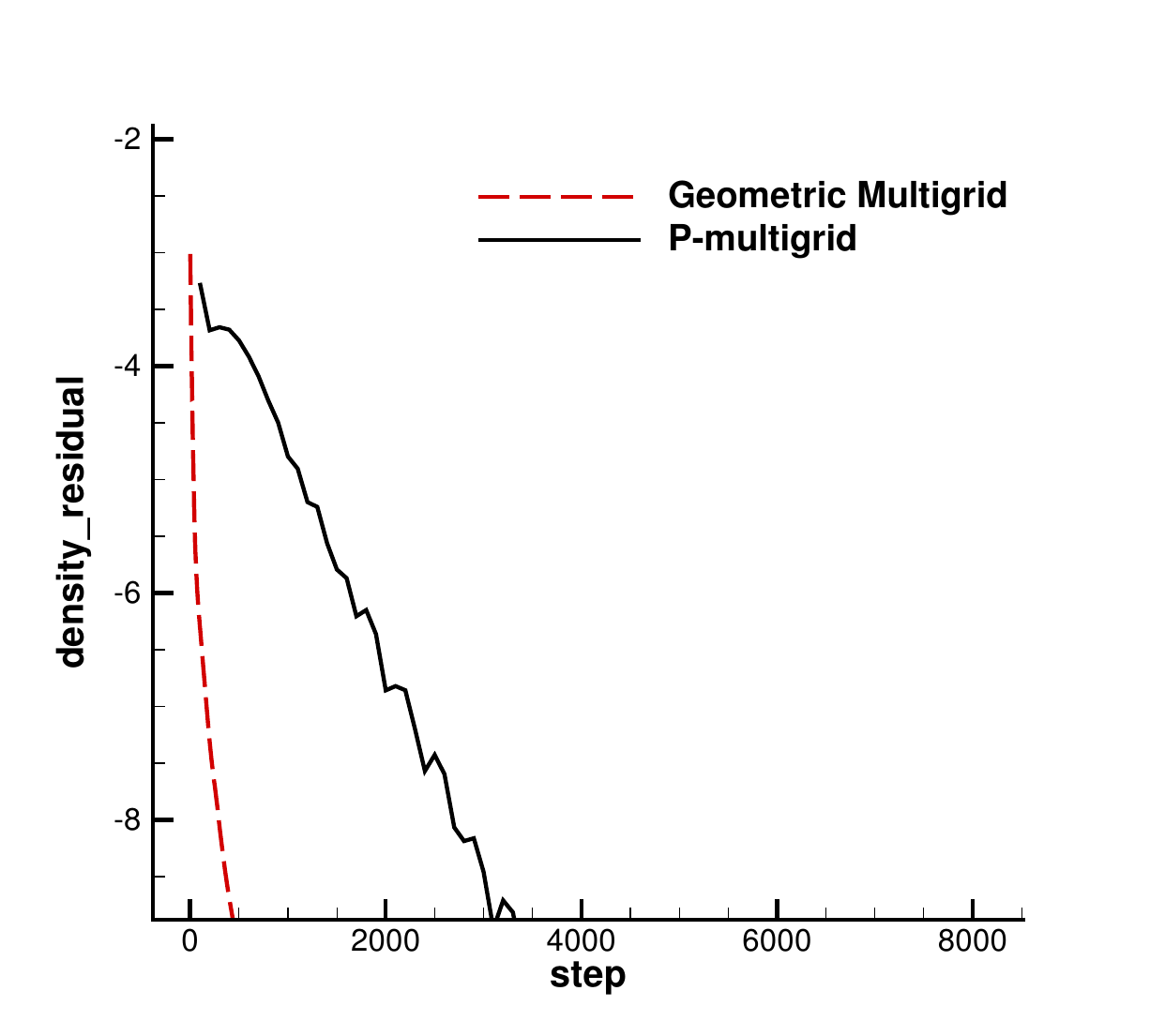}
	\caption{\label{viscous-sphere-residual}
		Residual compare. Left: Geometric-multigrid CGKS compared with explicit CGKS. Right: Geometric-multigrid CGKS compared with p-multigrid CGKS}
\end{figure}

\begin{figure}[htp]	
	\centering	
	\includegraphics[height=0.35\textwidth]{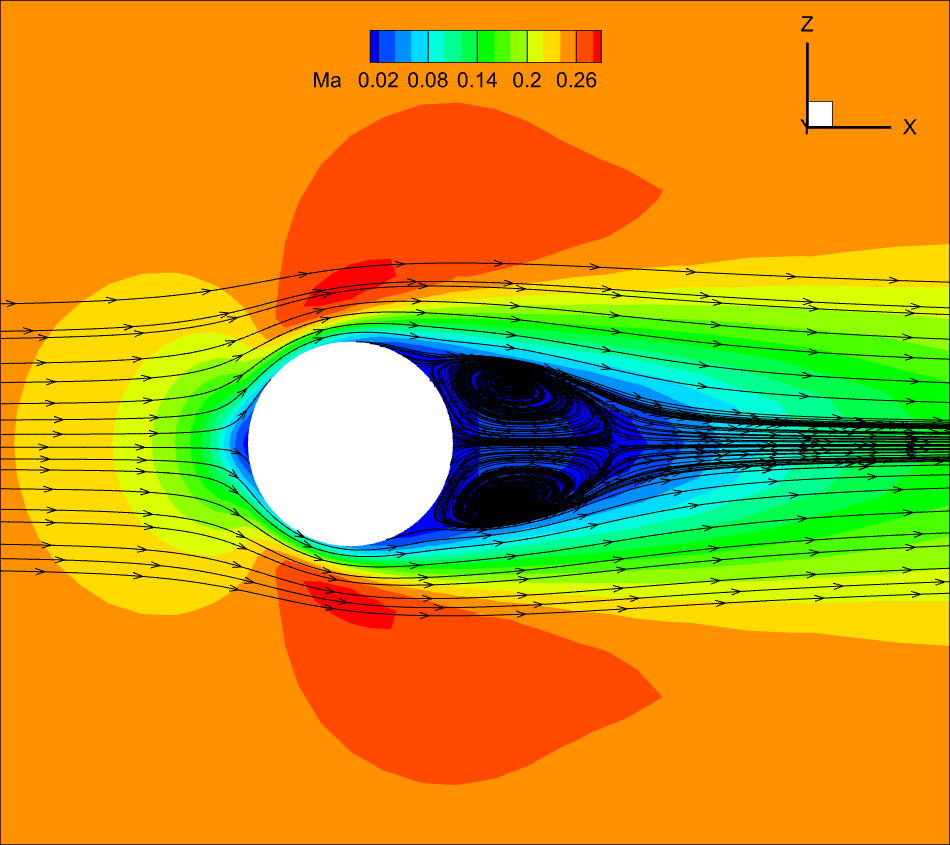}
	\includegraphics[height=0.35\textwidth]{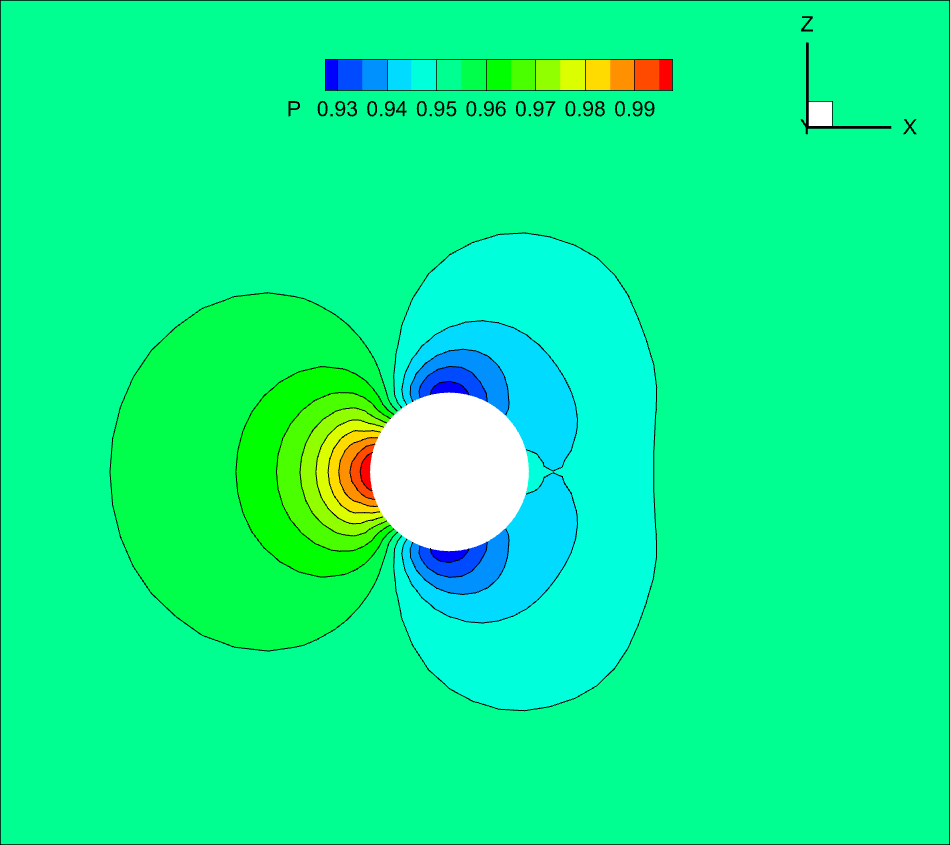}
	\caption{\label{viscous-sphere-density-contour}
		Subsonic flow around a sphere. Left: Mach contour and streamlines obtained by the geometric-multigrid CGKS. Right: Pressure contour obtained by geometric-multigrid CGKS}
\end{figure}

The residuals can be reduced to below $10^{-10}$ for both explicit CGKS and geometric-multigrid CGKS. The iteration number of the geometric-multigrid CGKS is about 1/14  of the number of iteration steps of the explicit CGKS at a residual level of $10^{-10}$, as shown in Fig.~\ref{viscous-sphere-residual}. The CPU time speedup for geometric-multigrid CGKS is about 6.67. The comparison result of convergence history between geometric-multigrid CGKS and p-multigrid CGKS \cite{JI2022105489} shown in Fig.~\ref{viscous-sphere-residual} indicates that the convergence rate of the geometric-multigrid CGKS is 5 times faster than that of p-multigrid CGKS at the residual level $10^{-8}$. The Mach number contour and streamline are also presented in Fig.~\ref{viscous-sphere-density-contour} to show the high resolution of the CGKS. Quantitative results are given in Table \ref{viscous subsonic sphere}, including the drag coefficient Cd, the separation angle $\theta$, and the closed wake length L, as defined in \cite{ji2021compact}.

\begin{table}[htp]
	\small
	\begin{center}
		\def\temptablewidth{1.0\textwidth}
		{\rule{\temptablewidth}{1pt}}
		\begin{tabular*}{\temptablewidth}{@{\extracolsep{\fill}}c|c|c|c|c|c}
			Scheme & Mesh number & Cd  & $\theta$  &L &Cl\\
			\hline
			Experiment \cite{taneda1956experimental}	&-- & 1.0  & 151 & 1.07 & -- \\ 	
			Third-order DDG \cite{cheng2017parallel} & 160,868 & 1.016 & 123.7 & 0.96 & --\\
			Fourth-order VFV \cite{wang2017thesis}  & 458,915 & 1.014 & --& -- & 2.0e-5\\
			Current & 50688 & 1.021  & 126.6 & 0.94 & 2.33e-5\\
		\end{tabular*}
		{\rule{\temptablewidth}{0.1pt}}
	\end{center}
	\vspace{-4mm} \caption{\label{viscous subsonic sphere} Quantitative comparisons among different compact schemes  for the subsonic flow around a sphere.}
\end{table}

\noindent{\sl{(b) transonic viscous flow around a sphere}}

\begin{figure}[htp]	
	\centering	
	\includegraphics[height=0.35\textwidth]{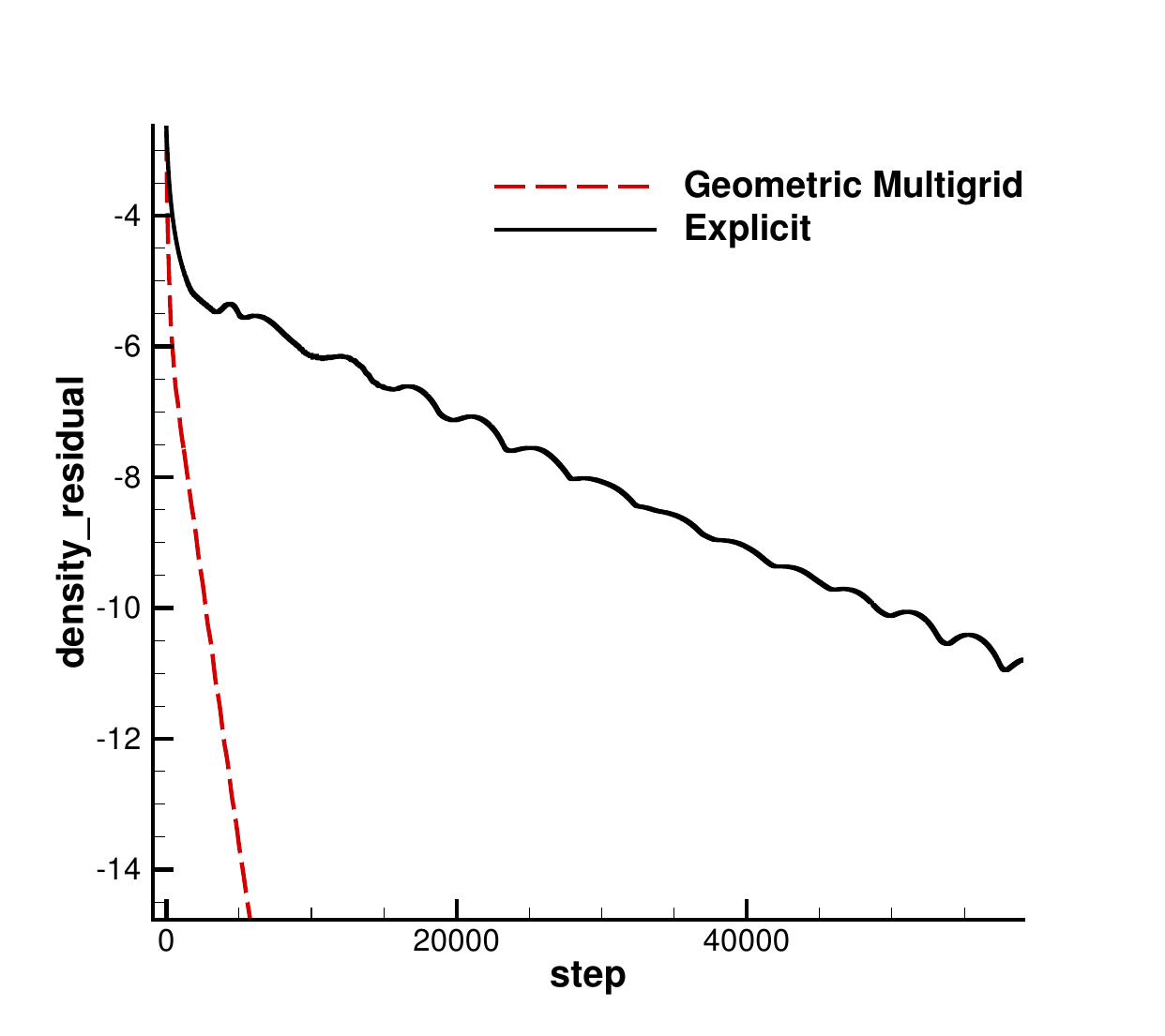}
	\includegraphics[height=0.35\textwidth]{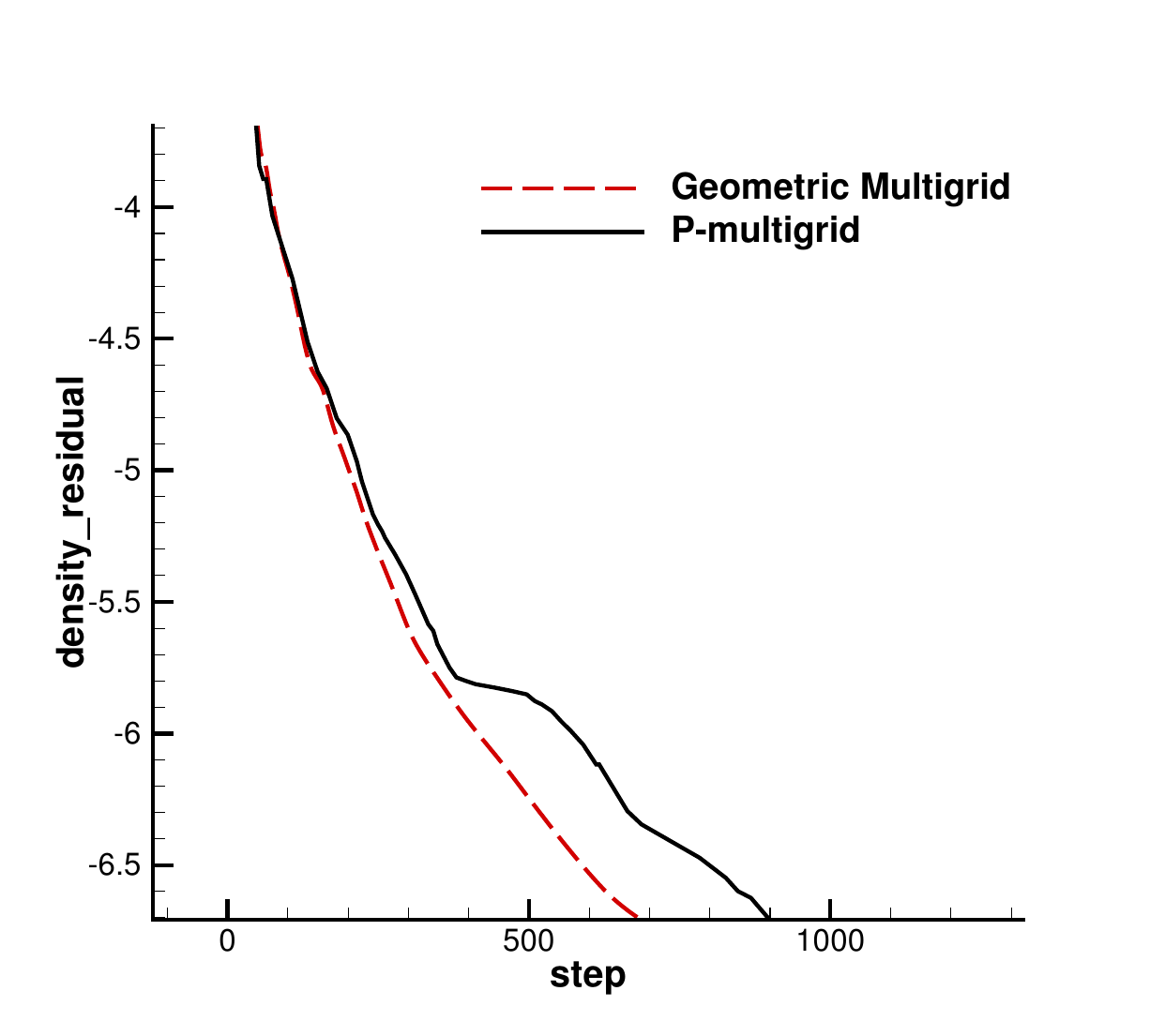}
	\caption{\label{viscous-sphere-residual-ma0.95}
		Residual compare. Left: Geometric  CGKS compared with explicit CGKS. Right: Geometric Multigrid CGKS compared with p-multigrid CGKS.}
\end{figure}

\begin{figure}[htp]	
	\centering	
	\includegraphics[height=0.35\textwidth]{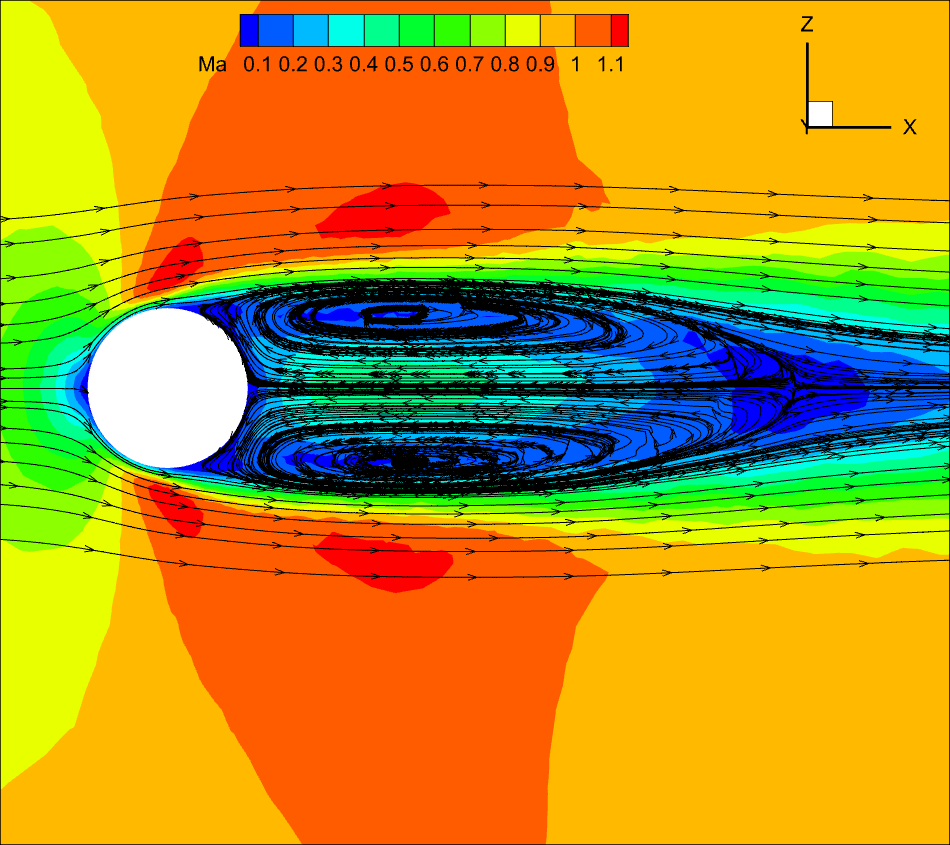}
	\includegraphics[height=0.35\textwidth]{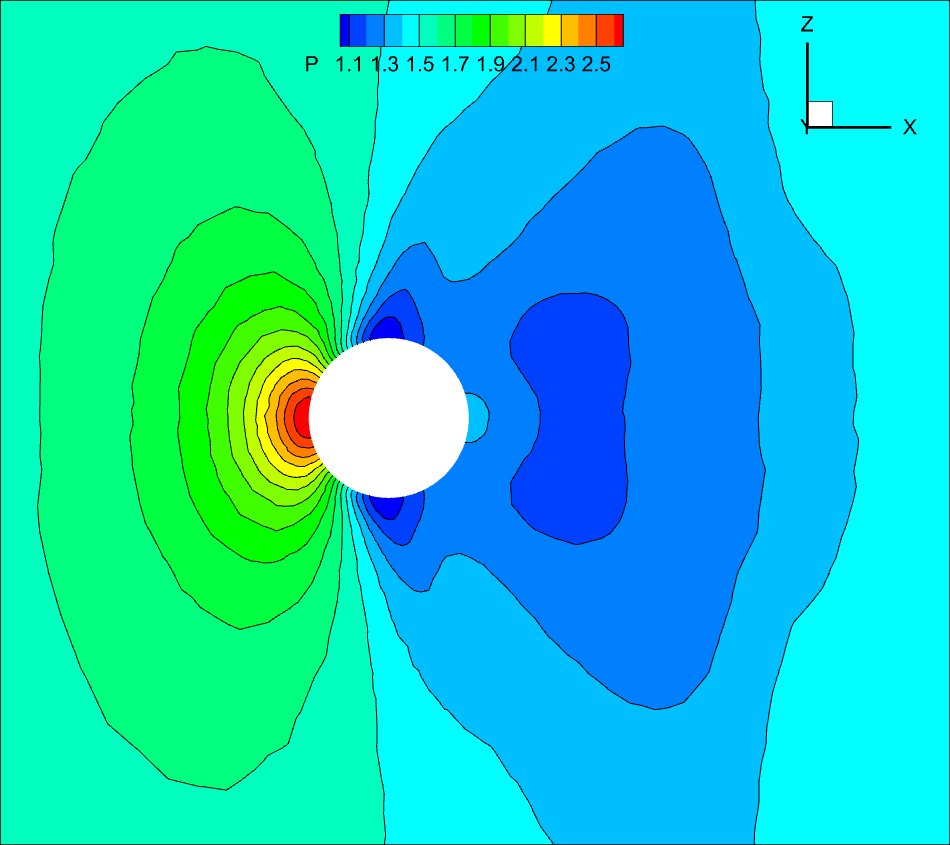}
	\caption{\label{viscous-sphere-ma0.95-contour}
		Transonic flow around a sphere. Left: Mach number contour with streamline through the sphere. Right: Pressure contour. }
\end{figure}

A transonic viscous flow around a sphere is simulated to show the performance of the geometric multigrid CGKS for transonic viscous flow. The Mach number is set to be 0.95 and Reynolds number is set to be 300.0. A three layer "V" cycle p-multigrid geometric multigrid is used to solve the case. The mesh used in this case is the same as the supersonic case. Fig.~\ref{viscous-sphere-residual-ma0.95} shows that the iteration steps of the geometric-multigrid CGKS is about 1/25  of the iteration steps of the explicit CGKS. The CPU time costed by explicit CGKS is 9.54 times more than geometric-multigrid CGKS. Comparison has also been made between geometric-multigrid CGKS and p-multigrid CGKS. The result in Fig.~\ref{viscous-sphere-residual-ma0.95} shows geometric-multigrid CGKS has a faster convergence rate. The numerical results of the Mach number contour and streamline around a sphere are shown in Fig.~\ref{viscous-sphere-ma0.95-contour}, which indicates the high resolution of CGKS.
\begin{table}[htp]
	\small
	\begin{center}
		\def\temptablewidth{1.0\textwidth}
		{\rule{\temptablewidth}{1pt}}
		\begin{tabular*}{\temptablewidth}{@{\extracolsep{\fill}}c|c|c|c|c}
			Scheme & Mesh Number & Cd  & $\theta$  &L\\
			\hline
			WENO6 \cite{Nagata2016sphere} 	&909,072 & 0.968  & 111.5 & 3.48\\ 	
			Current  & 665,914 & 0.963  & 110.0 & 3.44\\
		\end{tabular*}
		{\rule{\temptablewidth}{0.1pt}}
	\end{center}
	\vspace{-4mm} \caption{\label{transsonic-sphere} Quantitative comparisons between the current scheme and the reference solution for the transonic flow around a sphere.}
\end{table}

\noindent{\sl{(c) supersonic viscous flow around a sphere}}

\begin{figure}[htp]	
	\centering	
	\includegraphics[height=0.35\textwidth]{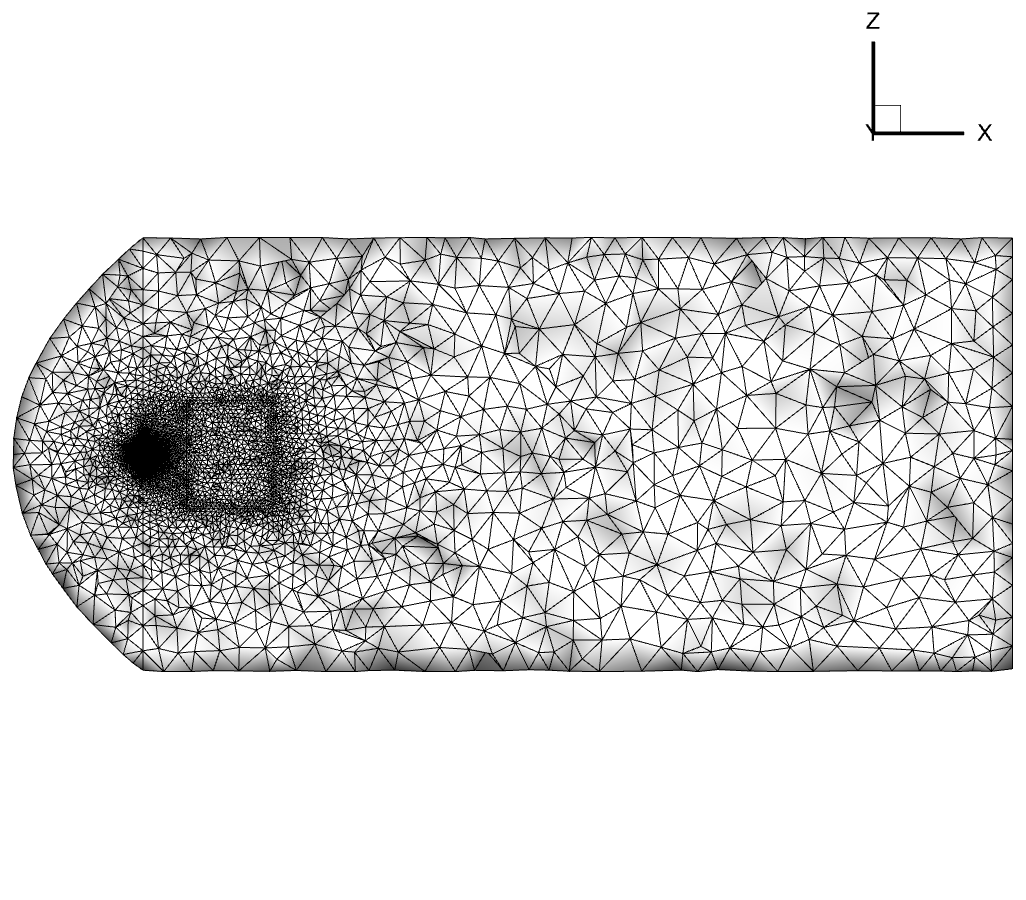}
	\includegraphics[height=0.35\textwidth]{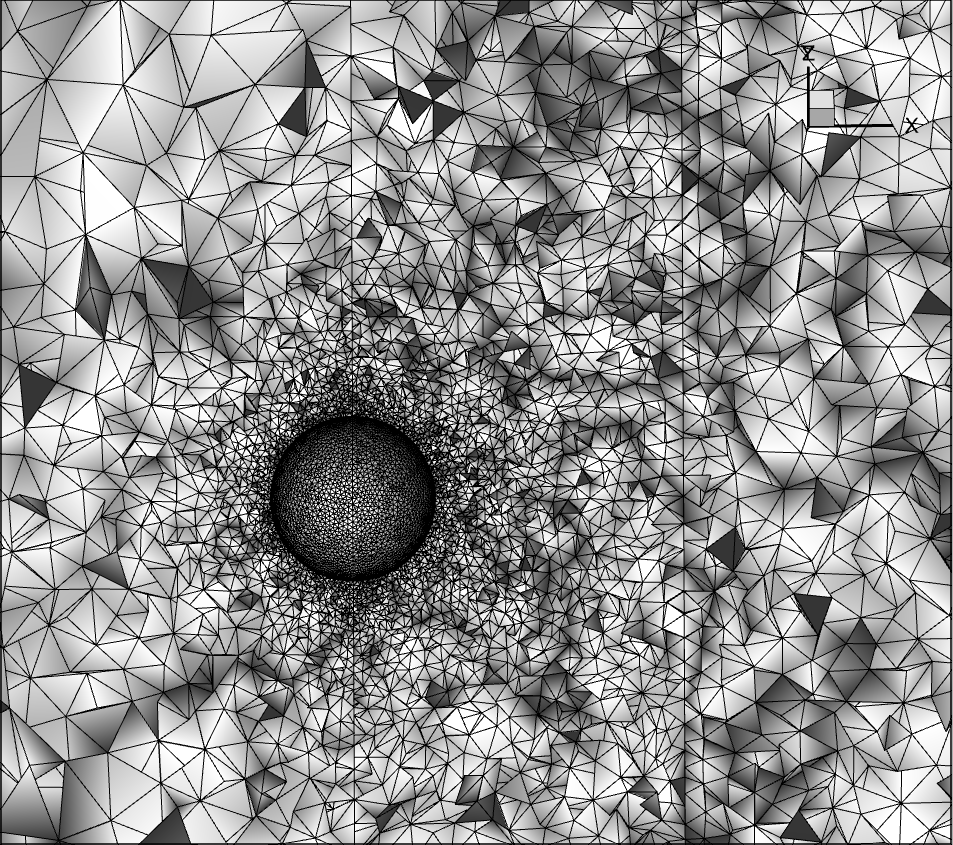}
	\caption{\label{viscous-sphere-ma1.2}
		Supersonic flow around  a sphere. Mesh sample. Left: Global mesh. Right: Local mesh}
\end{figure}

\begin{figure}[htp]	
	\centering	
	\includegraphics[height=0.25\textwidth]{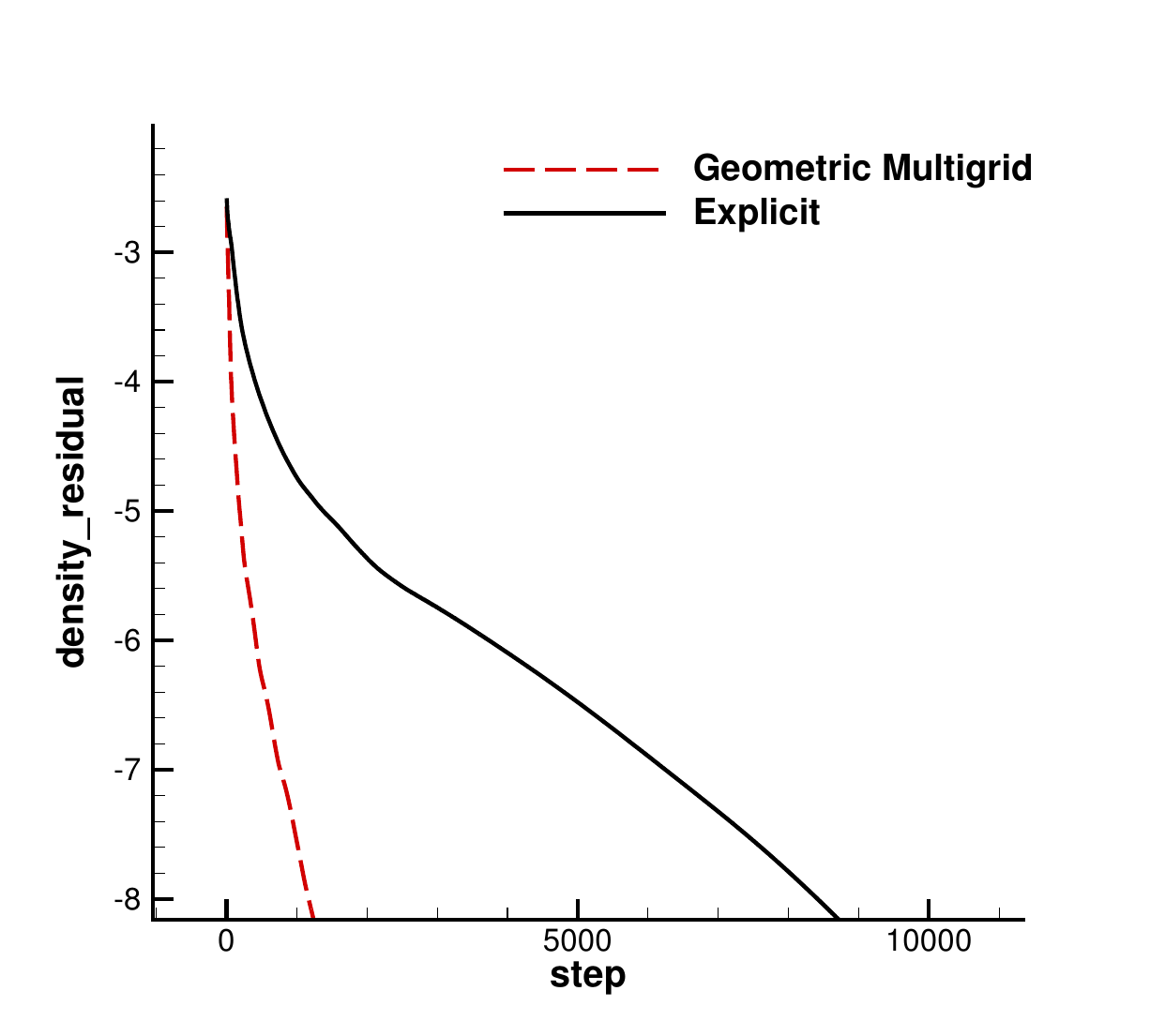}
	\includegraphics[height=0.25\textwidth]{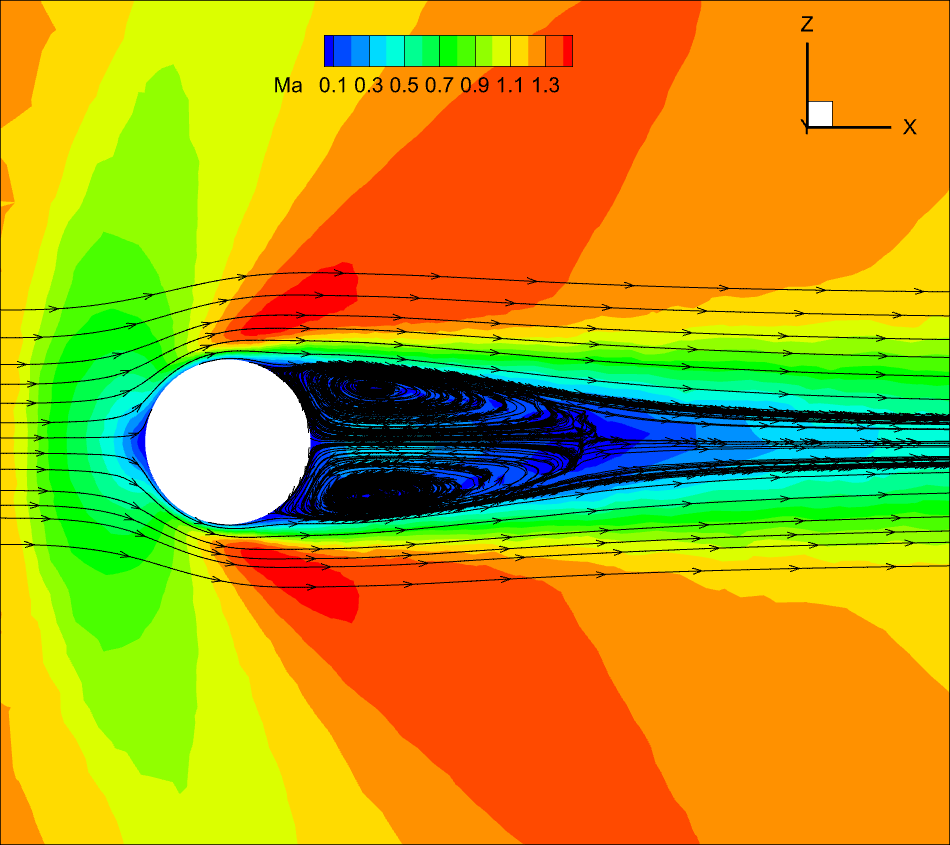}
	\includegraphics[height=0.25\textwidth]{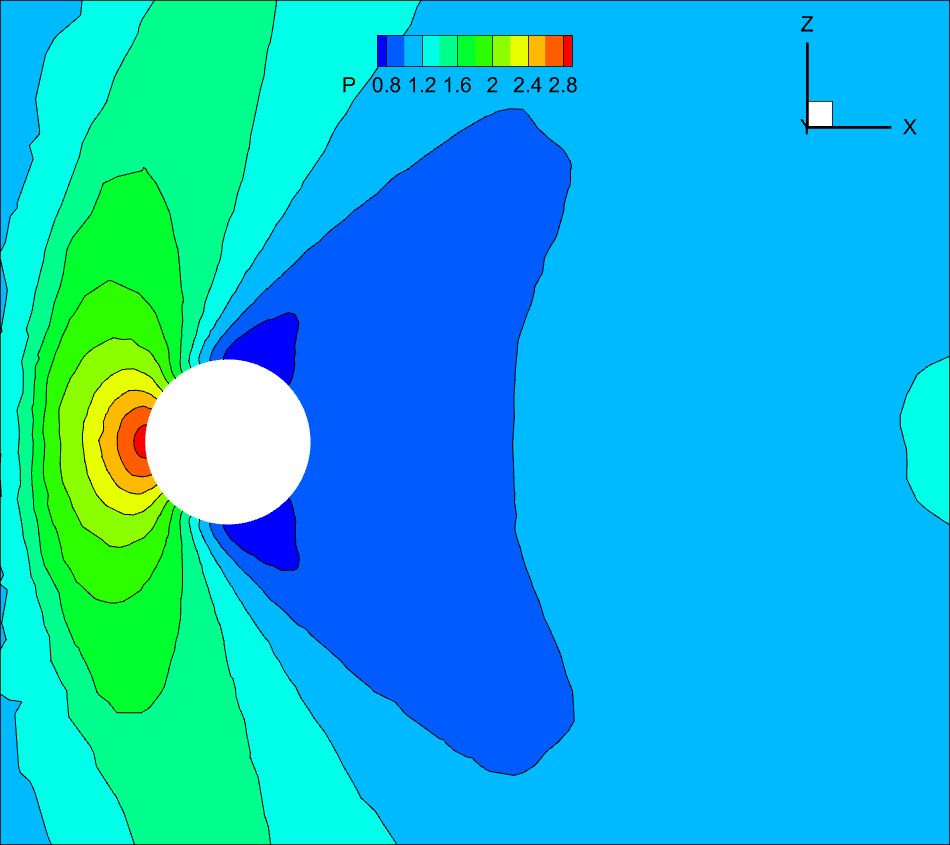}
	\caption{\label{viscous-sphere-ma1.2-residual}
			Numerical result of supersonic flow around a sphere. Left: Convergence history of geometric-multigrid CGKS and explicit CGKS. Mid: Mach number contour with streamline around a sphere. Right: Pressure contour.}
\end{figure}

To verify that geometric-multigrid CGKS can also have good performance in the supersonic flow region, a supersonic flow around a sphere is simulated. The Mach number is set to be 1.2 and the Reynolds number is set to be 300.  A three layer "V" cycle p-multigrid geometric multigrid is used to solve the case. The mesh used in this case is shown in Fig.~\ref{viscous-sphere-ma1.2}. The upstream length is 5 and the downstream length is 40. The first layer mesh at the wall has a thickness $2.3 \times 10^{-2}D$. The iteration steps of the geometric-multigrid CGKS is about 1/8  of the iteration steps of the explicit CGKS. Also, compared with explicit CGKS at the residual level $10^{-8}$, 2.67 times speedup of CPU time can be achieved, as Fig.~\ref{viscous-sphere-ma1.2-residual} shows. The result of Mach number with streamline around the sphere is also shown in Fig.~\ref{viscous-sphere-ma1.2-residual}, which indicates the high resolution of CGKS. Quantitative results are listed in Table \ref{supersonic-sphere}, which has good agreement with those given by Ref.\cite{Nagata2016sphere} .

\begin{table}[htp]
	\small
	\begin{center}
		\def\temptablewidth{1.0\textwidth}
		{\rule{\temptablewidth}{1pt}}
		\begin{tabular*}{\temptablewidth}{@{\extracolsep{\fill}}c|c|c|c|c|c}
			Scheme & Mesh Number & Cd  & $\theta$  &L & Shock stand-off\\
			\hline
			WENO6 \cite{Nagata2016sphere} 	&909,072 & 1.281  & 126.9 & 1.61 & 0.69 \\ 	
			Current  & 665,914 & 1.296  & 126.3 & 1.60 & 0.72 \\
		\end{tabular*}
		{\rule{\temptablewidth}{0.1pt}}
	\end{center}
	\vspace{-4mm} \caption{\label{supersonic-sphere} Quantitative comparisons between the current scheme and the reference solution for the supersonic flow around a sphere.}
\end{table}

\subsection{M6-wing}
\begin{figure*}[htp]	
	\centering	
	\includegraphics[height=0.35\textwidth]{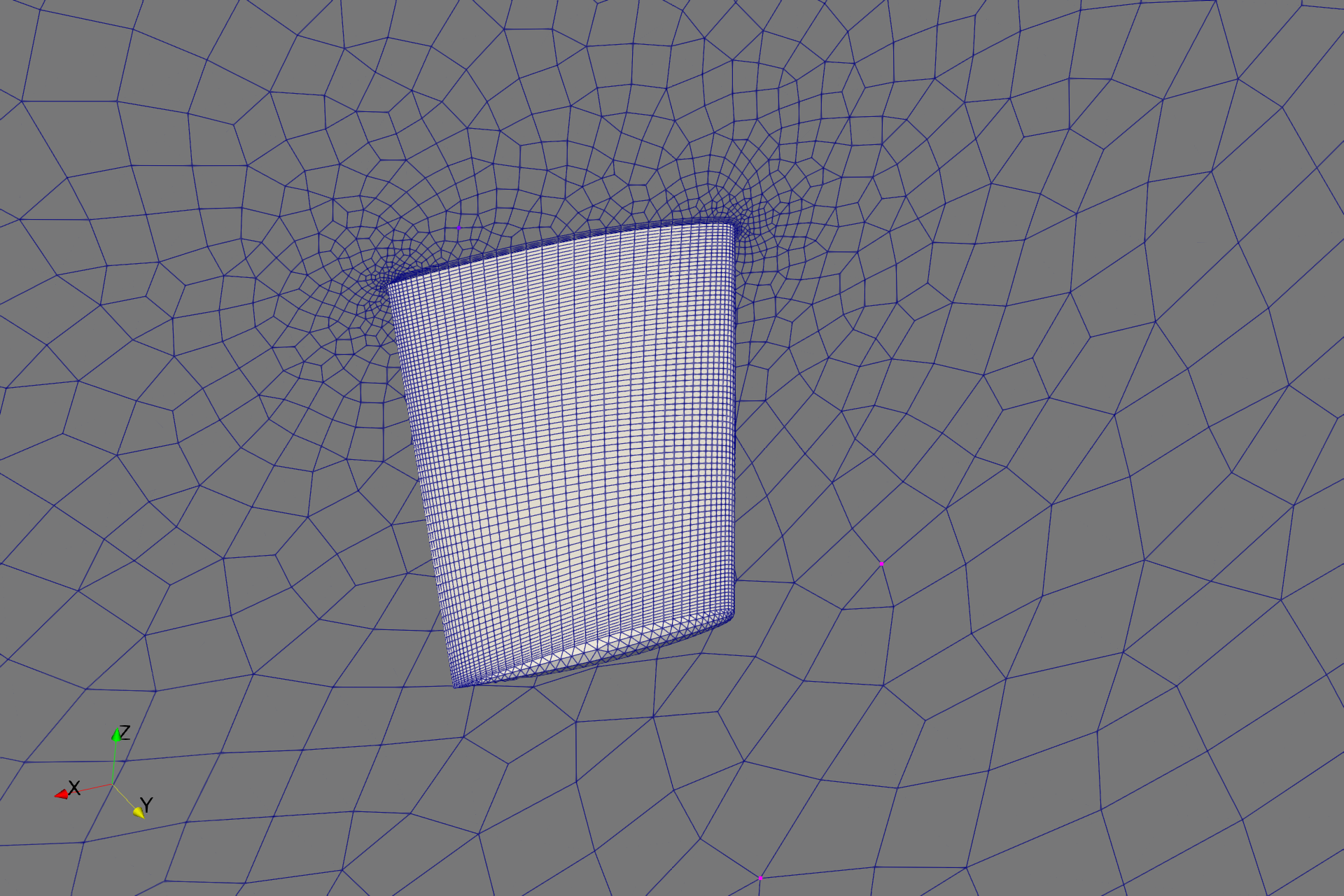}
	\includegraphics[height=0.35\textwidth]{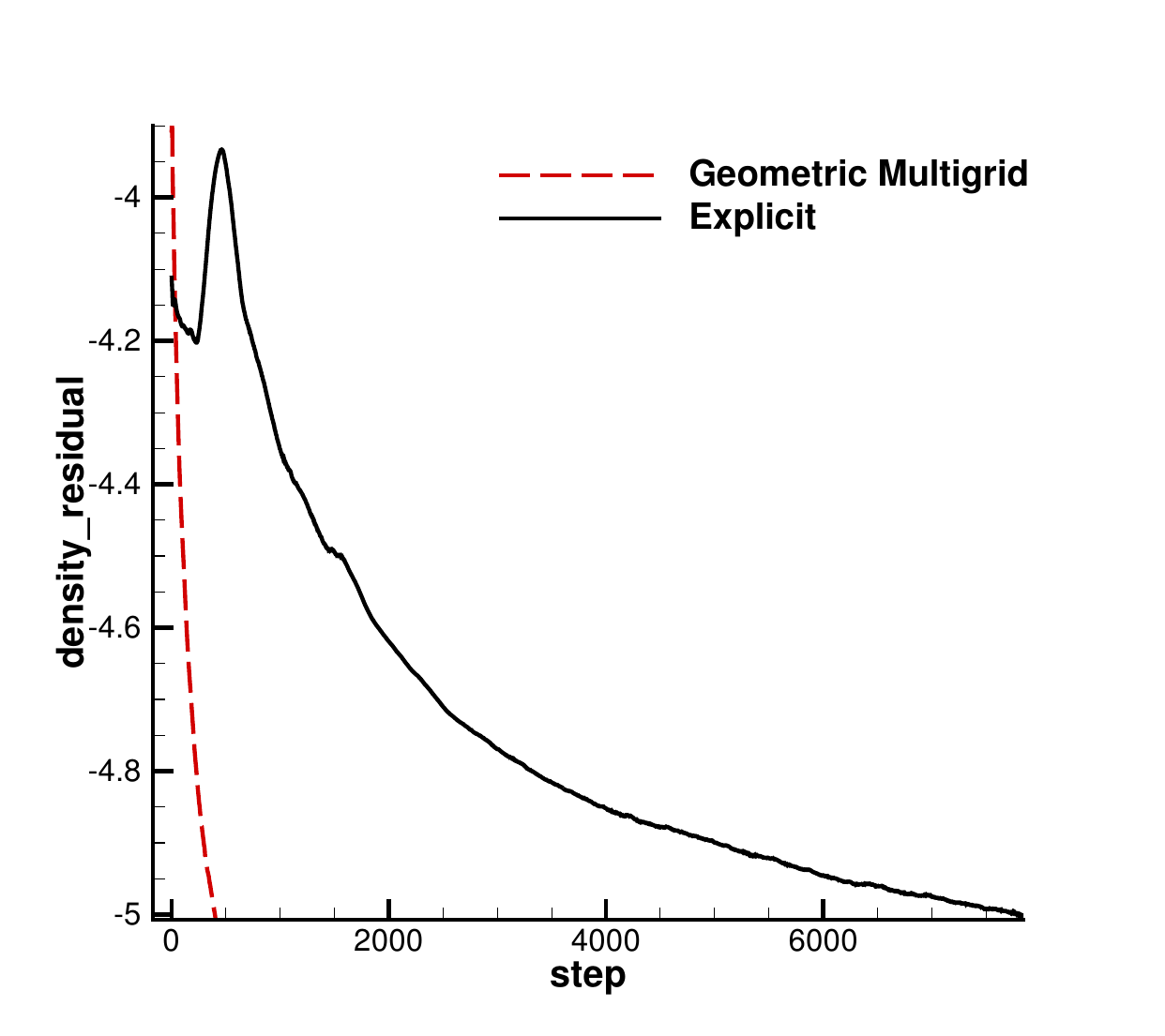}
	\caption{\label{M6 Mesh and convergence}
		Transonic flow around a M6 wing. Left: Hybrid unstructured mesh. Right: Convergence history.}
\end{figure*}

\begin{figure*}[htp]	
	\centering	
	\includegraphics[height=0.35\textwidth]{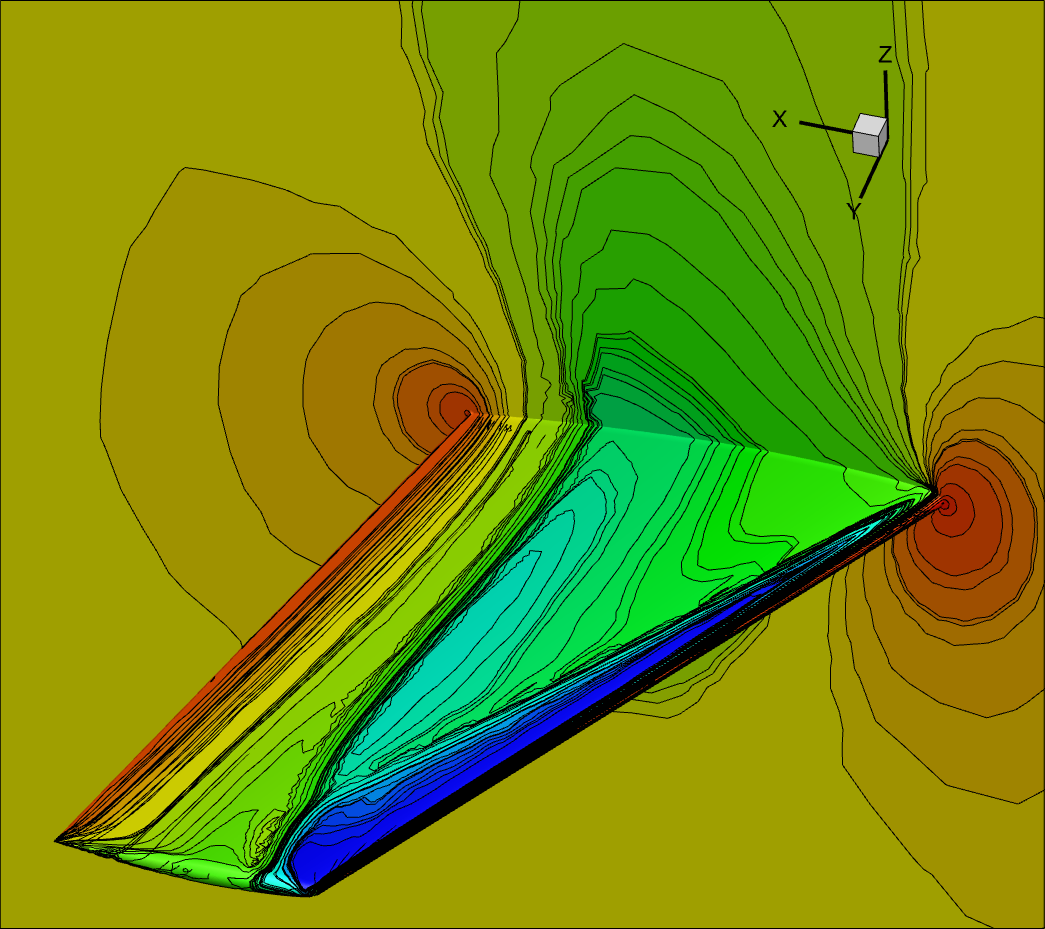}
	\includegraphics[height=0.35\textwidth]{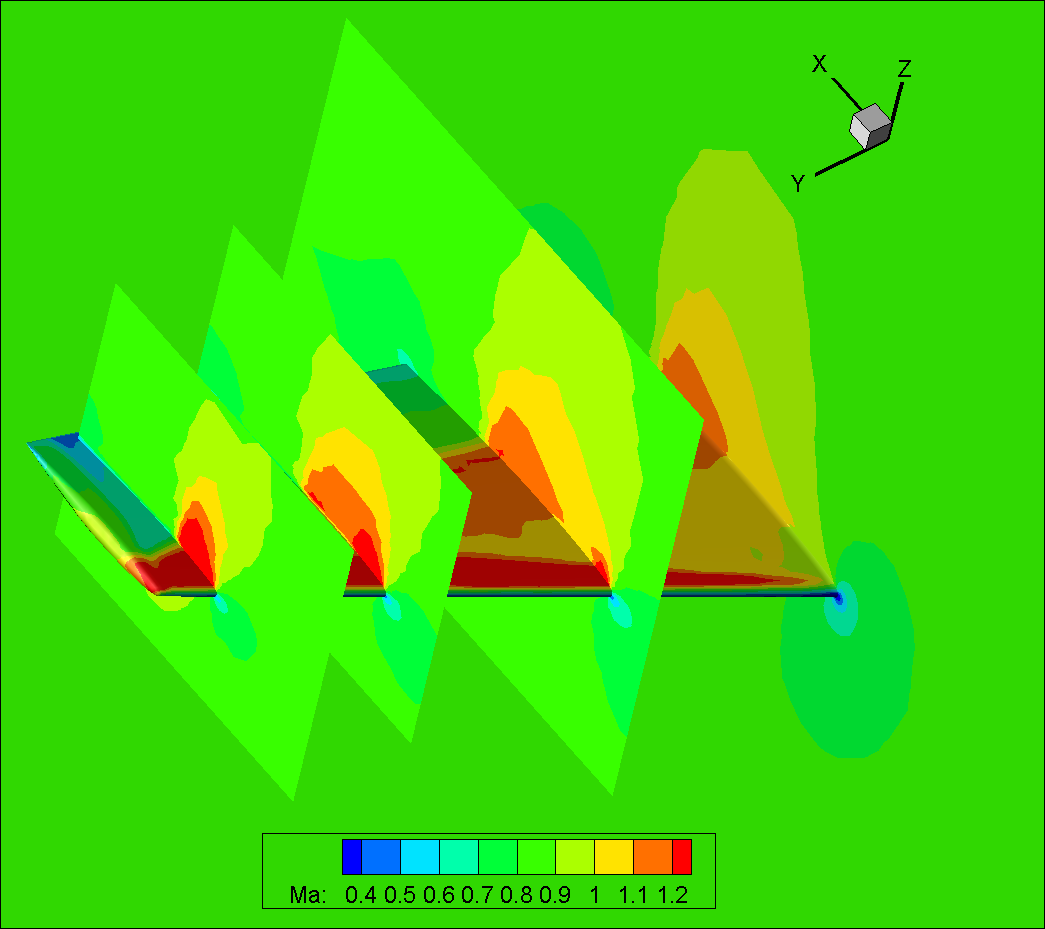}
	\caption{\label{M6 Contour}
		Transonic flow around a M6 wing. Left: Pressure contour. Right: Mach number contour.}
\end{figure*}

\begin{figure}[htp]	
	\centering	
	\includegraphics[height=0.35\textwidth]{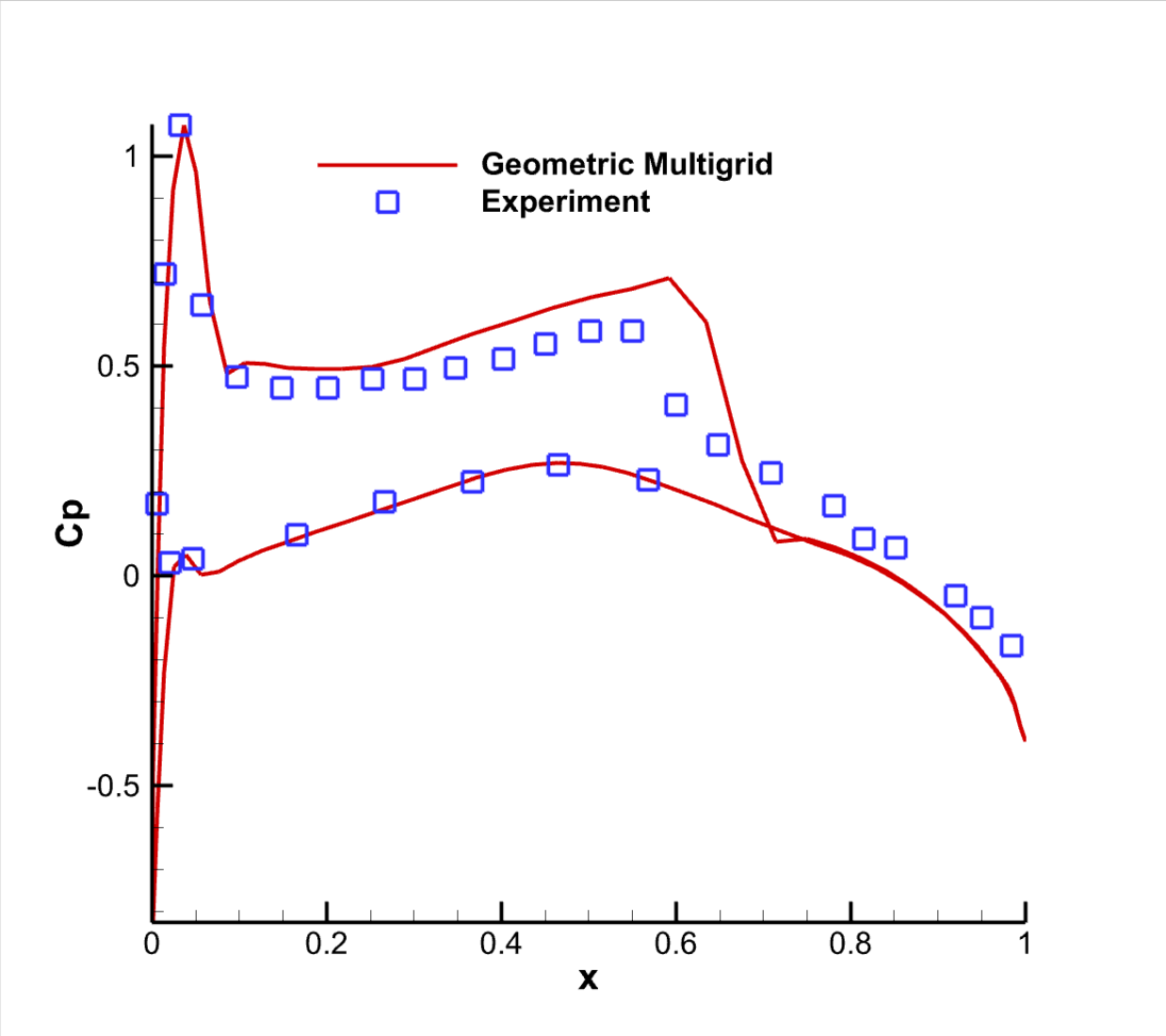}
	\includegraphics[height=0.35\textwidth]{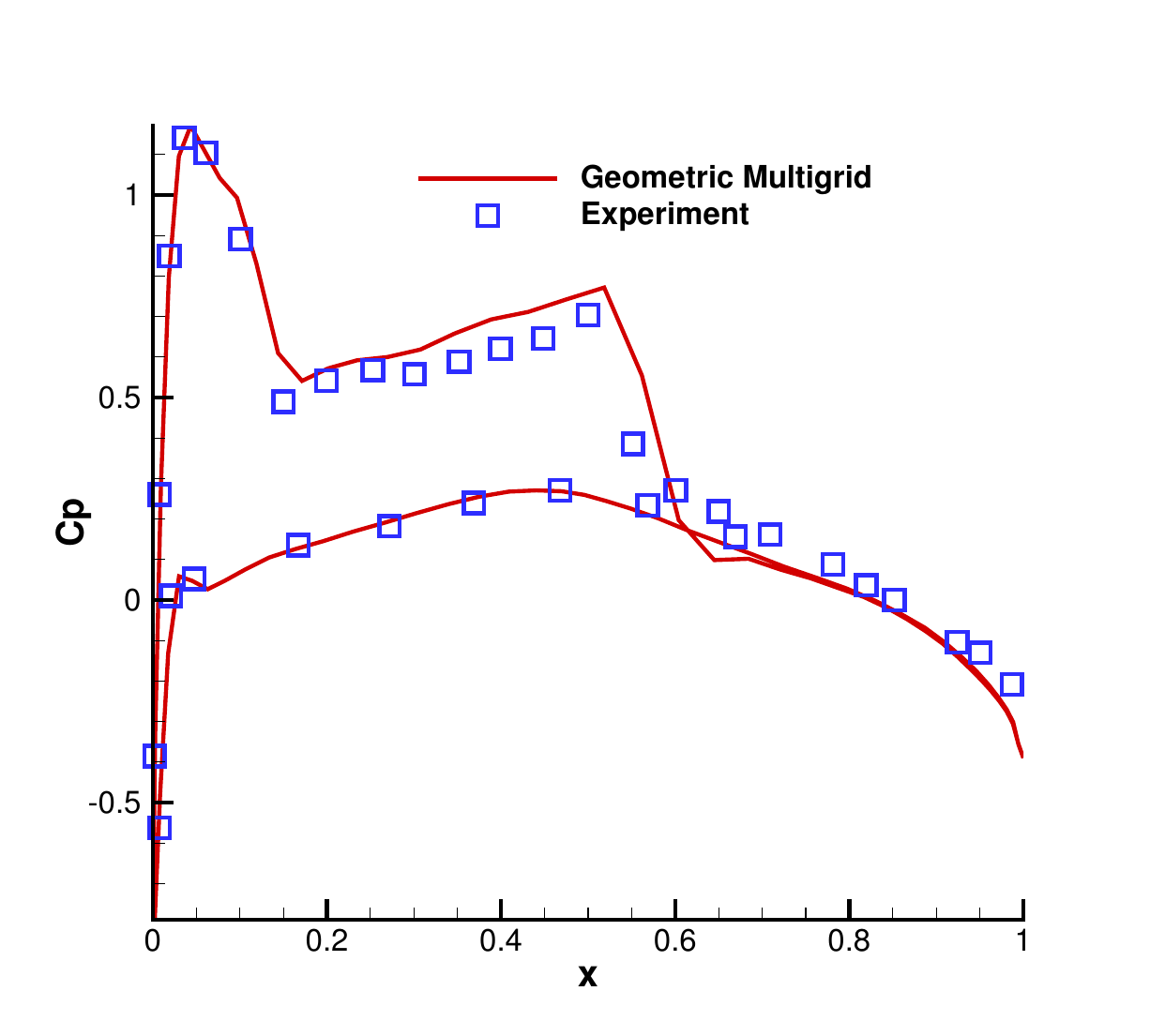}
	\includegraphics[height=0.35\textwidth]{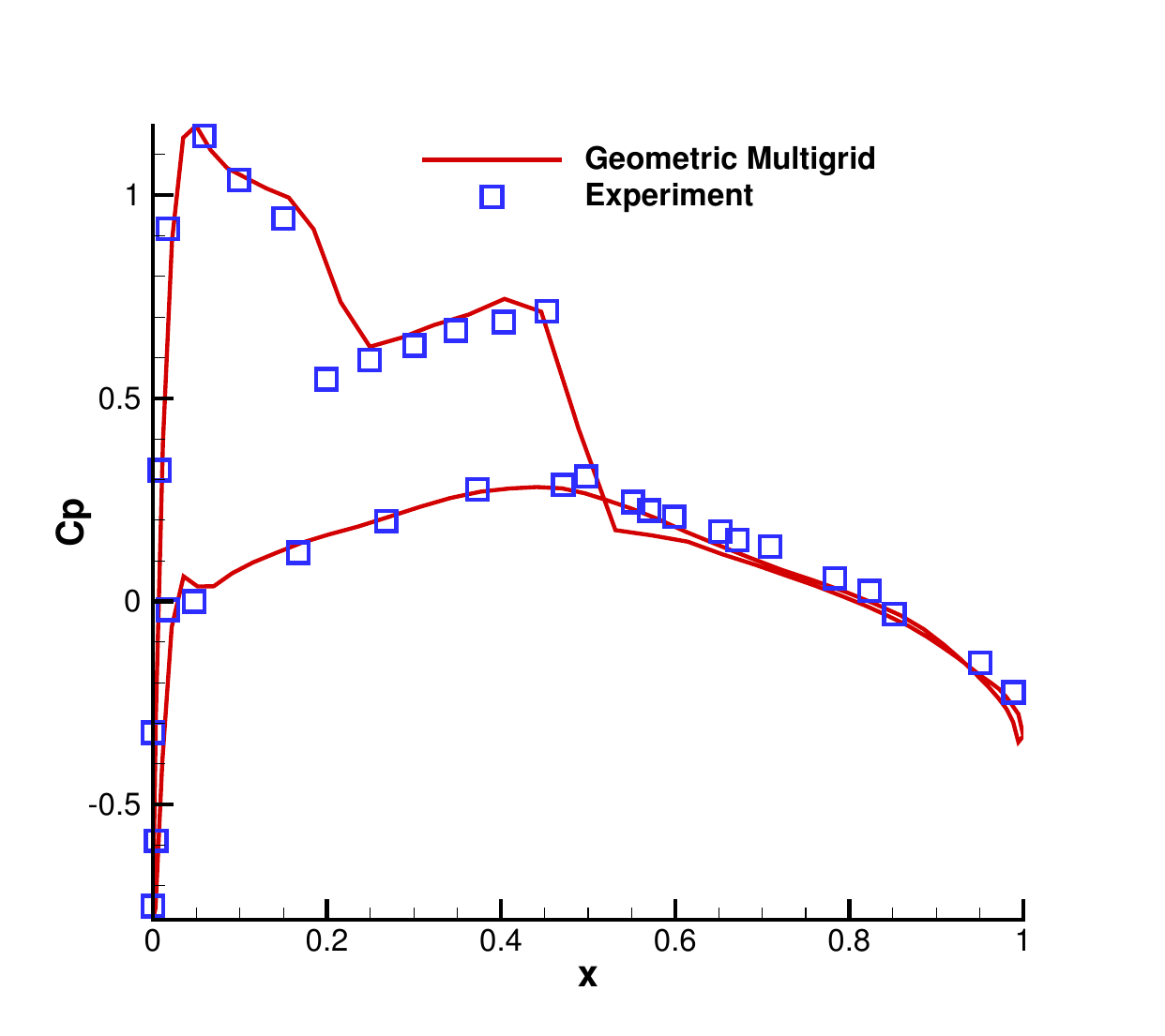}
	\includegraphics[height=0.35\textwidth]{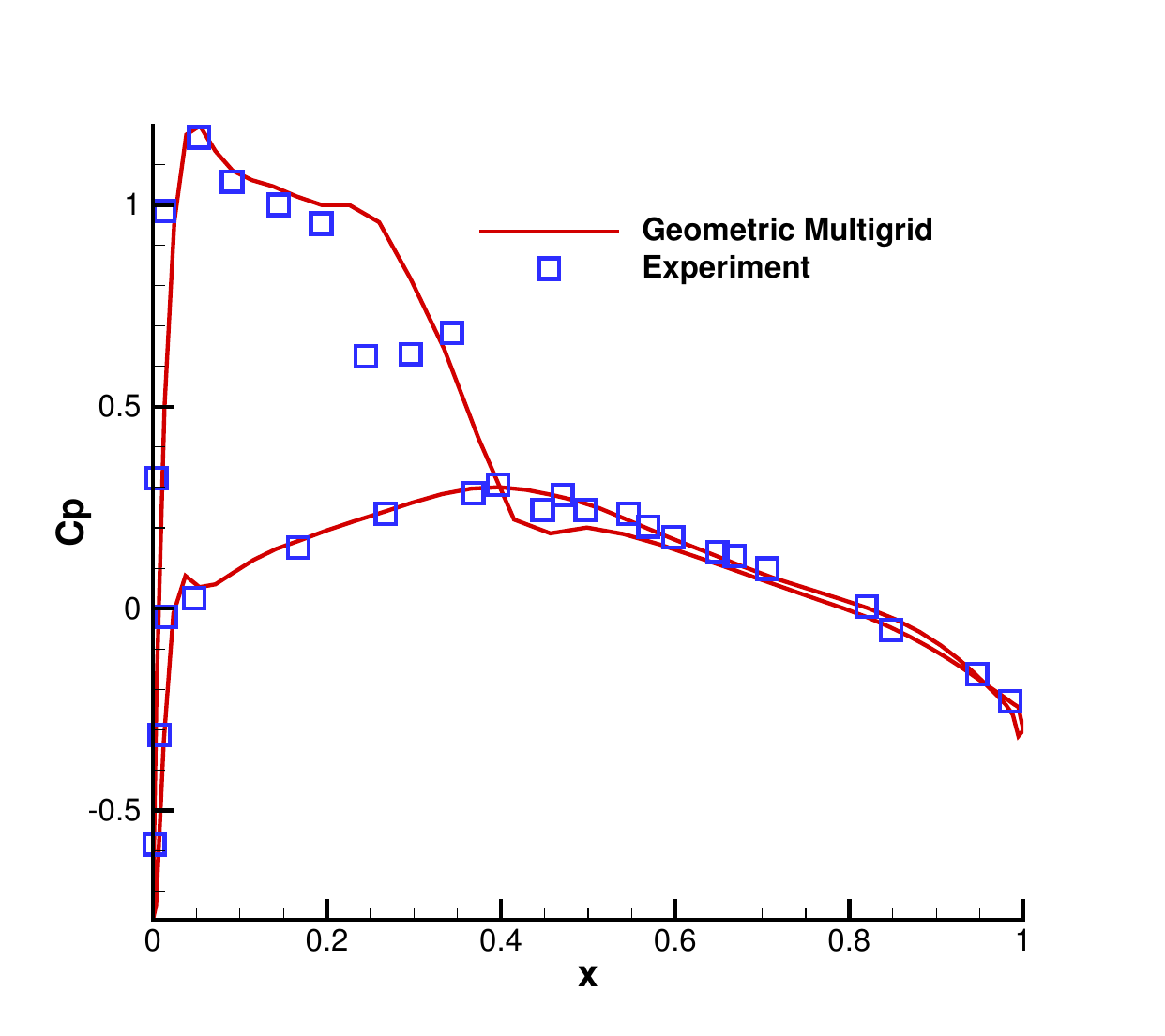}
	\includegraphics[height=0.35\textwidth]{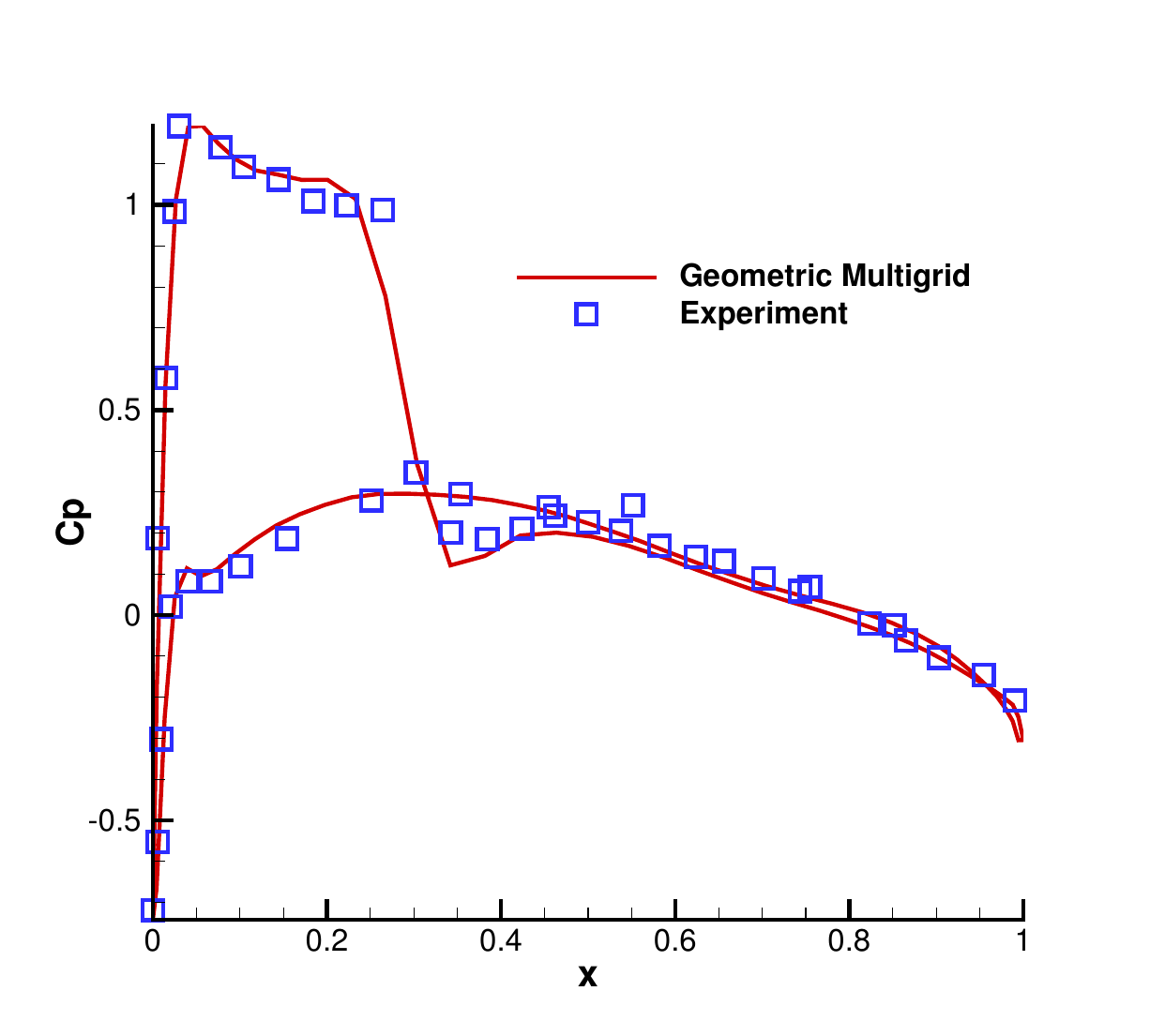}
	\includegraphics[height=0.35\textwidth]{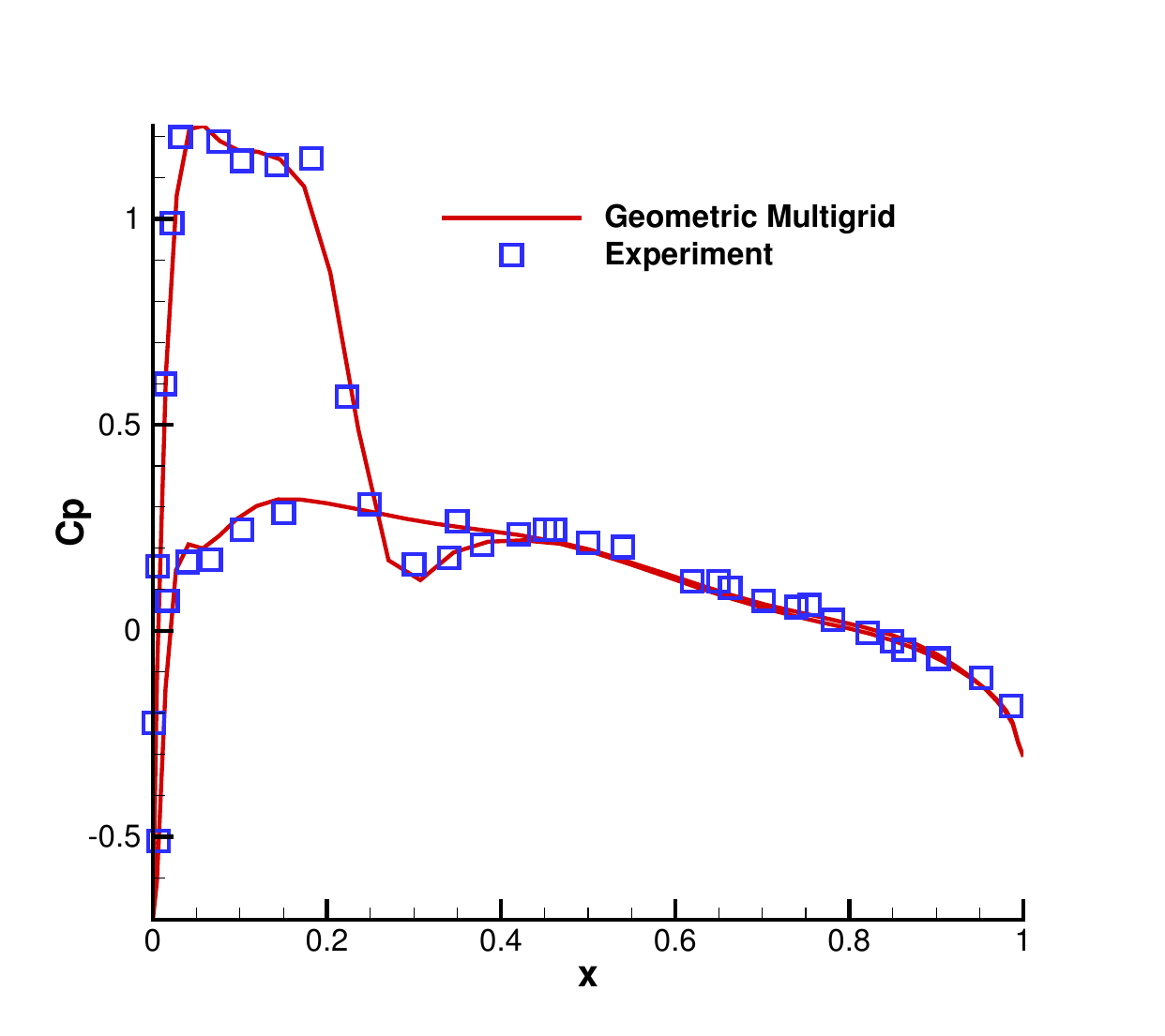}
	\caption{\label{M6 Contour-Cp}
		Transonic flow around a M6 wing. Cp distribution}
\end{figure}

Transonic flow around an ONERA M6 \cite{eisfeld2006onera} wing is a standard engineering case to verify the acceleration techniques used in CFD \cite{yang2023implicit}. The flow structure of it is complicated due to the interaction of shock and wall boundary. What is more, three-dimensional mixed unstructured mesh is also a challenge to high-order schemes with acceleration techniques. Thus, it is a appropriate test case to verify the efficiency and robustness of the geometric-multigrid CGKS. The farfield Mach number is set to be 0.8395 and the angle of attack is set to be 3.06$^{\circ}$. The adiabatic slip wall boundary is used on the surface of the ONERA M6 wing and the subsonic inflow boundary is set according to the local Riemann invariants.  A three layer "V" cycle p-multigrid geometric multigrid is used to solve the case. A hybrid unstructured mesh with a near-wall size $h\approx 2e^{-3}$ is used in the computation, as shown in Fig.~\ref{M6 Mesh and convergence}. The iteration steps of the geometric-multigrid CGKS is about 1/20  of the iteration steps of the explicit CGKS. Also, compared with explicit CGKS a 6.70 times speedup of CPU time is achieved, as shown in Fig.~\ref{M6 Mesh and convergence}. The pressure distribution on the wall surface and the Mach number slices at different wing sections are shown in Fig.~\ref{M6 Contour}. The pressure and Mach number contour in Fig.~\ref{M6 Contour} indicates that the p-multigrid geometric multigrid CGKS has captured the shock accurately. The quantitative comparisons on the pressure distributions at the semi-span locations Y /B = 0.20, 0.44, 0.65, 0.80, 0.90 and 0.95 of the wing are given in Fig.~\ref{M6 Contour-Cp}. The numerical results quantitatively agree well with the experimental data.

\subsection{Subsonic flow around a NACA0012 airfoil}

\noindent{\sl{(a) viscous flow with zero angle of attack}}

\begin{figure}[htp]	
	\centering	
	\includegraphics[height=0.35\textwidth]{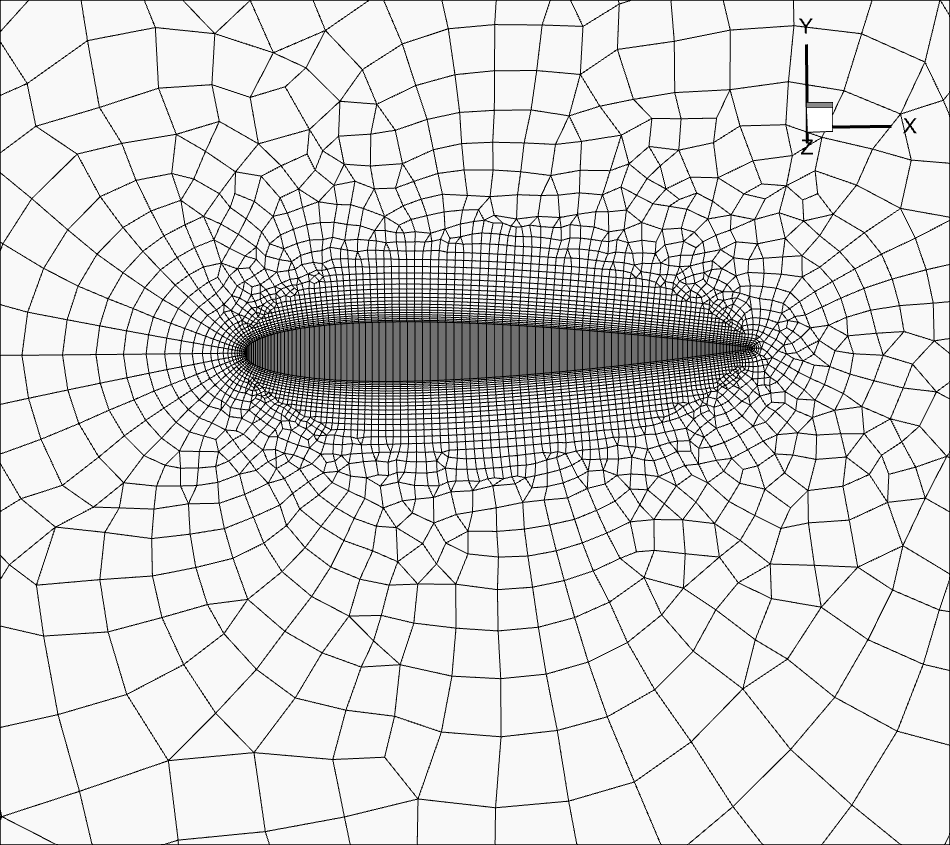}
	\includegraphics[height=0.35\textwidth]{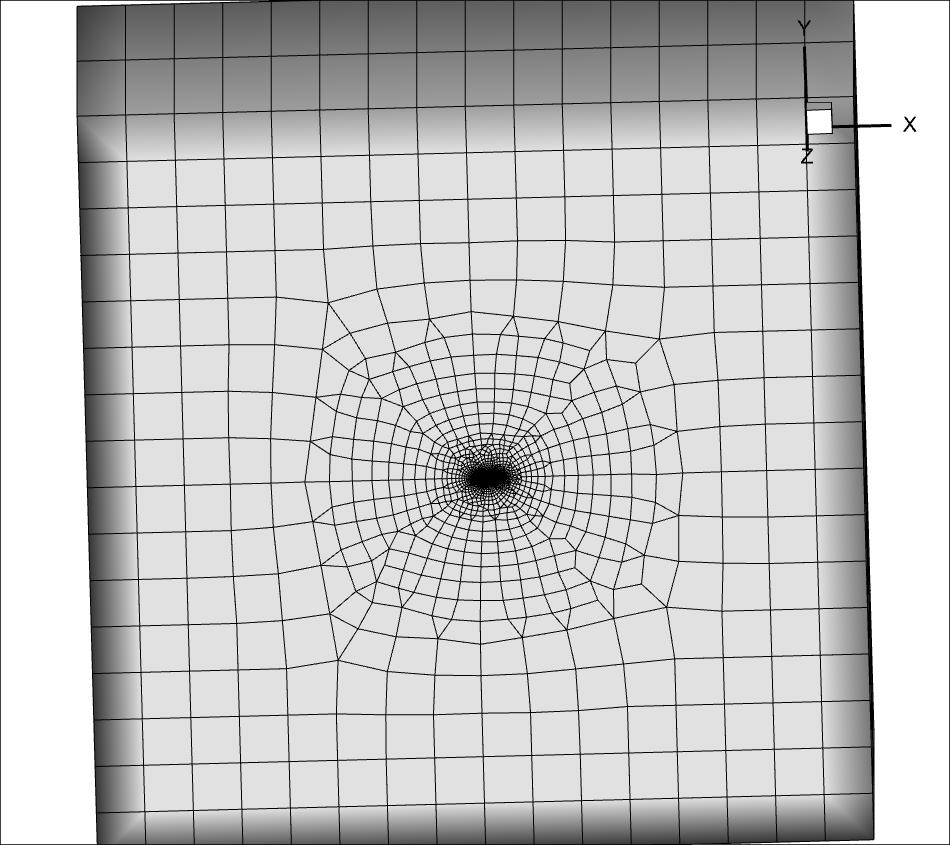}
	\caption{\label{naca0012 mesh}
		NACA0012 airfoil Mesh.}
\end{figure}

\begin{figure}[htp]	
	\centering	
	\includegraphics[height=0.35\textwidth]{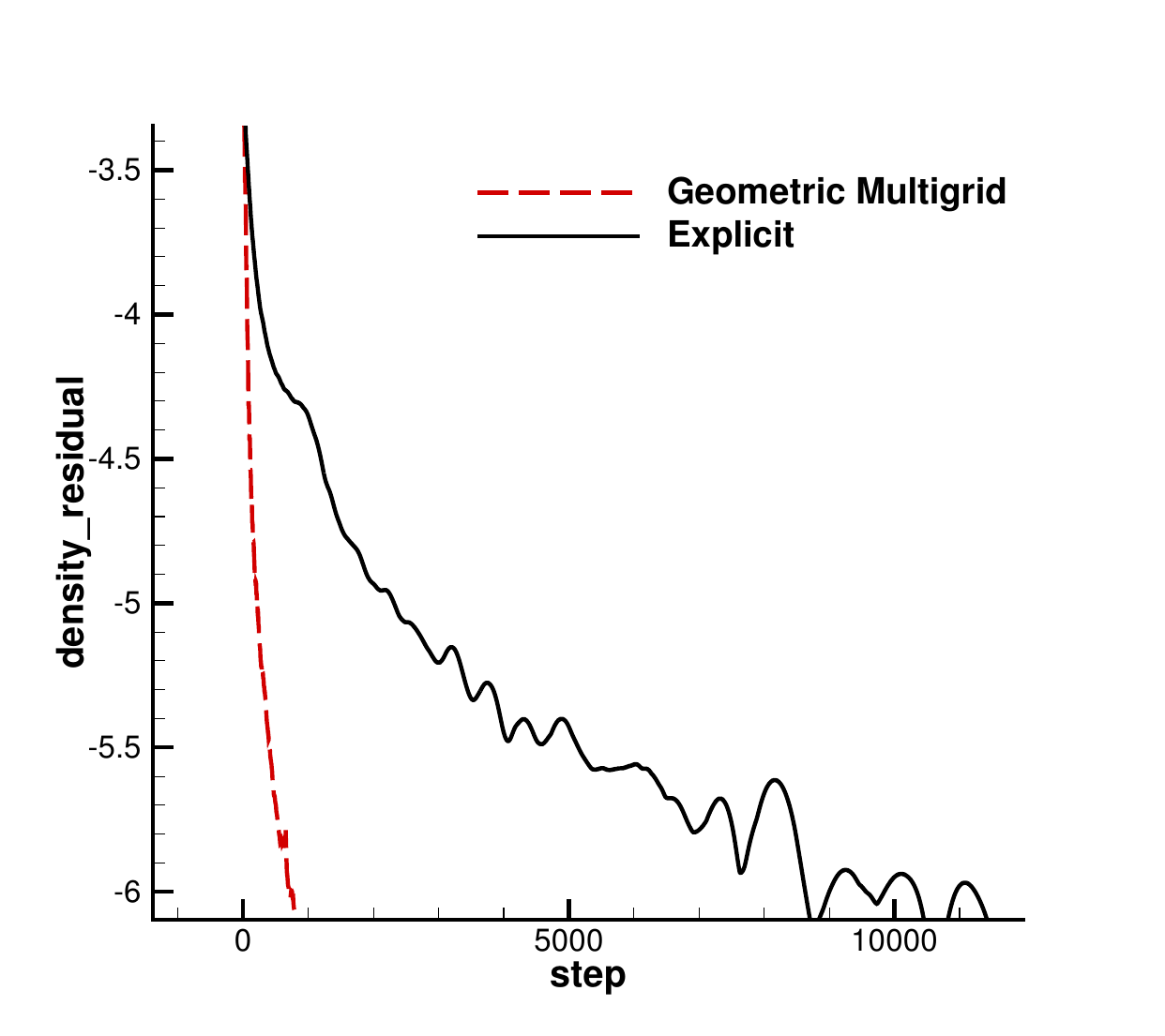}
	\includegraphics[height=0.35\textwidth]{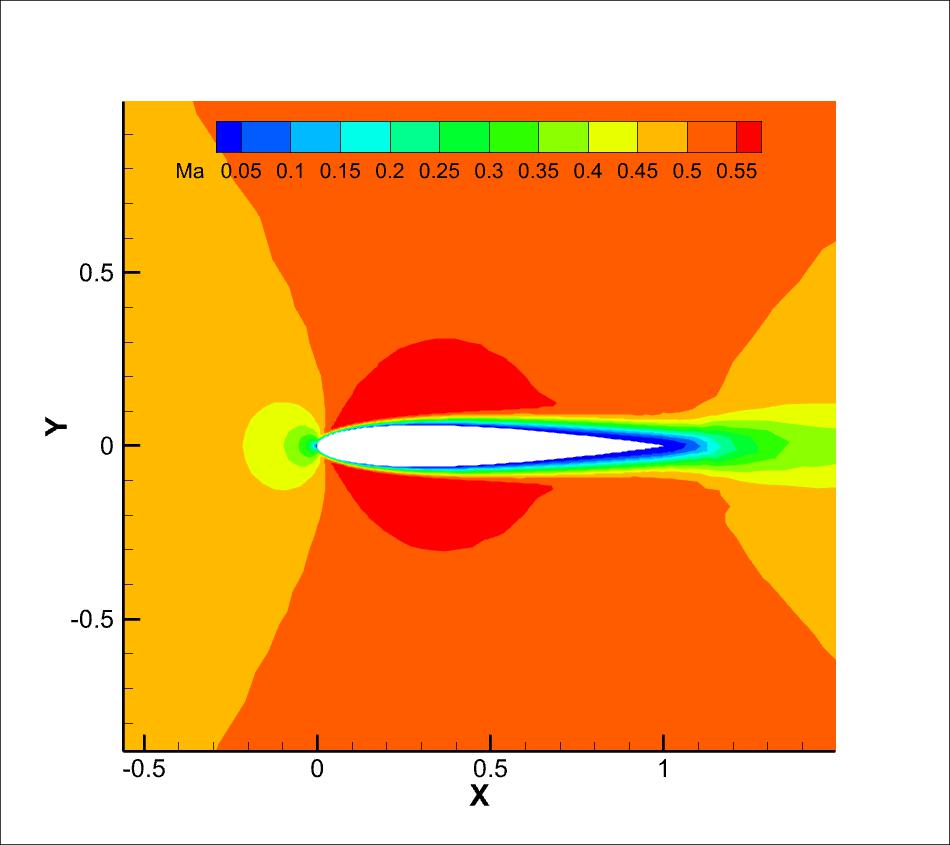}
	\caption{\label{naca0012 result}
		Subsonic flow around a NACA0012 airfoil. Left: convergence history. Right: Mach number contour.}
\end{figure}

\begin{figure}[htp]	
	\centering	
	\includegraphics[height=0.35\textwidth]{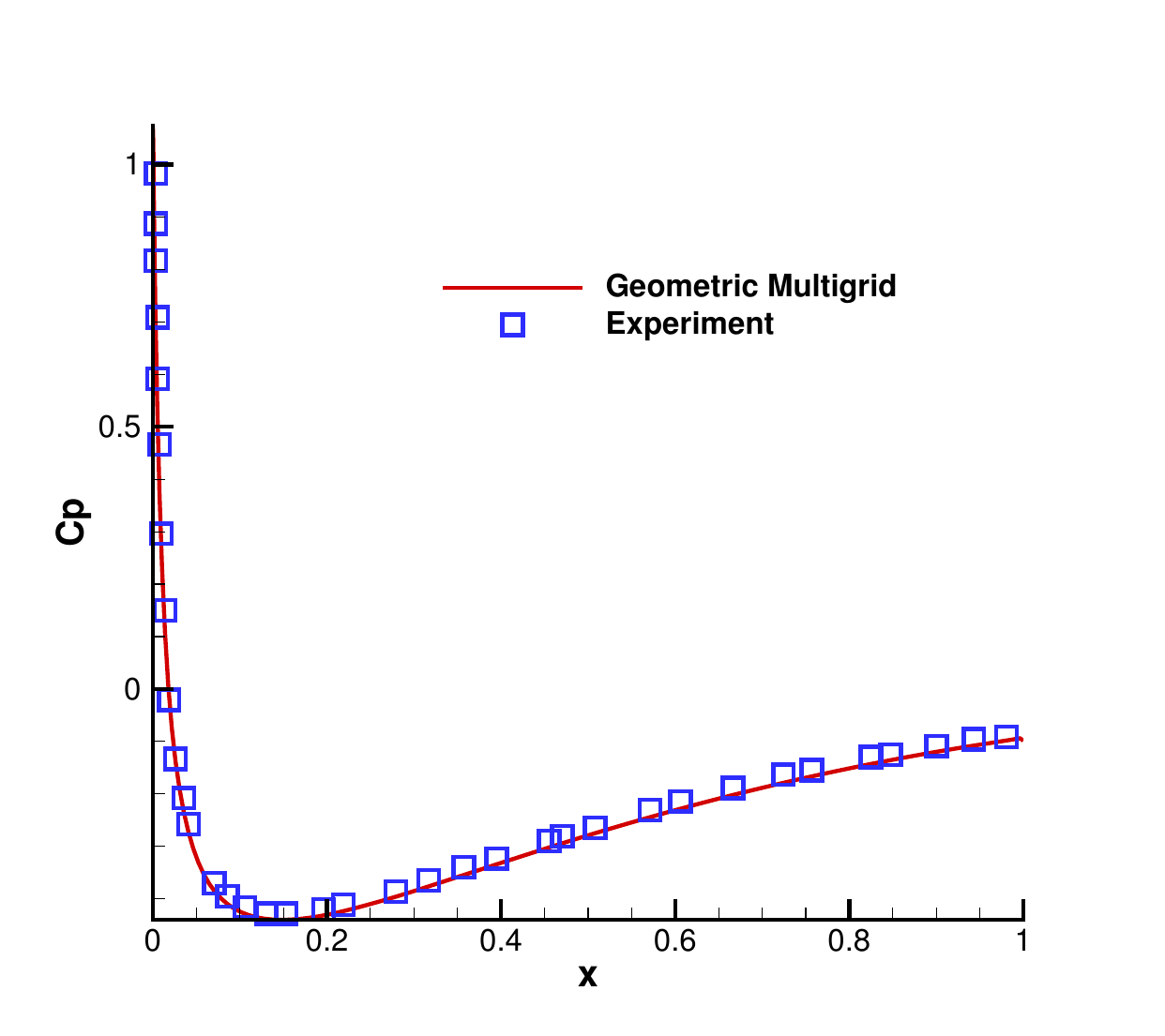}
	\caption{\label{naca0012 cp}
		Subsonic flow around a NACA0012 airfoil. Surface pressure coefficient distribution.}
\end{figure}

In this section, viscous flow around a NACA0012 airfoil is simulated. The incoming Mach number is set to be 0.5 and the incoming Reynolds number is set to be 5000 based on the chord length L=1. The subsonic farfield is calculated by Riemann invariants and the solid wall of the airfoil is set to be a adiabatic non-slip wall. Total 6538 $\times$ 2 hybrid prismatic cells are used in a cuboid domain [-15, 15] $\times$ [15, 15] $\times$ [0, 0.1]. The hybrid unstructured mesh is shown in Fig.~\ref{naca0012 mesh}.  A three layer "V" cycle p-multigrid geometric multigrid is used to solve the case. From the residual result shown in Fig.~\ref{naca0012 result}, the iteration steps of the geometric-multigrid CGKS is about 1/12  of the iteration steps of the explicit CGKS. Also, a 3.89 times speedup in CPU time is achieved compared with explicit CGKS. The Mach number contour is shown in Fig.~\ref{naca0012 result}. Quantitative result including the surface pressure coefficient is extracted and plotted in Fig.~\ref{naca0012 cp}, which highly agrees with the Ref \cite{bassi1997ns}.

\noindent{\sl{(b) viscous flow with an angle of attack}}

\begin{figure}[htp]	
	\centering	
	\includegraphics[height=0.25\textwidth]{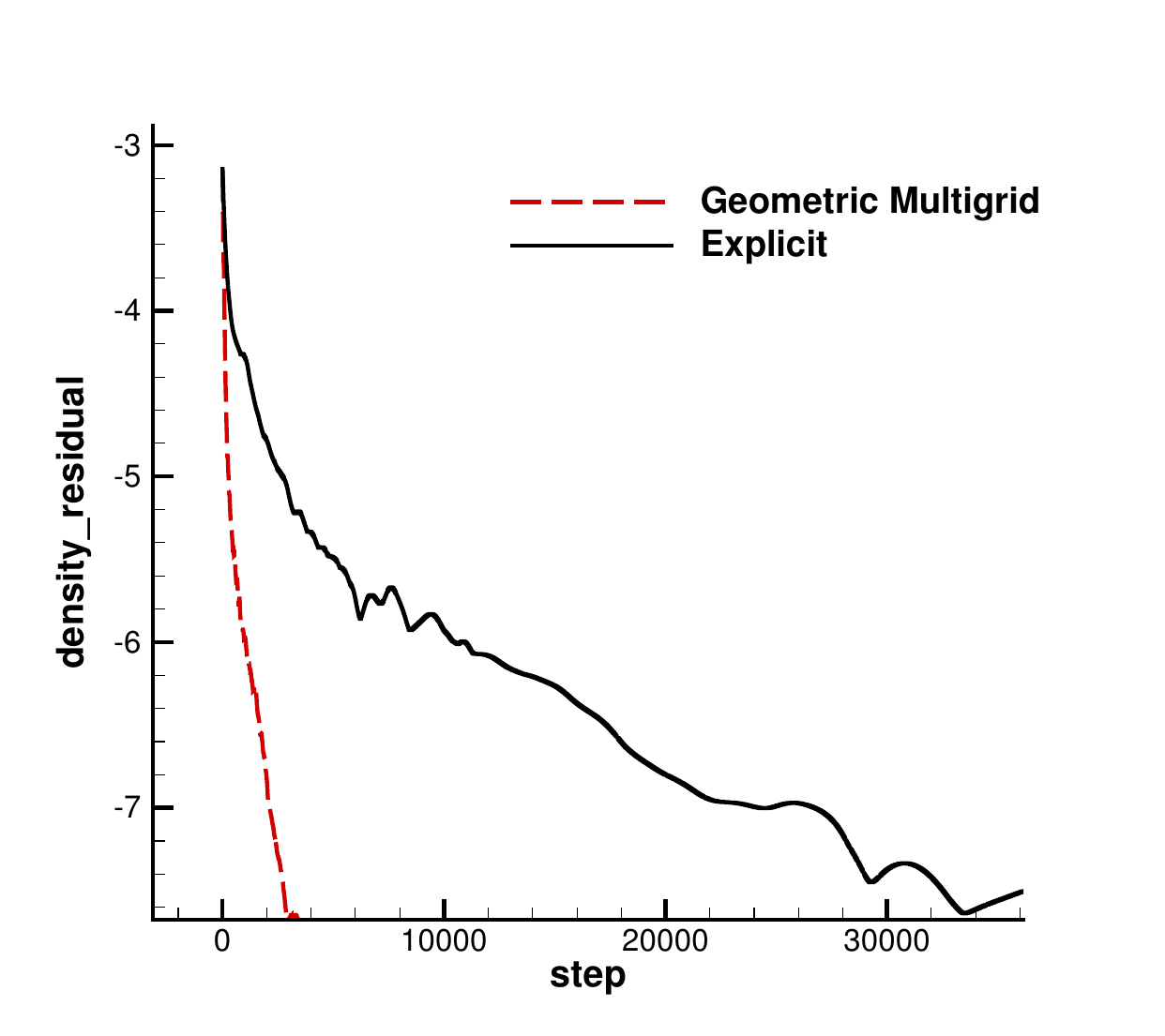}
	\includegraphics[height=0.25\textwidth]{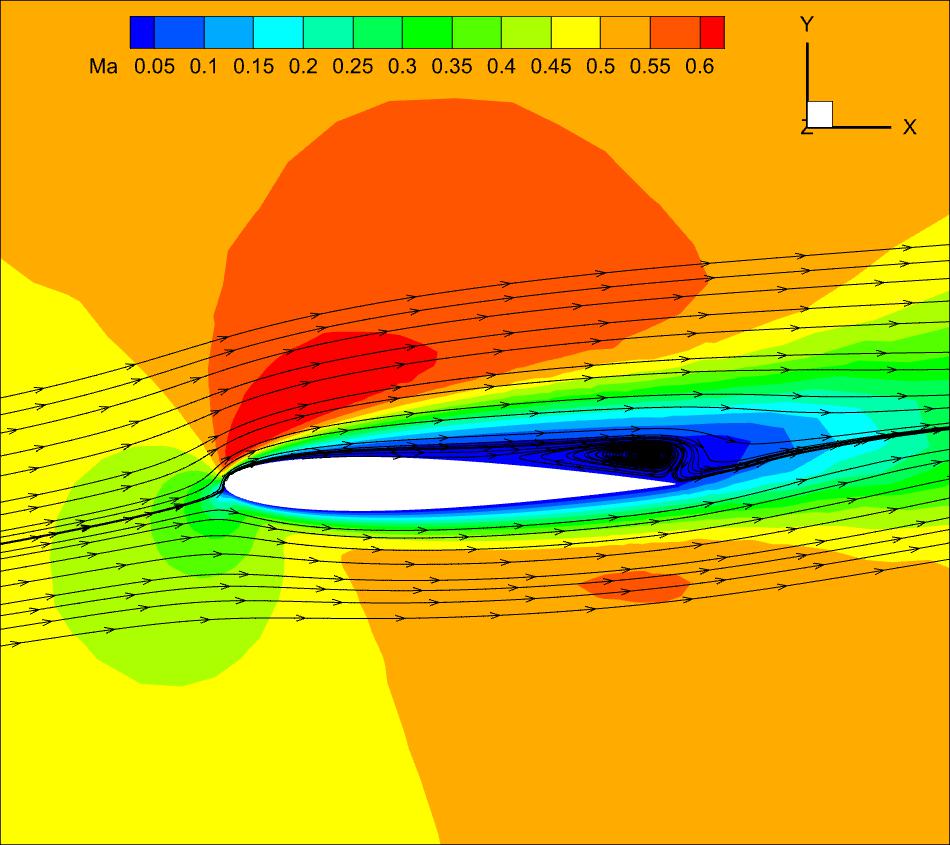}
	\includegraphics[height=0.25\textwidth]{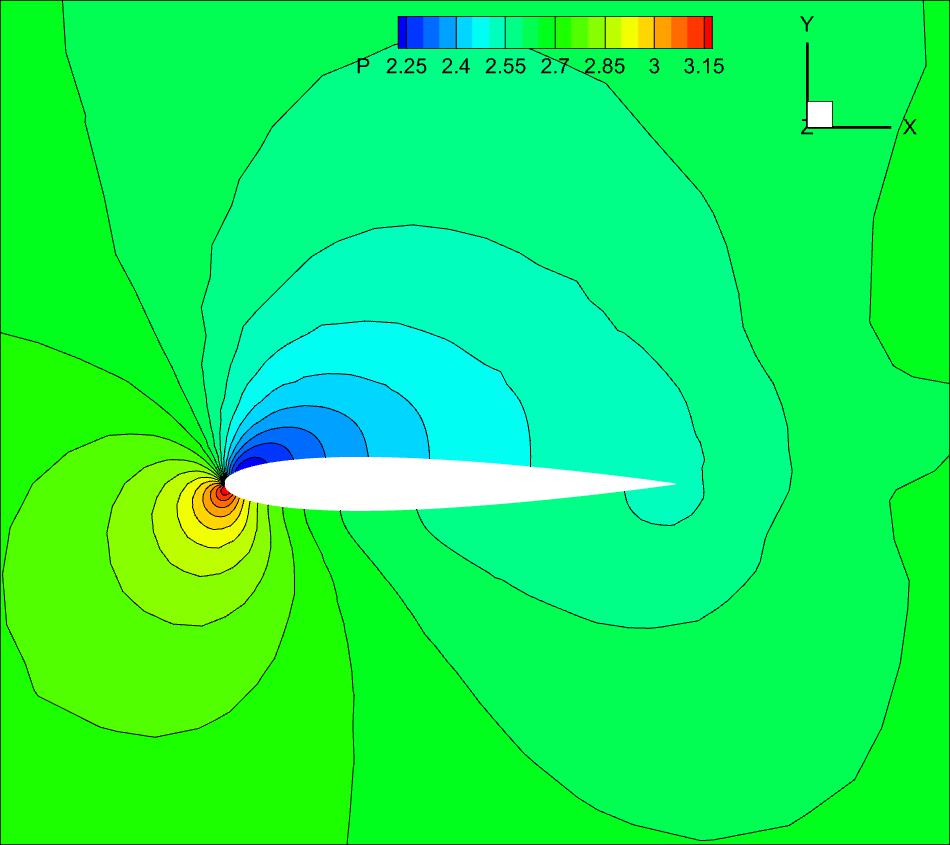}
	\caption{\label{naca0012 aoa result}
		Subsonic flow around a NACA0012 airfoil with an angle of attack. Left: Convergence history. Mid: Mach number contour. Right: Pressure contour.}
\end{figure}

In this case, except the angle of attack is set to be 8 $^{\circ}$, flow conditions are the same with the case of zero angle.  A three layer "V" cycle p-multigrid geometric multigrid is used to solve the case. From the convergence history shown in Fig.~\ref{naca0012 aoa result}, the iteration steps of the geometric-multigrid CGKS is about 1/11  of the iteration steps of the explicit CGKS. Also, a 3.8 times speedup in CPU time is achieved compared with explicit CGKS. The Mach number contour with streamline through the airfoil and the pressure contour are shown in Fig.~\ref{naca0012 aoa result}.

\subsection{Transonic flow around dual NACA0012 airfoils}
\begin{figure}[htp]	
	\centering	
	\includegraphics[height=0.35\textwidth]{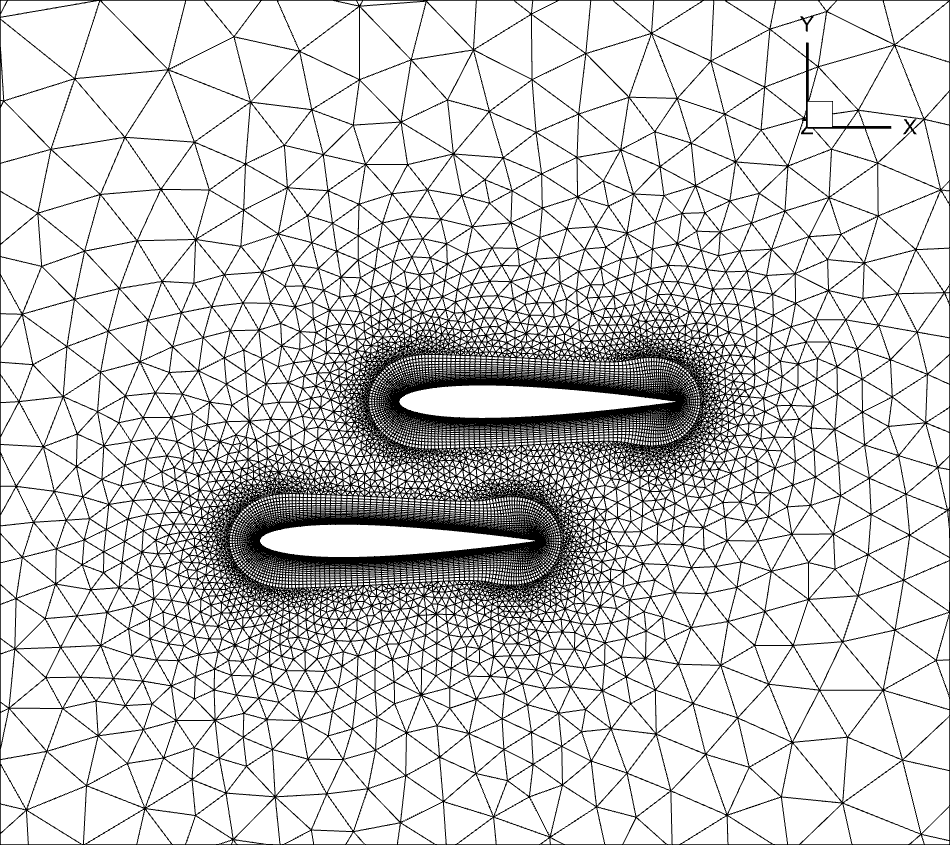}
	\includegraphics[height=0.35\textwidth]{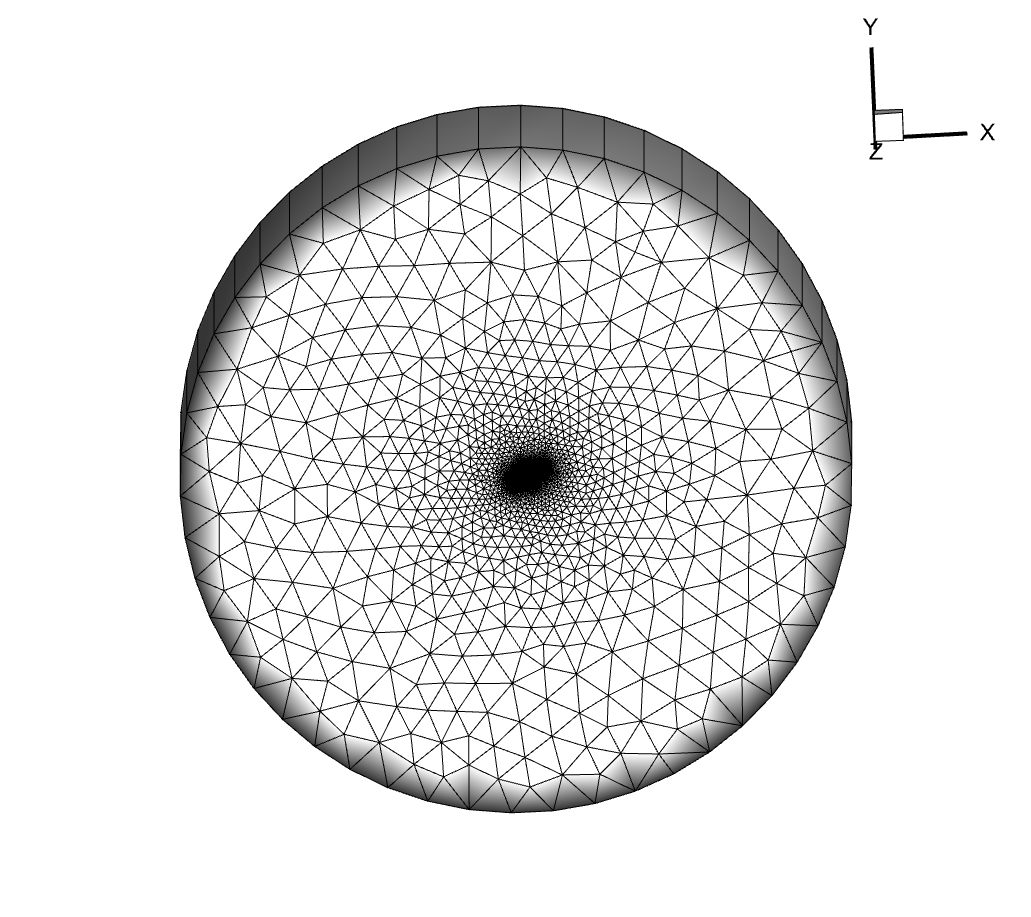}
	\caption{\label{dual naca0012 mesh}
		Transonic flow around dual NACA0012 airfoils. Mesh.}
\end{figure}

\begin{figure}[htp]	
	\centering	
	\includegraphics[height=0.35\textwidth]{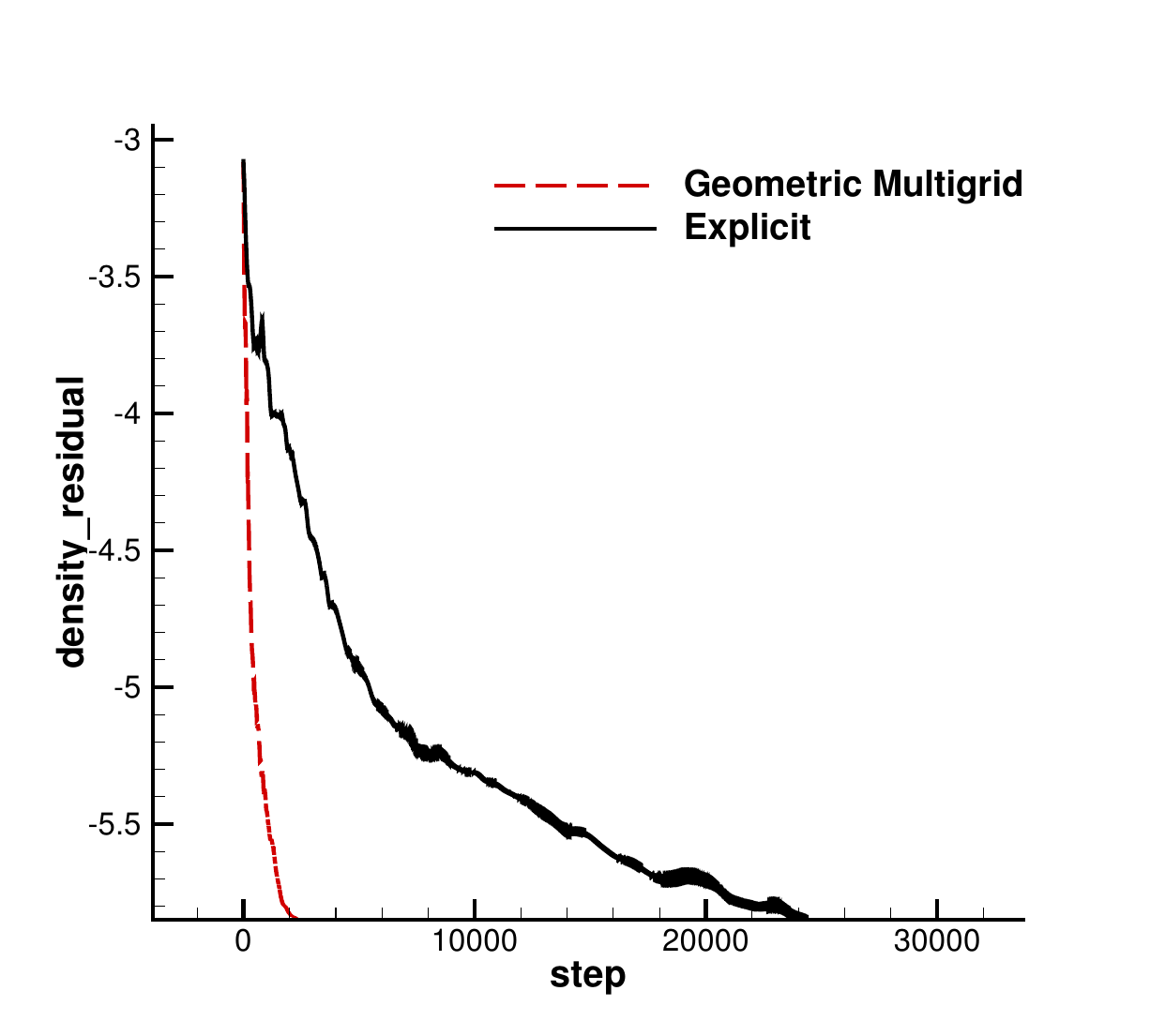}
	\caption{\label{dual naca0012 residual}
		Transonic flow around dual NACA0012 airfoils. Convergence history.}
\end{figure}

\begin{figure}[htp]	
	\centering	
	\includegraphics[height=0.35\textwidth]{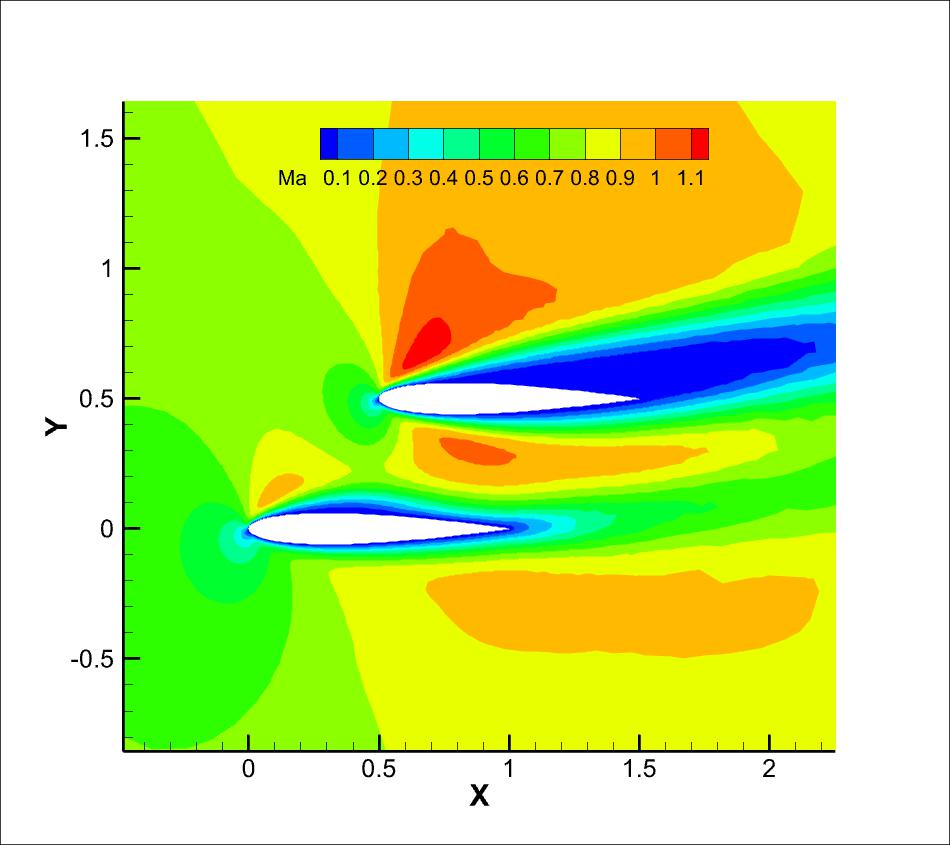}
	\includegraphics[height=0.35\textwidth]{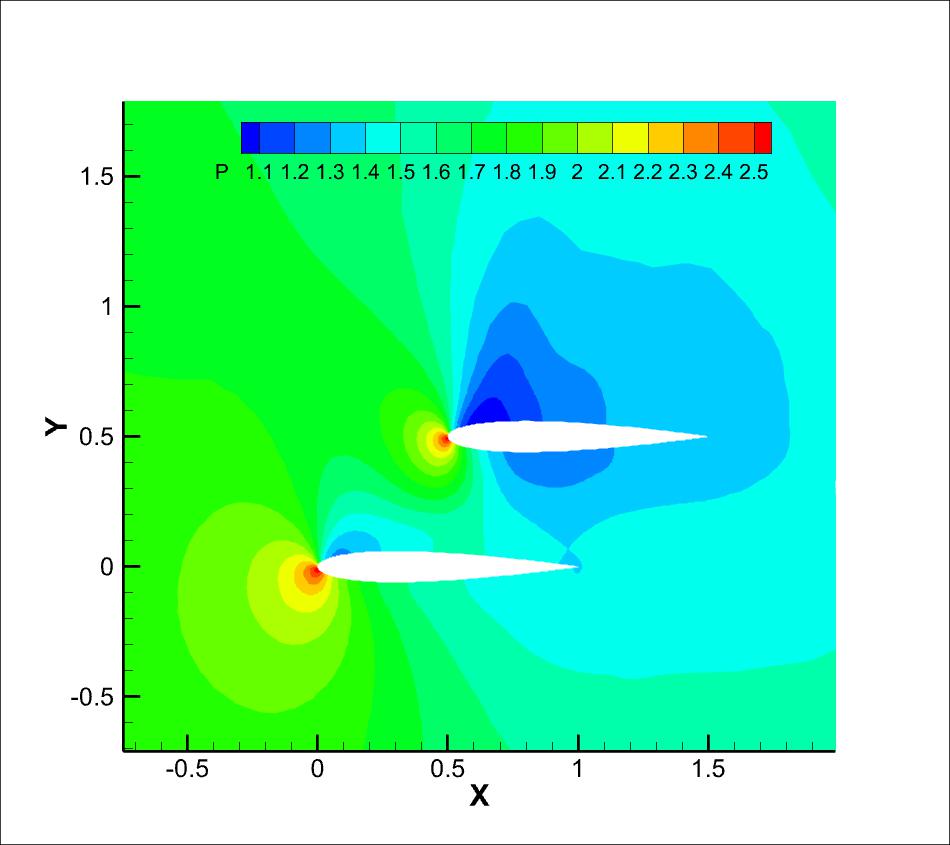}
	\caption{\label{dual naca0012 result}
		Transonic flow around dual NACA0012 airfoils. Left: Mach number distribution. Right: Pressure distribution.}
\end{figure}

\begin{figure}[htp]	
	\centering	
	\includegraphics[height=0.35\textwidth]{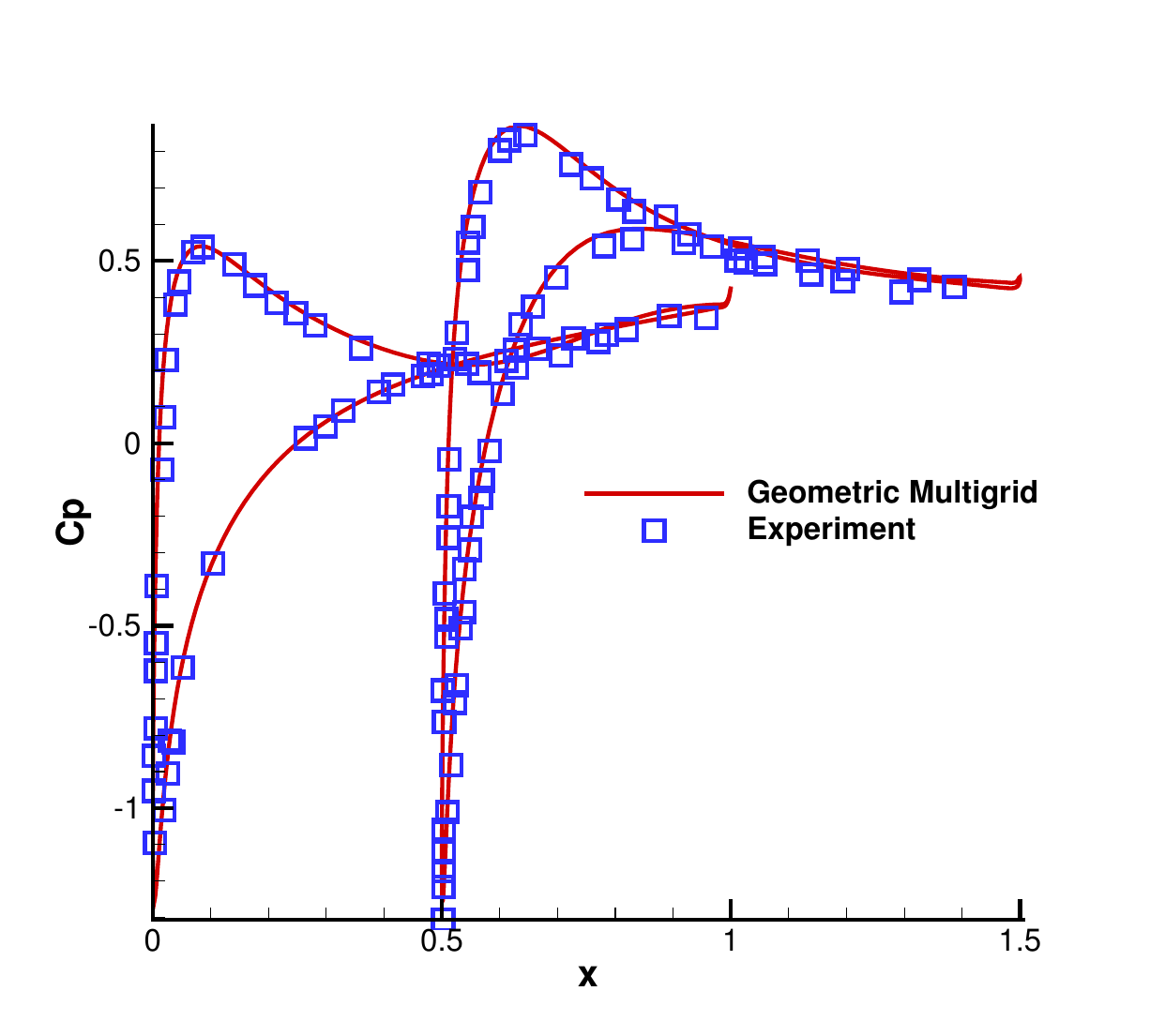}
	\caption{\label{dual naca0012 cp}
		Transonic flow around dual NACA0012 airfoils. Surface pressure coefficient distribution.}
\end{figure}

To further verify that the geometric-multigrid CGKS has a good performance both on complex geometry and discontinuity, transonic flow around dual NACA0012 airfoils is simulated. The head of the first airfoil is located at (0, 0) and the second one is located at (0.5, 0.5). Both airfoils are put in parallel with the x-axis. The incoming Mach number is set to be 0.8 with an angle of attack AOA = 10 $^{\circ}$ and the Reynolds number is set to be 500 based on the chord length L =1. The mesh consists 28678 mixed elements. The near wall size of the mesh is set to be h = 2 $\times$ $10^{-3}$, which indicates that the grid Reynolds number is 2.5 $\times$ $10^{5}$. The farfield boundary condition is set to be subsonic inflow using Riemann invariants and the wall is set to be a non-slip adiabatic wall. The mesh is presented in Fig.~\ref{dual naca0012 mesh}. In this case,  the iteration steps of the geometric-multigrid CGKS is about 1/24  of the iteration steps of the explicit CGKS. Also, a 8 times speedup of CPU time is achieved compared with explicit CGKS shown in Fig.~\ref{dual naca0012 residual}. The Mach number distribution and the pressure distribution are shown in Fig.~\ref{dual naca0012 result}. The oblique shock wave can be observed at the front of the top airfoil. The surface pressure coefficient is also extracted and compared with the reference data \cite{jawahar2000high}, as shown in Fig.~\ref{dual naca0012 cp}. The result obtained by geometric-multigrid CGKS agrees well with the experimental data.


\section{Conclusions}
In this study, we integrate a geometric multigrid approach with the compact gas-kinetic scheme (CGKS) to enhance computational efficiency. A rapid coarse mesh generation technique minimizes the time typically required for mesh coarsening. We employ a three-layer V-cycle algorithm to hasten the scheme's convergence, using a single-step third-order CGKS for the finest mesh, while a first-order kinetic flux vector splitting scheme with local time stepping addresses low-frequency errors on coarser levels.
The parallel communication strategy for the coarse mesh is both straightforward and time-efficient. We demonstrate the method's performance on 3-D hybrid unstructured meshes, achieving a three to tenfold reduction in CPU time compared to the conventional explicit CGKS. Notably, the method preserves result consistency across serial and parallel computations, suggesting its suitability for large-scale applications, including multi-GPU acceleration. Our analysis indicates that the geometric multigrid method offers greater acceleration in transonic flows compared to supersonic ones, with quantitative results closely matching those from explicit methods. Future work will explore incorporating implicit schemes like the GMRES method into the geometric multigrid framework to further elevate the convergence rate.

\section*{Acknowledgments}

The current research is supported by National Science Foundation of China (12172316, 12302378, 92371201, 92371107),
Hong Kong Research Grant Council (16208021,16301222).
\color{black}

\section*{References}
\bibliographystyle{plain}%
\bibliography{jixingbib}

\end{document}